\documentclass{emulateapj}
\usepackage{bm}

\usepackage{graphicx}
\def\unit #1{\,{\rm #1}}

\def\angst{\unit{\AA}}

\begin {document}
\slugcomment{to appear in the Astrophysical Journal}
\shorttitle{ }
\shortauthors{ }



\title{Rest-frame UV versus optical morphologies of galaxies using S\'ersic profile fitting: the importance of morphological K-correction}

\author{Abhishek Rawat,\altaffilmark{1}
Yogesh Wadadekar,\altaffilmark{2}
Duilia De Mello,\altaffilmark{3,4,5}
}

\altaffiltext{1}{Inter University Centre for Astronomy and
Astrophysics, 
Post Bag 4, Ganeshkhind, Pune 411007, India: rawat@iucaa.ernet.in}
\altaffiltext{2}{National Centre for Radio Astrophysics, Post Bag 3, 
Ganeshkhind, Pune 411007, India.}
\altaffiltext{3}{Observational Cosmology Laboratory, Code 665, Goddard Space Flight Center, Greenbelt, MD 20771, USA}
\altaffiltext{4}{Catholic University of America, Washington, DC 20064, USA}
\altaffiltext{5}{Johns Hopkins University, Baltimore, MD 21218, USA}

\begin{abstract}

We show a comparison of the rest-frame UV morphologies of a sample of
162 intermediate redshift ($z_{\rm median}=1.02$) galaxies with their
rest-frame optical morphologies. We select our sample from the deepest
near-UV image obtained with the Hubble Space Telescope (HST) using the
WFPC2 (F300W) as part of the parallel observations of the Hubble Ultra
Deep Field campaign overlapping with the HST/ACS GOODS dataset. We
perform single component S\'ersic fits in both WFPC2/F300W (rest-frame
UV) and ACS/F850LP (rest-frame optical) bands and deduce that the
S\'ersic index $n$ is estimated to be smaller in the rest-frame UV
compared to the rest-frame optical, leading to an overestimation of
the number of merger candidates by $\sim$40\%-100\% compared to the
rest-frame optical depending upon the cutoff in $n$ employed for
identifying merger candidates.  This effect seems to be dominated by
galaxies with low values of $n(F300W)\leq0.5$ that have a value of
$n(F850LP)\sim1.0$. We argue that these objects are probably clumpy
starforming galaxies or minor mergers, both of which are essentially
contaminants, if one is interested in identifying major mergers.

In addition we also find evidence that the axis ratio $b/a$ is lower, 
i.e. ellipticity $(1-b/a)$ is higher in rest-frame UV compared to the rest-frame optical.
Moreover, we find that in the rest-frame UV, 
the number of high ellipticity ($e\geq0.8$) objects 
 are higher by a factor of $\sim2.8$ compared to the rest-frame optical.
This indicates that the reported dominance of elongated morphologies among high-{\it z} LBGs
might just be a bias related to the use of rest-frame UV datasets in high-{\it z} studies. 
\end{abstract}

\keywords{galaxies: evolution - galaxies: formation - galaxies: statistics}

\section{Introduction}
\label{intro}
Most of our knowledge about the properties of high redshift galaxies
has been acquired by studying Lyman Break Galaxies (LBGs) at redshift
$\sim$2.0-5.0 (e.g. Giavalisco 2002; Steidel et
al. 2003; Ravindranath et. al. 2006; Wadadekar, Casertano \& de Mello 2006). Since
most of these studies are in the optical, they 
probe the rest-frame UV light of LBGs at these redshifts. Deriving
morphological information about a galaxy from its rest-frame UV light 
can be misleading as the UV light is mainly dominated by patchy
star forming regions that are not representative of the underlying 
galaxy mass distribution (e.g. Teplitz et al. 2006). Even so, people have used a variety of techniques to derive the morphologies of such high-z objects, including the non-parametric CAS (Conselice et al. 2003) and Gini/M20 coefficients (Lotz et al. 2004) and the parametric light profile fitting such as {\it Galfit} (Peng et al. 2002). While each of these methods have their pros and cons, all of them have been used in the past (and continue to be used) by the community for studying quantitative morphologies of high redshift galaxies. In light of this fact, it becomes imperative to be aware of the biases that are expected, as one probes progressively bluer rest-frame wavelengths at higher redshifts using each of these techniques. 

Papovich et al.~\cite{papovich2005} 
and Conselice et al.~\cite{conselice2005} have carefully examined the 
extent of this {\it morphological K-correction} using WFPC2 and NICMOS observations of the Hubble Deep Field North (HDF-N).
More recently, Conselice et al.~\cite{conselice2008} have examined the
effects of morphological K-correction on the estimation of
non-parametric CAS (Conselice et al. 2003) and Gini/M20 coefficients (Lotz et al. 2004), using ACS and NICMOS
imaging of the Hubble Ultra Deep Field. They find that CAS and
Gini/M20 parameters give similar results, except that the
latter tend to find larger number of mergers than the former. 
Despite the fact that {\it Galfit} has been used extensively for quantifying galaxy morphologies at high redshifts (eg. Ravindranath et al. 2006), there has been no systematic study evaluating the effects of morphological K-correction on the structural parameters derived using {\it Galfit}. It is this problem that we are trying to address in the current paper.
In this work, we quantify the biases that are encountered in 
estimating the morphological parameters of galaxies in the rest-frame
UV as compared to rest-frame optical light using {\it{Galfit}}, a structural decomposition code that 
fits 2-D parameterized, axisymmetric, functions directly to galaxy images.


\section {The Data and Sample selection}
\label{data}

We have utilized the HST/WFPC2 F300W band observations (orients 310
and 314) obtained in parallel with the HST/ACS HUDF that fall within
the GOODS-S area and constitute the deepest image of the near UV sky
ever obtained with the HST (total exposure $\sim$323.1 ksec).  The
dataset consists of a total of 409 WFPC2 F300W parallel images, with
exposure times ranging from 700 sec - 900 sec each.  A
cosmic-ray-rejected, drizzled image with a pixel scale of 0.06
arcsec/pixel was constructed by combining the 409 individual exposures
(de Mello et al. 2006; Wadadekar, Casertano \& de Mello 2006), using a
drizzle-based technique developed for data processing by the WFPC2
Archival Parallels Project (Wadadekar et al. 2006). This drizzled
image, which was accurately registered with respect to the GOODS
images, was then used for performing the morphological analysis
presented in this paper.

For the optical imaging, we have used the publicly available version
v1.0 of the reduced, calibrated images of the Chandra Deep Field South (CDFS) acquired with
HST/ACS as part of the {\it{Great Observatories Origins Deep Survey}}, 
GOODS (Giavalisco et al. 2004). The overlap between the HST/WFPC2 F300W image footprint and the HST/ACS GOODS image footprint is shown in Fig. 1 of de Mello et al.~\cite{demello2006}. 

\begin{figure}[t]
\includegraphics[height=7.0cm,clip]{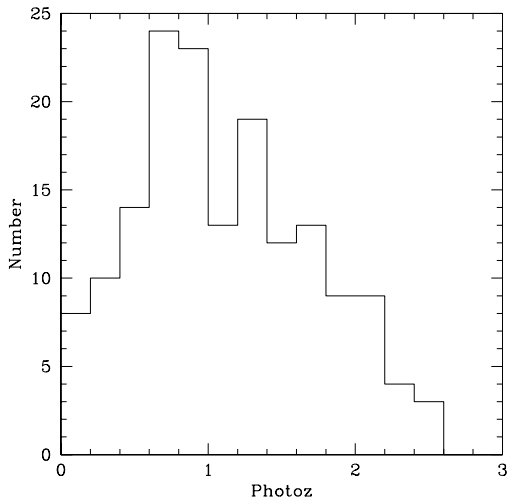}
\caption{The photometric redshift distribution of our 162 U-band selected galaxies
  with optical detection ($z_{\rm mag}\leq25.5$). The median photometric redshift is 1.02.}
\label{photoz}
\end{figure}

A SExtractor (Bertin \& Arnouts 1996) based source
catalog was prepared for the WFPC2/F300W image by de Mello et
al.~\cite{demello2006} containing 415 sources. We positionally matched this F300W-band based
catalog with the publicly available HST/ACS
version r1.1 multi-band source catalog to find optical counterparts
for the 415 UV bright sources. We only included optical counterparts of UV sources
brighter than $z_{\rm mag}=25.5$ and having a photometric redshift (photo-{\it z}) as 
determined by de Mello et al.~\cite{demello2006}. 
This yielded a sample of 162 sources having photometric redshift distribution as shown in Fig.~\ref{photoz}   

\section {The methodology}
\label{methodology}
Our aim is to quantify any biases in the derived galfit parameters, on account of using rest-frame UV light of program objects, as opposed to their rest-frame optical light.
In order to do this, we performed single component S\'ersic fits to our sample of 162 objects in both UV as well as optical. The goal is to look for systematic differences in the derived galaxy parameters as a function of their rest-frame wavelength.

\subsection{The F300W band fits}
\label{f300w}

We used the software {\it Galfit} to
carry out two-dimensional modeling for our sample of galaxies. 
We performed single component S\'ersic fits, where the intensity profile of the galaxy is modeled with the S\'ersic law (S\'ersic 1968), 




\begin {equation}
\label{sersic}
\Sigma(r)=\Sigma_e \mbox{exp}\Big[{-2.303b_{n} \Big[\Big({\frac{r}{r_e}}\Big)^{1/n} - 1\Big]}\Big]
\end {equation}

with $r_{e}$ being the half light radius, $\Sigma_e$ the surface
brightness at $r_{e}$ and $n$ being the S\'ersic index
and $b_n$ is related to $n$ such that $P(2n,2.303b_n) = 0.5$, where $P(a,b)$ is the incomplete gamma function. 
 
\begin{figure*}[t]
\centering
\includegraphics[height=5.5cm,clip]{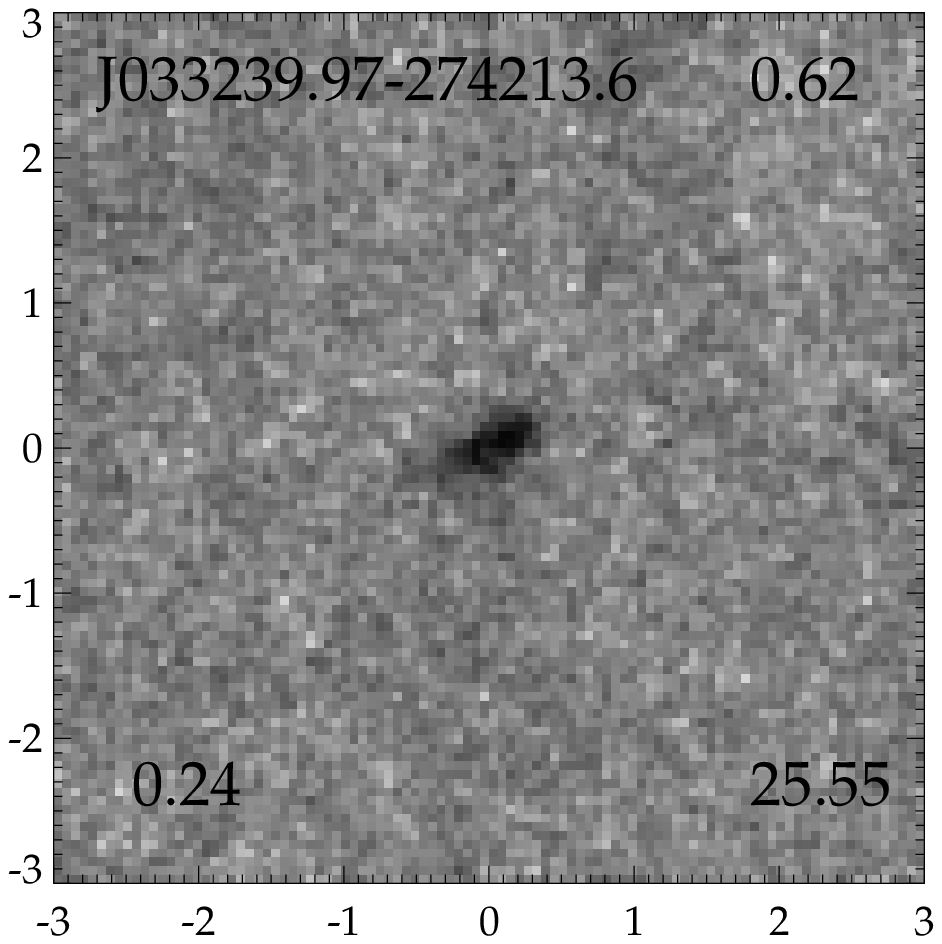}%
\includegraphics[height=5.5cm,clip]{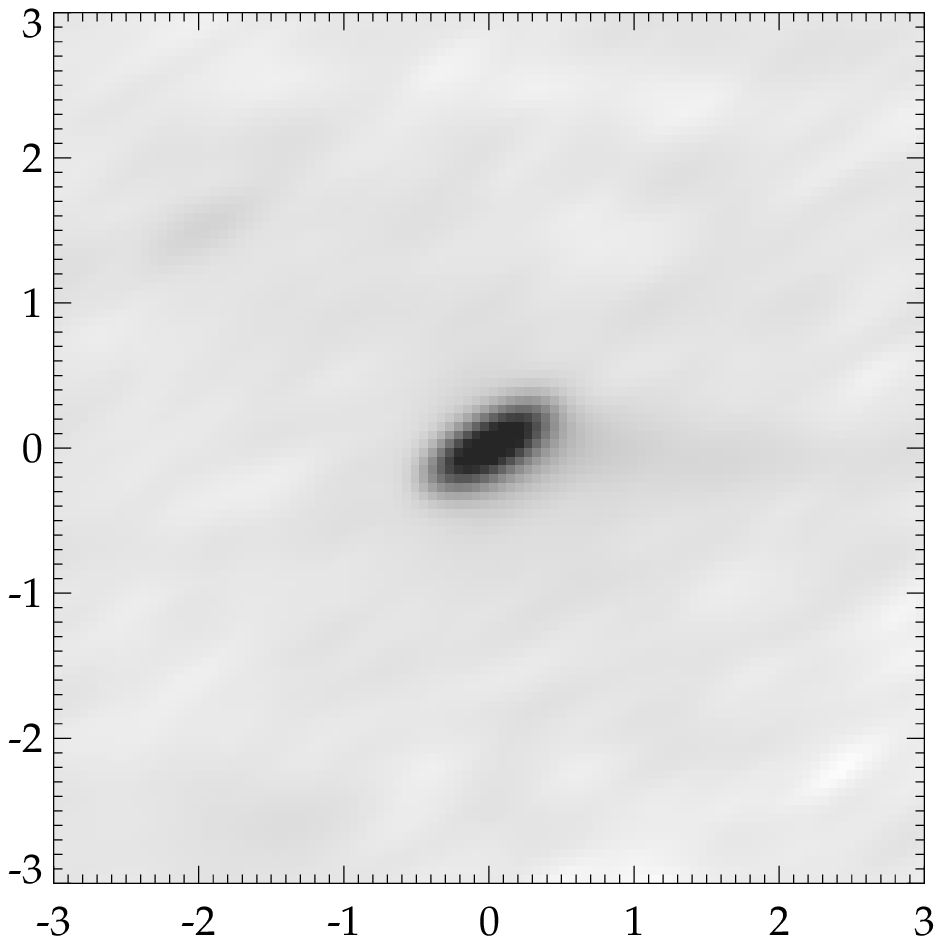}%
\includegraphics[height=5.5cm,clip]{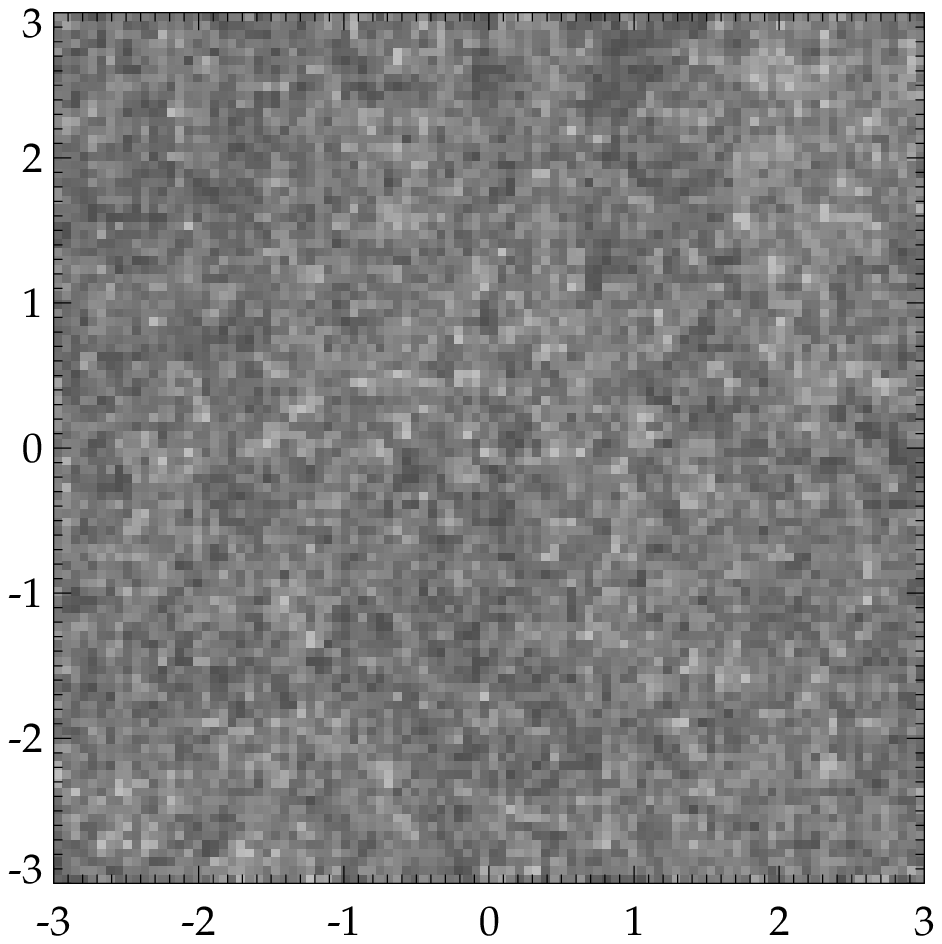}
\caption{An example of single component S\'ersic fit in the F300W band. 
From left to right are shown the F300W band image, the {\it Galfit} S\'ersic
best fit model and the residual after subtracting the model from the
image. The IAU name of the object is indicated at top left, the 
photometric redshift of the object is given at top right, the fitted S\'ersic
index $n$ is given at bottom left and the magnitude of the object 
is given at bottom right of the object frame. Each image is 6 arcsec $\times$ 6 arcsec.}
\label{fits}
\end{figure*}

The model galaxy is constructed as a combination of a
S\'ersic component and the sky background
value. Other free parameters in the model are the ellipticity of the S\'ersic component and its position angle. The
SExtractor based photometric catalog of our sources provided us with
valuable starting values for most of the above mentioned parameters. This analytic model of the galaxy light distribution is then convolved with the Point Spread Function (PSF) of the observation and compared with the observed galaxy image. 
The values of the free parameters are determined iteratively by 
minimizing the reduced $\chi^{2}$.
In Fig.~\ref{fits} we show an 
example fit for one of the targets in the F300W band.

In the above scheme, an accurate estimation of the PSF is crucial to
the robust determination of the structural parameters of a
galaxy. However, there
are only 4 identifiable stars in the F300W image 
which can be used to characterize the PSF. One of these
was saturated and another was too close to the edge of the frame. We 
used the other two stars (say X and Y) in turn, to establish
that the values of the derived parameters are not severely dependent on the choice 
of the PSF star.

   

\begin{figure}[t] \centering
   
   \includegraphics[angle=0,height=4.0cm,clip]{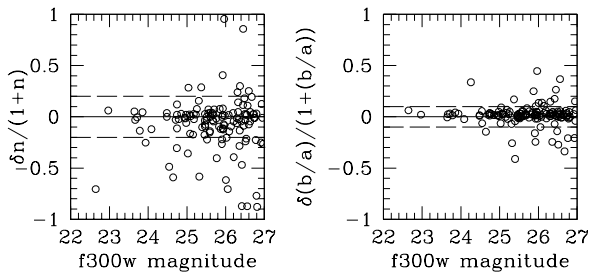}
\caption{The difference between the structural parameters $n$ and $b/a$ obtained in two separate {\it Galfit} runs, using two different stars as the PSF in the F300W band as a function of the F300W magnitude of the objects. As expected, the difference {\it flares} out at fainter magnitudes. The dashed lines in the two plots encompass $\sim80\%$ of all the objects.}
   \label{compare2}
\end{figure}

Here we use the methodology of Rawat et al.~\cite{rawat2007} to show that the derived parameters are invariant under the change of the PSF star.
For our sample of 162 objects in the F300W band, we used {\it Galfit} in batch mode to derive the structural parameters {\it {twice}} for each object. In the first run, we used one of the stars (X) as the PSF. In the second run, we used the second star (Y) as the PSF, everything else remaining the same as in the first run.
In Fig.~\ref{compare2} we show the difference between the structural parameters ($n$ and $b/a$) obtained in the two cases as a function of the F300W band magnitude of the objects. 
As is evident, our estimation of the structural parameters is quite robust irrespective of which star is used
as the PSF. The derived structural parameters in the two runs agree quite well for most of the objects, 
with a symmetrical scatter around the {\it zero} line, pointing towards random errors as the cause for any dispersion in the structural parameters obtained in the two runs rather than any systematic effect. For fainter objects, the random photon noise is higher and may overwhelm any systematic effect caused by the use of two different PSF stars. 
 Also, the S\'ersic index $n$, which is well known to be difficult to constrain, exhibits a greater amount of scatter than the axis ratio $b/a$. The dashed lines in the two panels encompass $\sim80\%$ of all the objects. 


\subsection{The F850LP band fits}
\label{zband_fits}

\begin{figure*}[t!]
\centering
\includegraphics[height=5.5cm,clip]{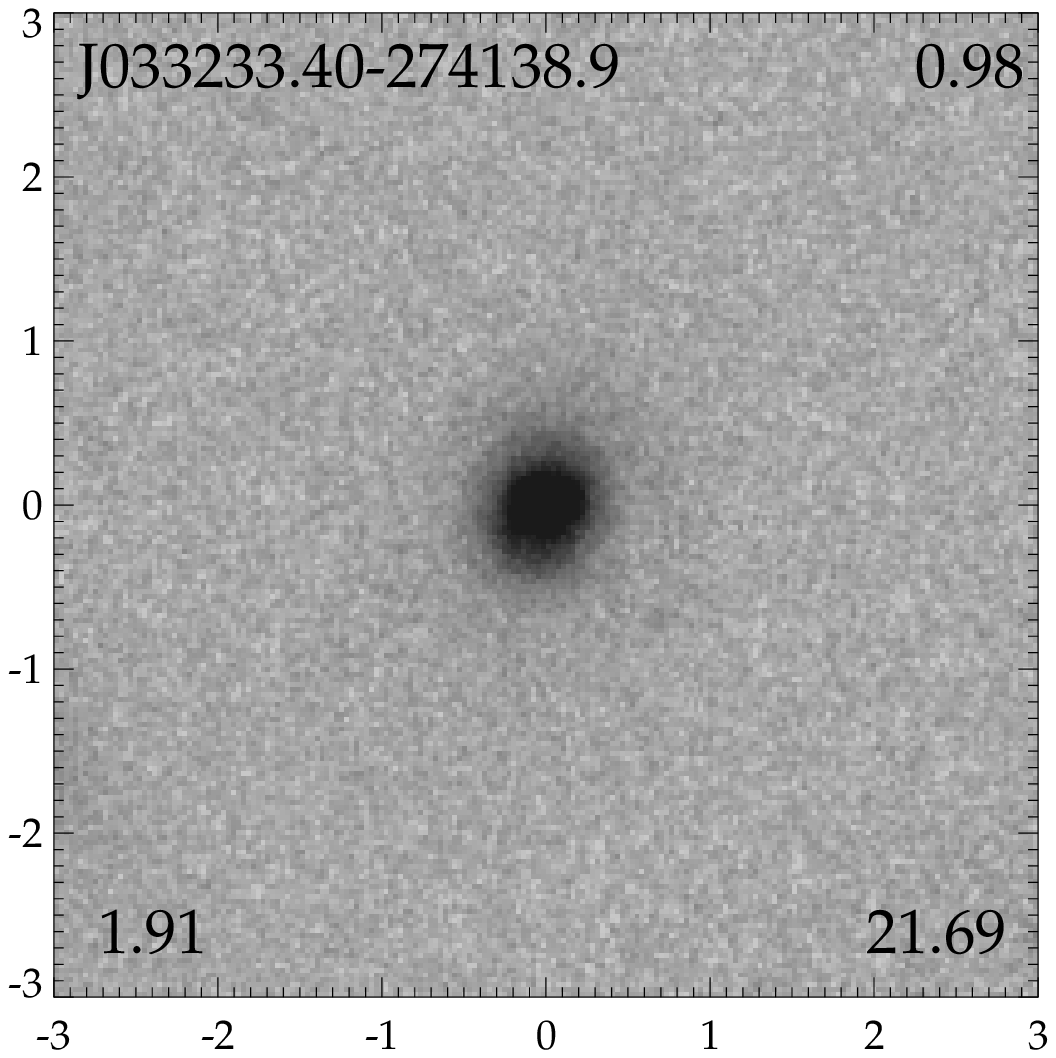}%
\includegraphics[height=5.5cm,clip]{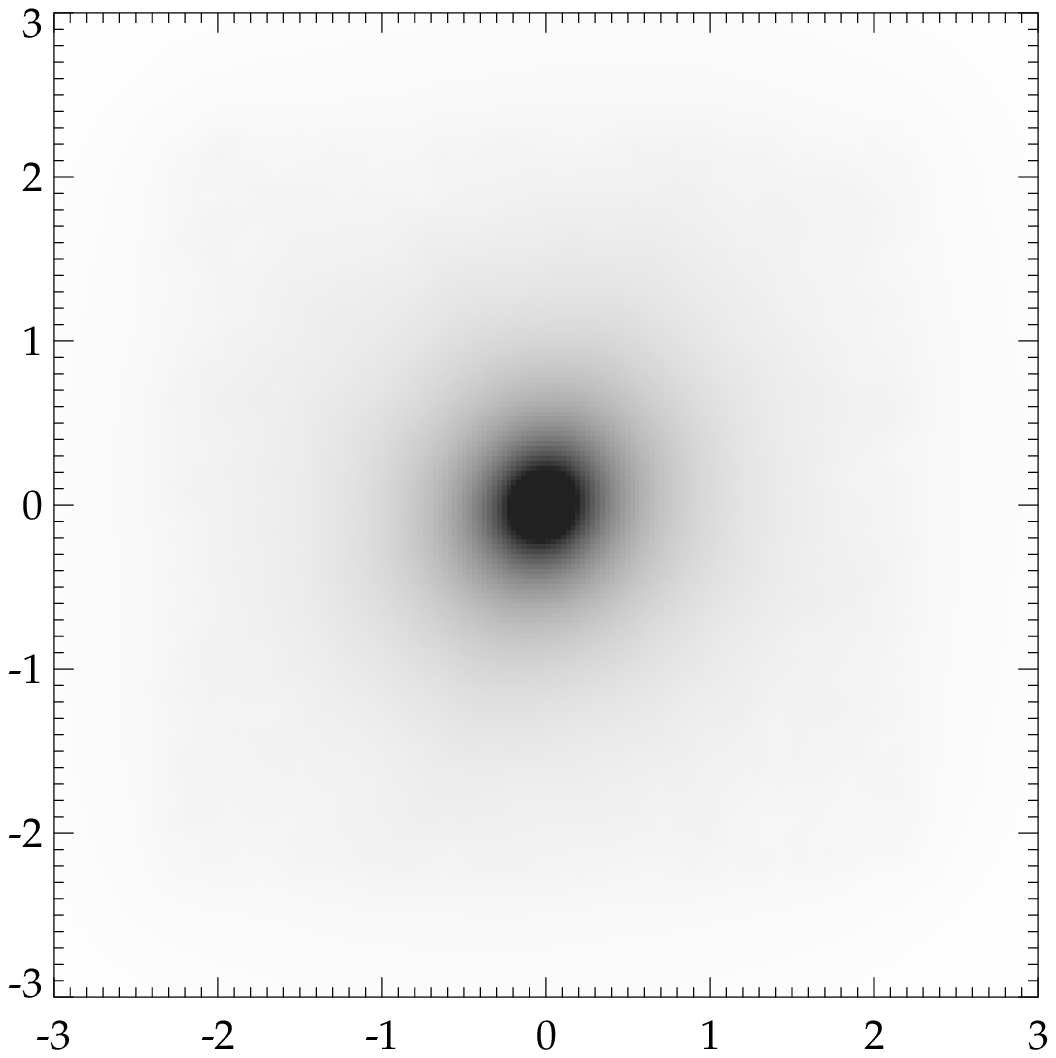}%
\includegraphics[height=5.5cm,clip]{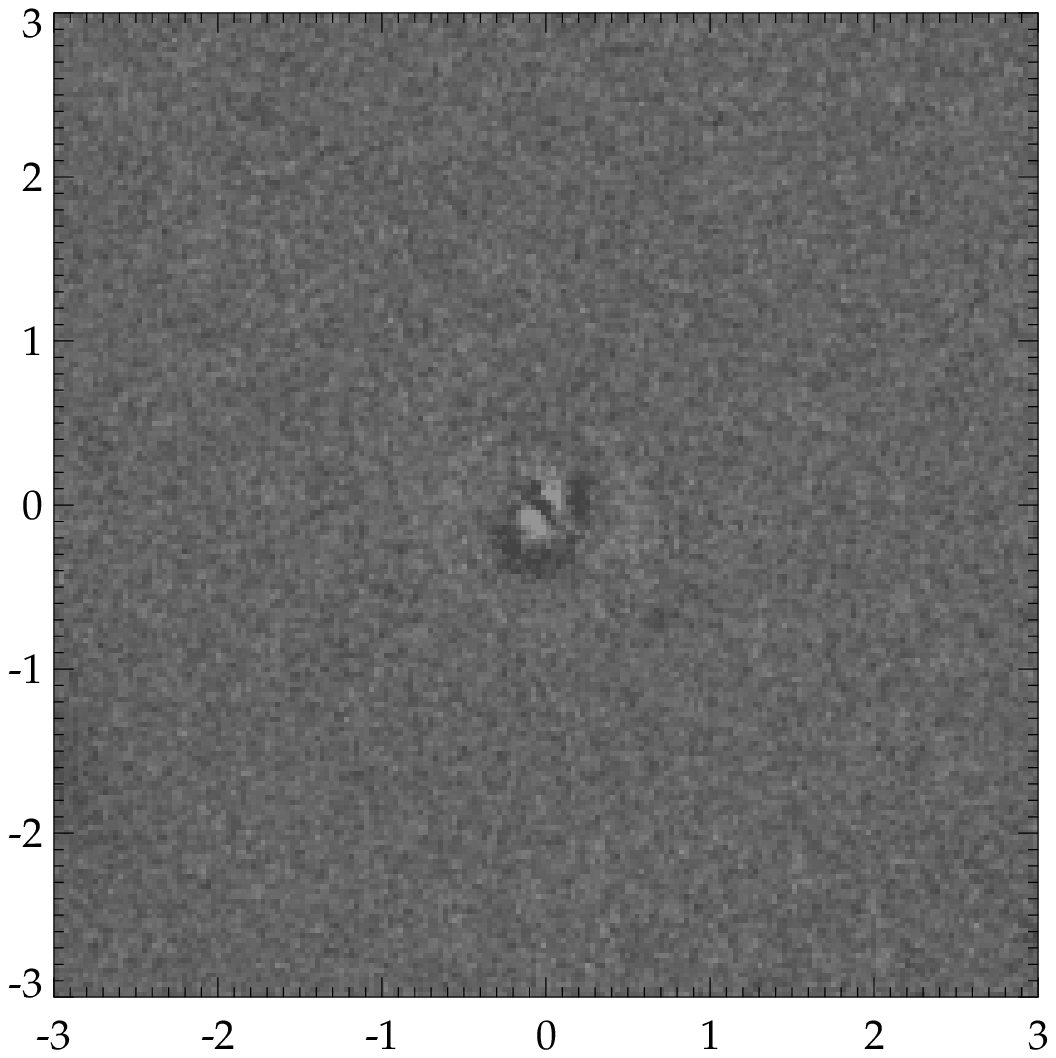}
\caption{An example of single component S\'ersic fit in the F850LP band. From left to right are shown the F850LP band image, the {\it Galfit} S\'ersic best fit model and the residual after subtracting the model from the image. Labels are same as in Fig.~\ref{fits}. Each image is 6 arcsec $\times$ 6 arcsec }
\label{zfits}
\end{figure*}

Single component S\'ersic fits were performed on the HST/ACS F850LP
image of the 162 objects.  The fits were performed in a manner
consistent with the F300W-band fitting procedure. Since in the z-band we have a large number of identifiable stars which can be used for 
characterizing the PSF, we have used the star closest to a given
galaxy as the model PSF. This choice is not critical; as demonstrated by 
Rawat et al.~\cite{rawat2007} the derived {\it Galfit} parameters 
are stable to within 10\% irrespective of which star is used as the
PSF. One of the fits that we obtained in the F850LP band is shown in
Fig.~\ref{zfits} for an object at photo-{\it z}=0.98.

\section{Comparison between F300W and F850LP fits}
\label{compare_u_z}
\begin{figure*}[b] \centering

\resizebox{0.25\textwidth}{!}{\includegraphics*[0.6cm,0.6cm][10.5cm,10.5cm]{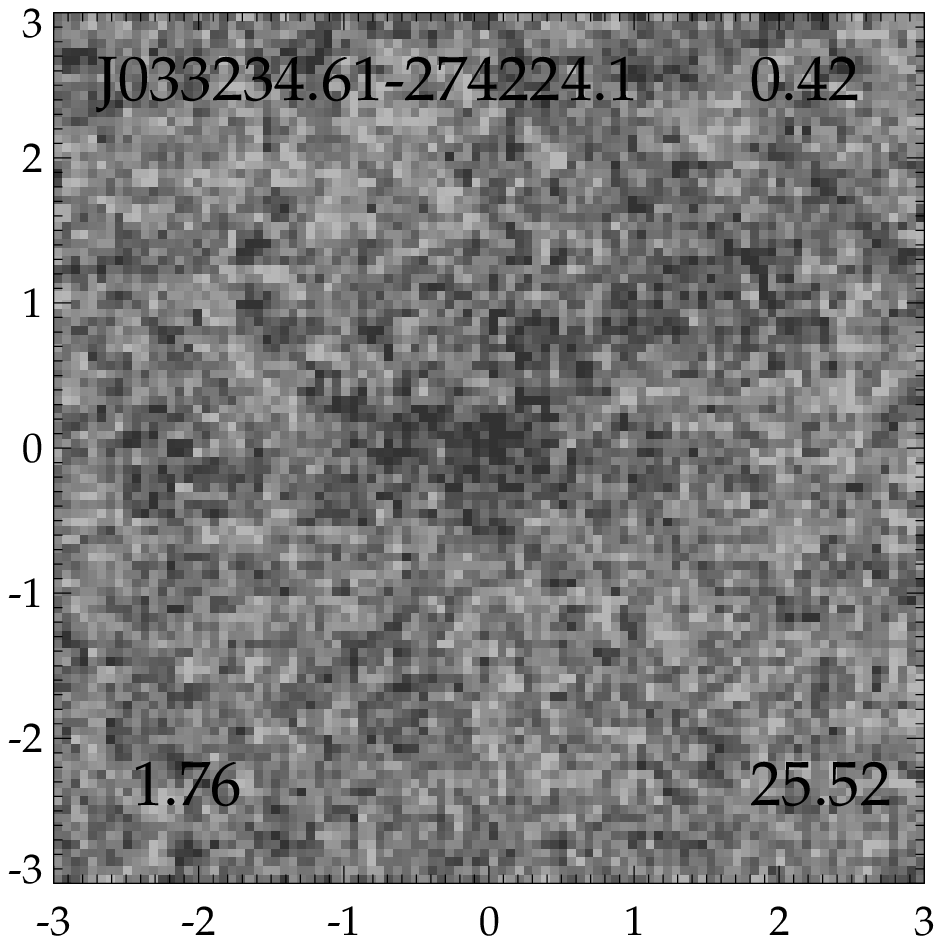}}%
\resizebox{0.25\textwidth}{!}{\includegraphics*[0cm,0cm][11cm,11cm]{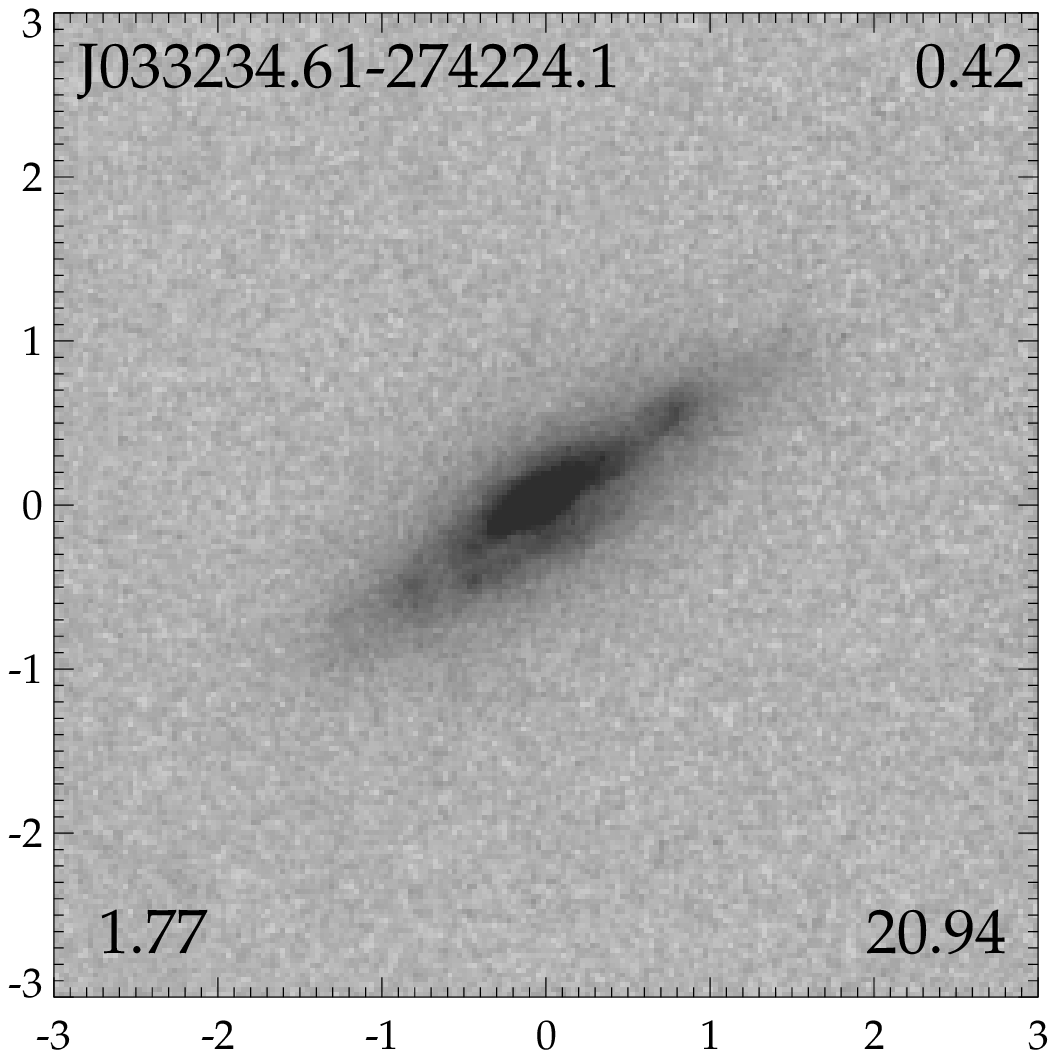}}%
\resizebox{0.25\textwidth}{!}{\includegraphics*[0.6cm,0.6cm][10.5cm,10.5cm]{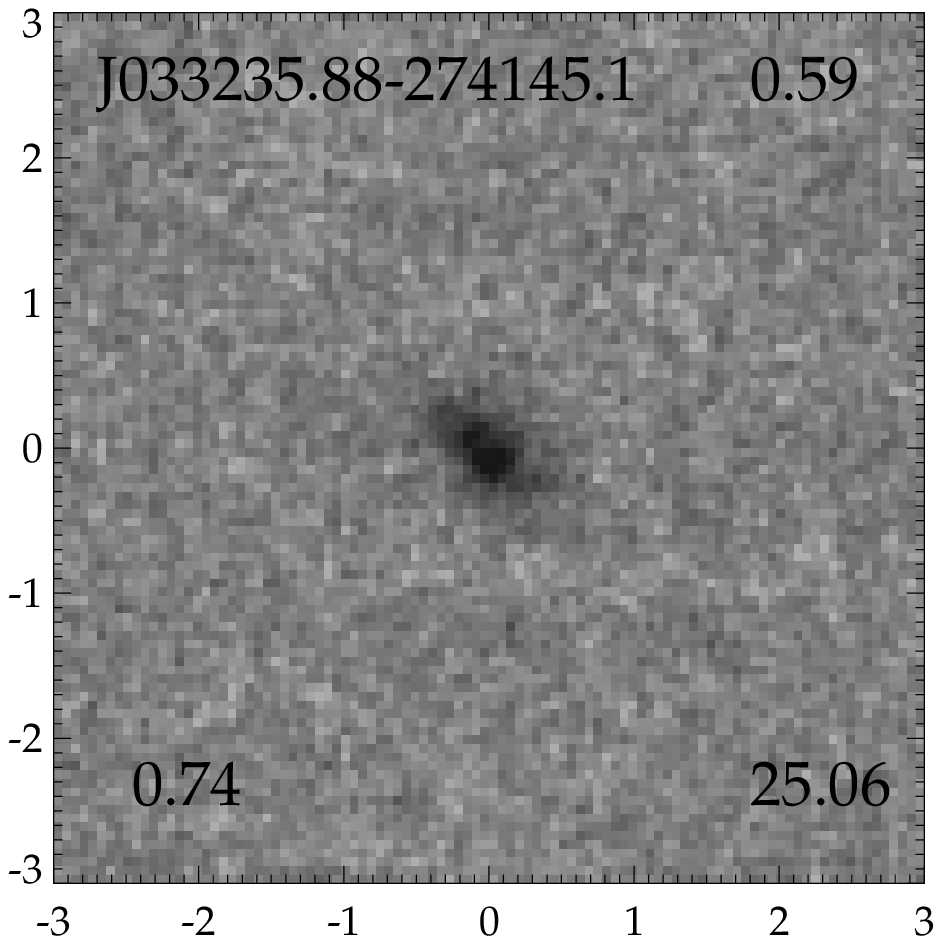}}%
\resizebox{0.25\textwidth}{!}{\includegraphics*[0cm,0cm][11cm,11cm]{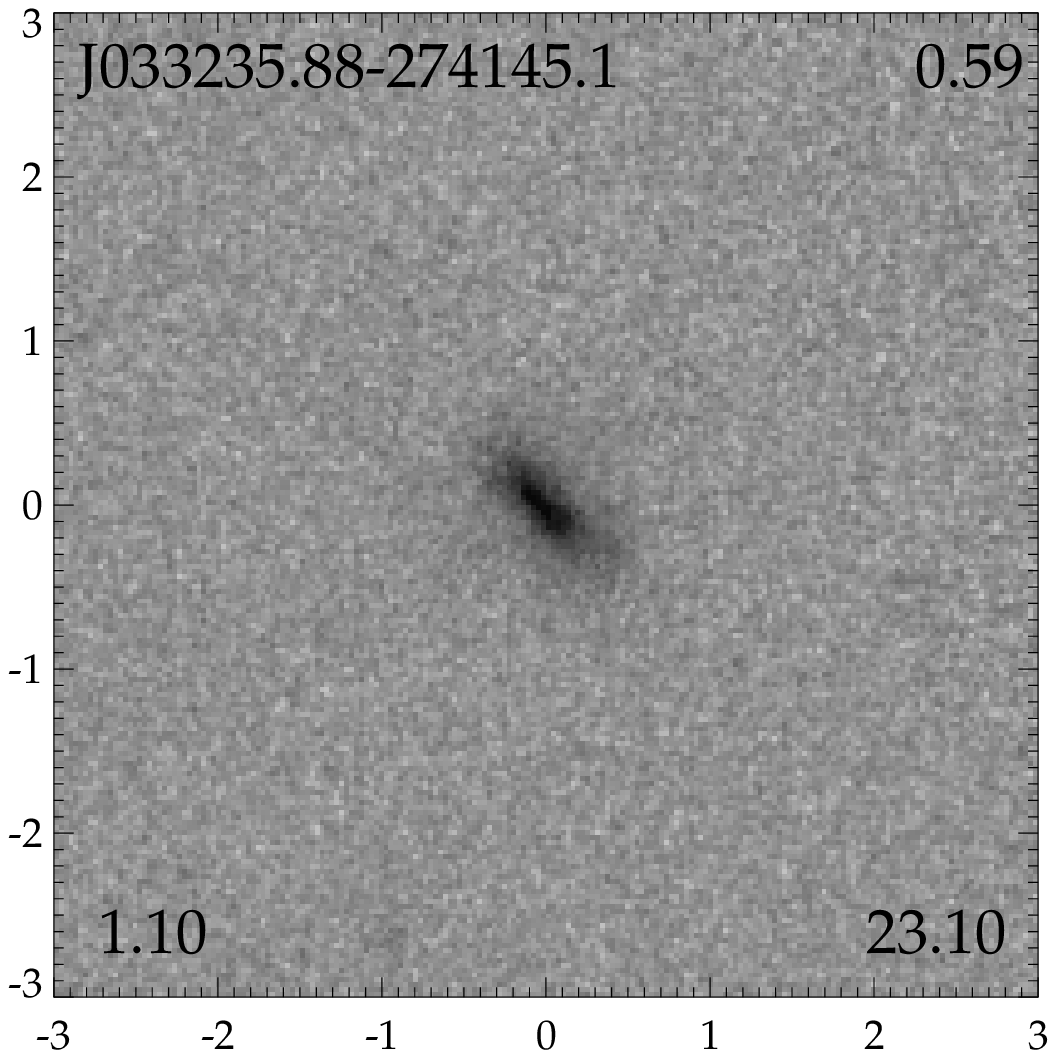}}

\resizebox{0.25\textwidth}{!}{\includegraphics*[0.6cm,0.6cm][10.5cm,10.5cm]{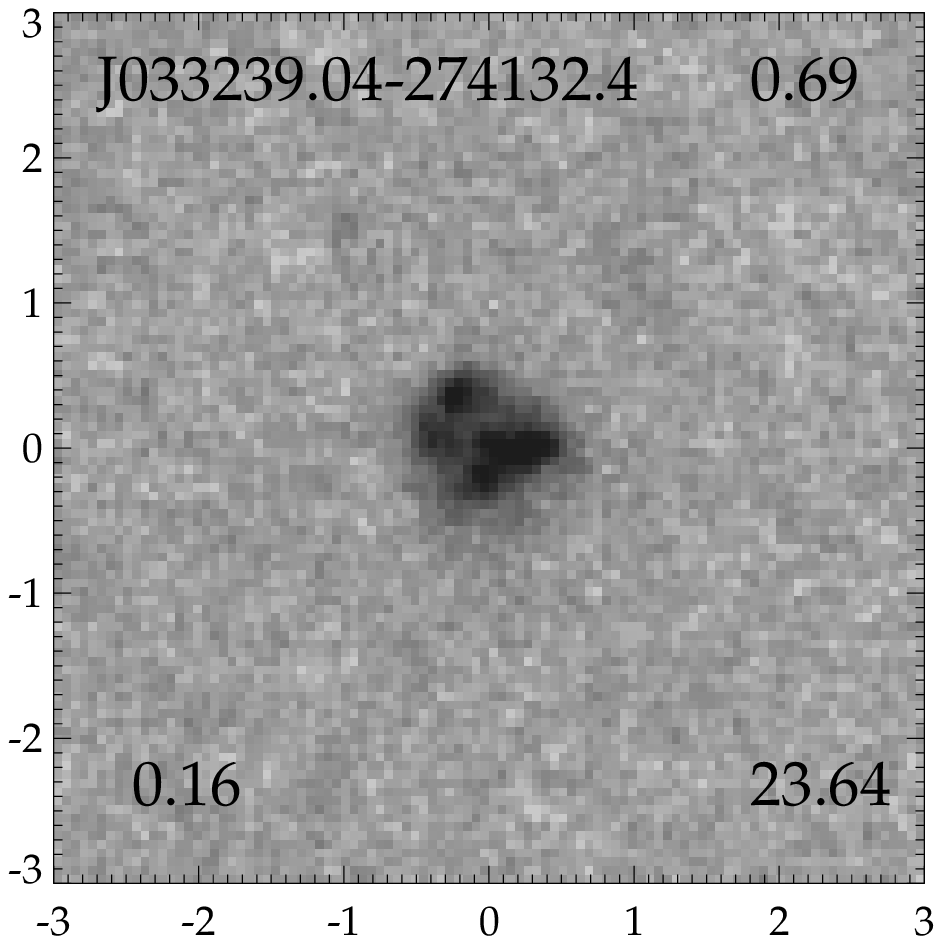}}%
\resizebox{0.25\textwidth}{!}{\includegraphics*[0cm,0cm][11cm,11cm]{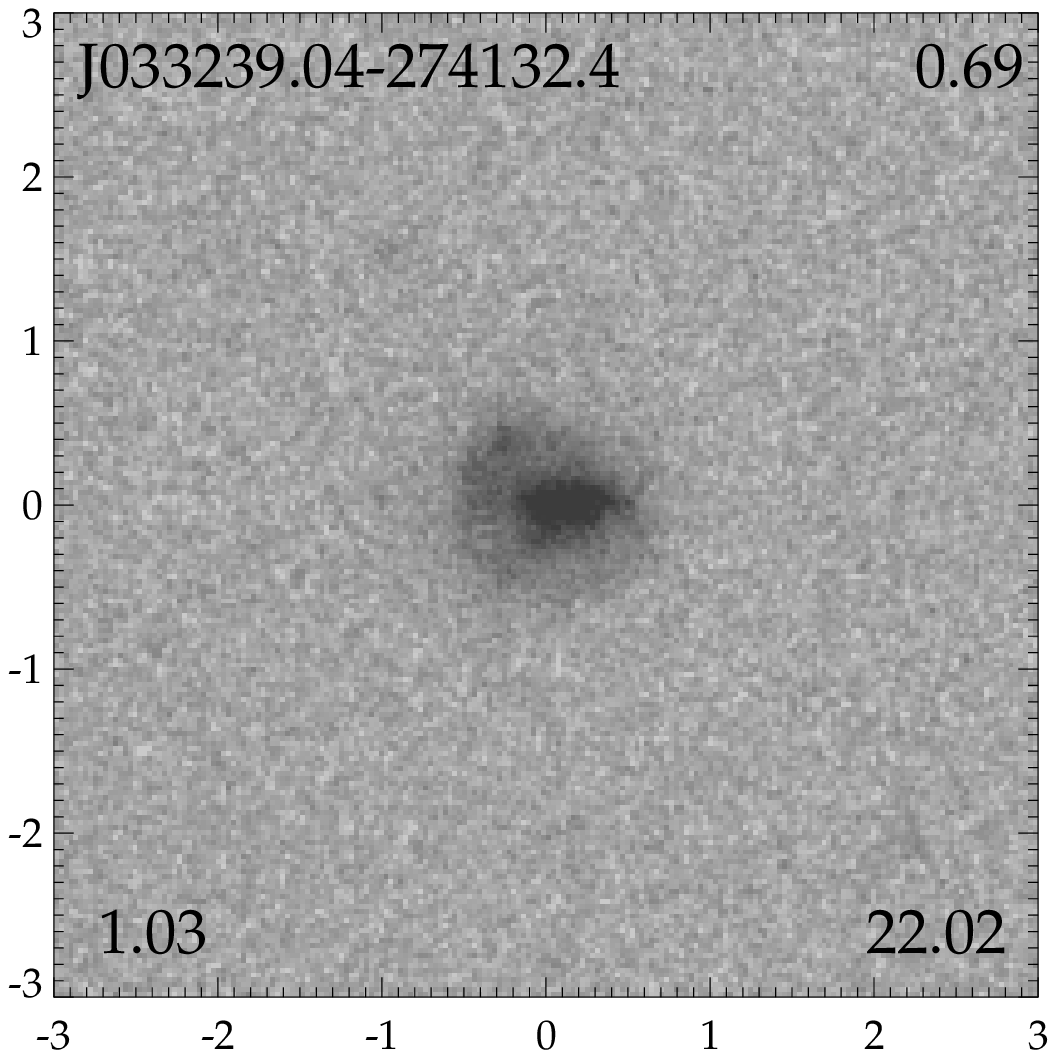}}%
\resizebox{0.25\textwidth}{!}{\includegraphics*[0.6cm,0.6cm][10.5cm,10.5cm]{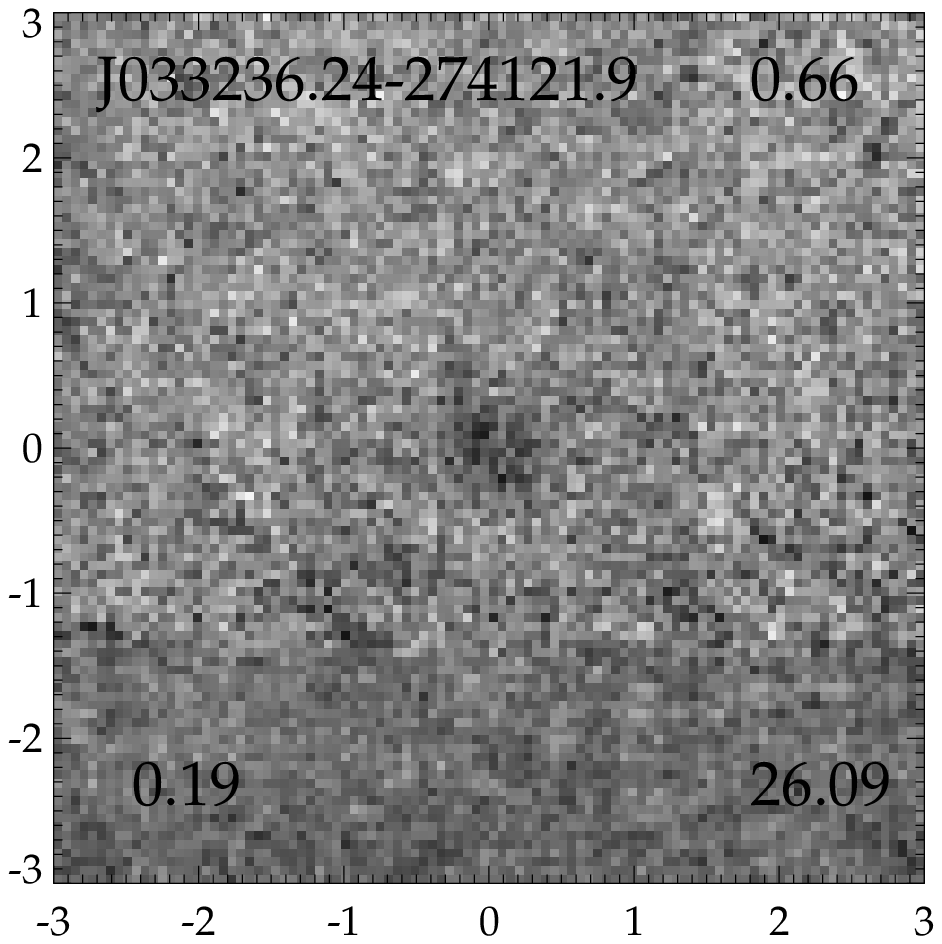}}%
\resizebox{0.25\textwidth}{!}{\includegraphics*[0cm,0cm][11cm,11cm]{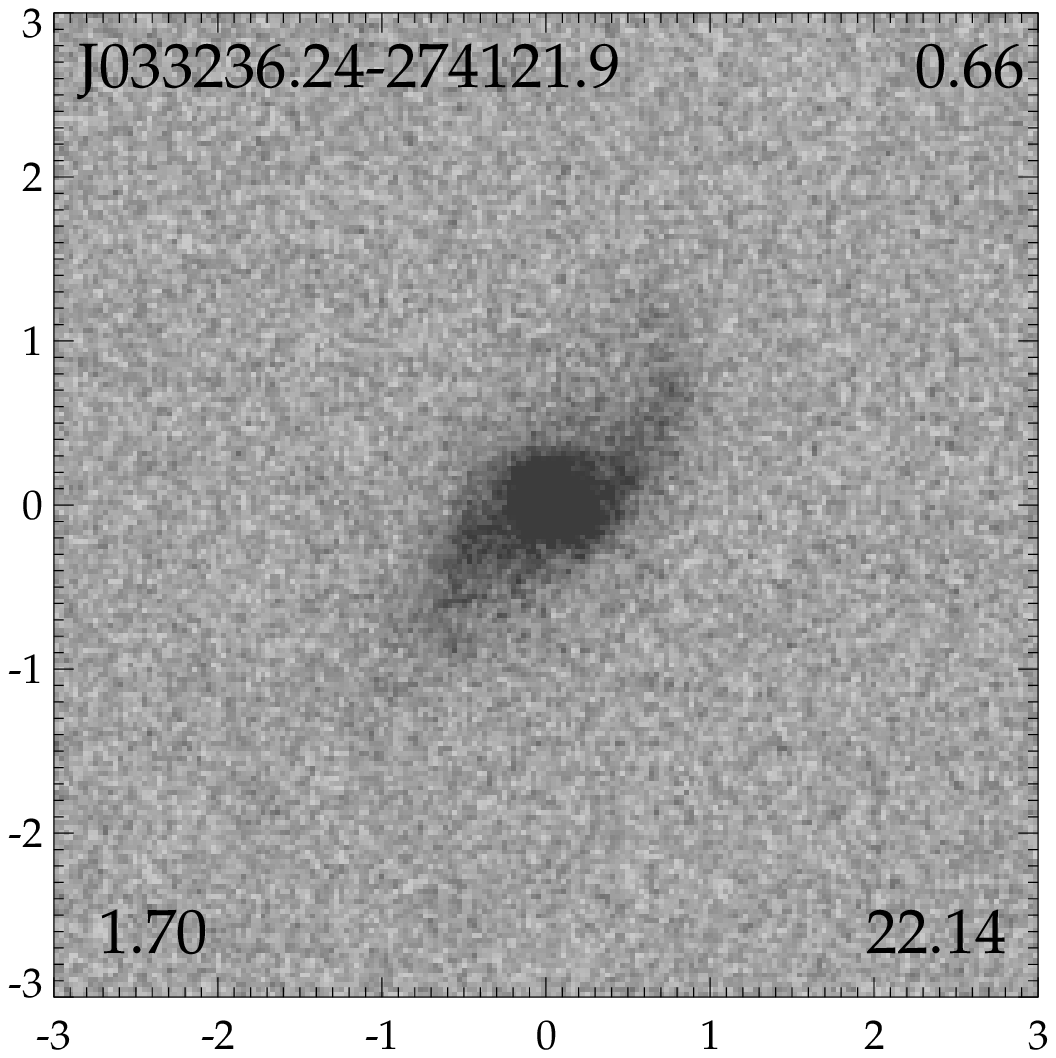}}

 \caption{Some examples of the difference in the appearance of a
   galaxy observed in the UV (left) in comparison with its appearance
   in the optical (right). The left cutout is a F300W filter HST/WFPC2 image from the Hubble-UDF
   parallels, while the right cutout is the F850LP HST/ACS image of
   the same object from the GOODS survey. Labels are same as in
   Fig.~\ref{fits}. Each image is 6 arcsec $\times$ 6 arcsec.}   
\label{montage}
\end{figure*}

Fig.~\ref{montage} shows some examples of the morphological differences in the
appearance of a galaxy observed in the UV in comparison to its
appearance in the optical. In each case, the left panel is a F300W HST/WFPC2 image while the
right panel is an F850LP HST/ACS image. The first object is a red, edge on disk galaxy. The second and third objects are classified as mergers ($n\leq0.8$) in the UV (see below) but not in the optical. The last object is classified as merger in the UV, but shows clear spiral structure in the optical.

Before we compare the derived quantitative parameters for our 162 objects in the F300W and F850LP band, it is important to check for possible biases due to the varying Signal-to-Noise Ratio (SNR) in the two filter datasets. The F300W band observation suffers from the poor UV sensitivity of the WFPC2 detector ($\sim$1.9\% for WFPC2/F300W filter)$\footnote[1]{Efficiency near filter pivot wavelength, including HST+instrument+filter as quoted from Table 1.2:{\it{Comparison of WFPC2 and ACS Filters}} of WFPC2 Instrument Handbook v9.0}$, while the F850LP band observations gain from the exquisite red sensitivity of the ACS/WFC detector ($\sim$25\% for ACS/WFC F850LP filter)$\footnotemark[1]$. This vast difference in the UV vs optical senstitvities is compensated to a large extent by the exceptionally deep exposure in the WFPC2/F300W ($\sim323$ ksec), compared to $\sim$10 ksec in the ACS/F850LP. 
In order to quantify the expected SNR of a typical object in the UV vs optical, we made use of the exposure time calculators for the two detectors$\footnote[2]{WFPC2 ETC for Extended Sources - v4.0 \url{http://www.stsci.edu/hst/wfpc2/software/wfpc2-etc-extended-source-v40.html}}$$^{,}$$\footnote[3]{ACS Imaging ETC\\ \url{http://etc.stsci.edu/webetc/acsImagingETC.jsp}}$. 
The median surface brightness for our 162 sample objects in the F300W band is $24.03~mag/arcsec^{2}$ compared to $22.95~mag/arcsec^2$ in the F850LP band. We used the known exposure times in each of the two filters to estimate the expected SNR for the above median surface brightness objects using the HST/ACS and WFPC2 exposure time calculators, yielding an average SNR of $\sim$21 per pixel in the F300W band and $\sim$73 per pixel in the F850LP band. 

Ravindranath et al.~\cite{ravindranath2006} have performed extensive Monte Carlo simulations to estimate the reliability of the morphological parameters derived from 2D galaxy fitting algorithm {\it Galfit}, as a function of the SNR of the galaxy images. Their simulations suggest that the morphological parameters can be well recovered for objects with SNR$\geq$10-15. Since we have used the methodology similar to that of Ravindranath et al.~\cite{ravindranath2006} for deriving morphological parameters and the same code ({\it Galfit}) for our sample galaxies, we believe that even though the SNR in the F300W band images for our sample objects is significantly lower compared to the F850LP band, it is still sufficient (SNR$\geq$10-15) to allow an accurate estimation of their structural parameters. Therefore, we do not expect any bias in the derived structural parameters in the F300W band compared to F850LP band due to the difference in their SNR.
%

\begin{figure*}[t!]
\includegraphics[height=8.5cm,clip]{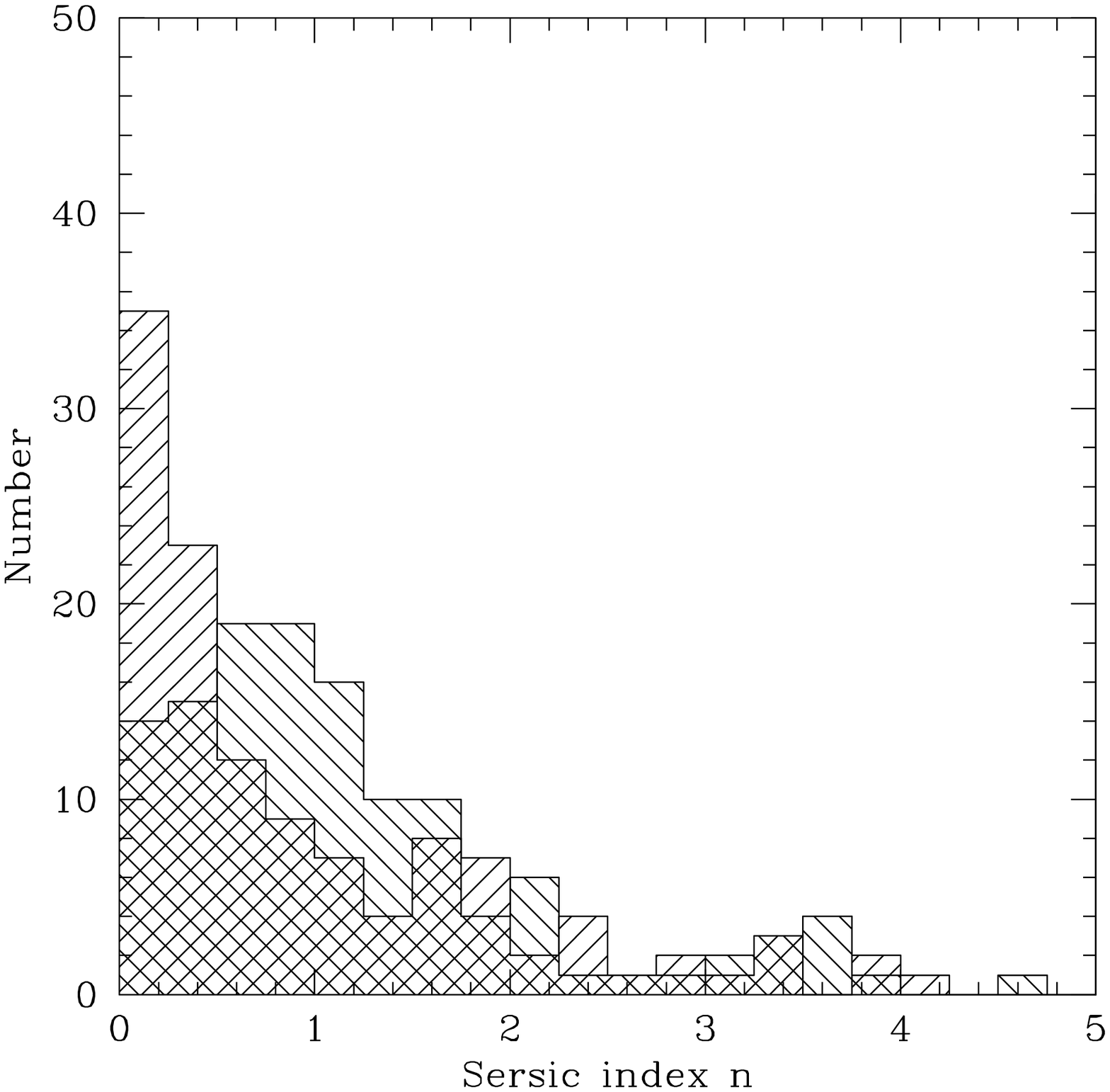}%
\includegraphics[height=8.5cm,clip]{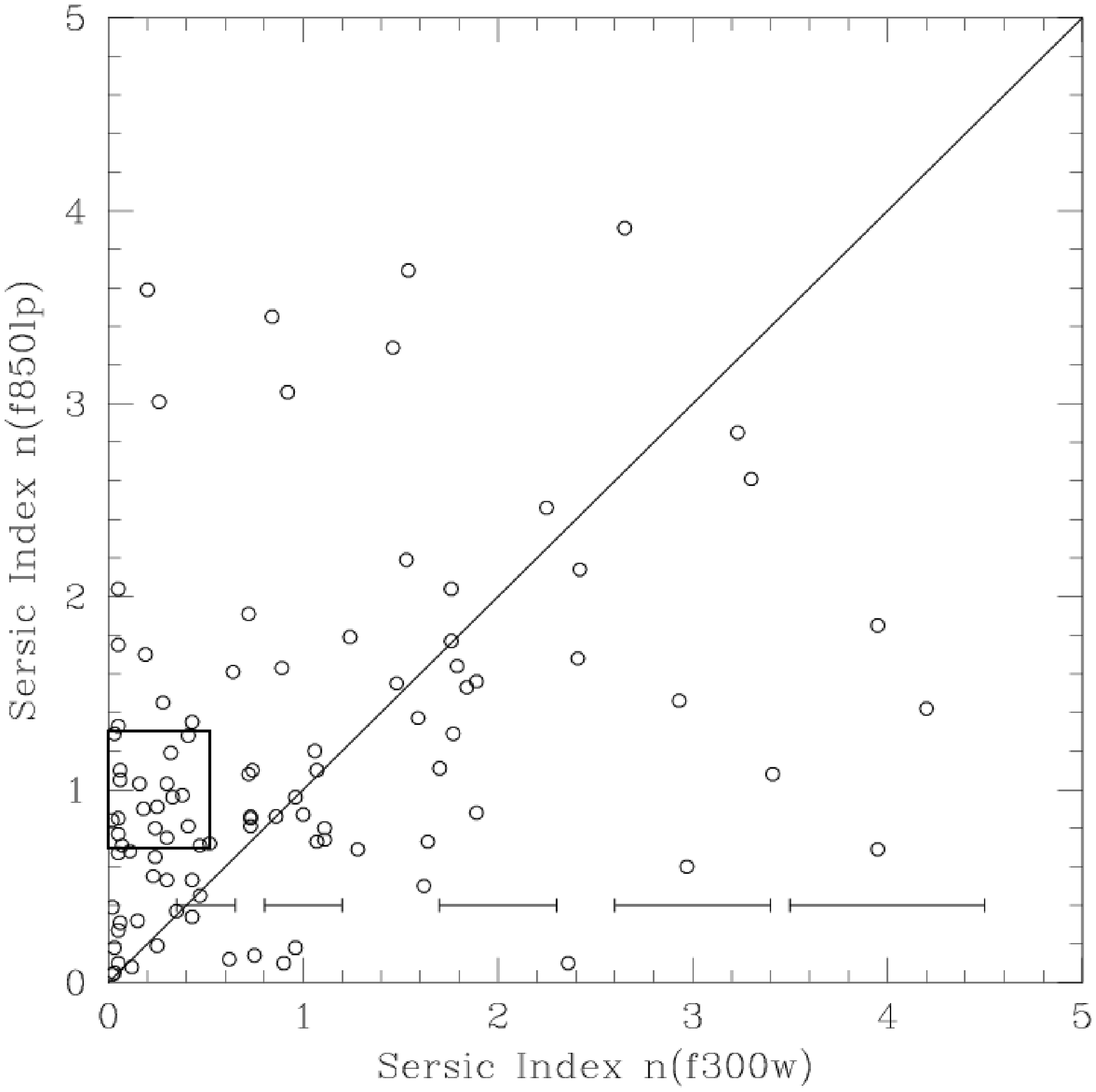}
\caption{ (left) A comparison between the S\'ersic index
  $n$ derived using {\it Galfit}, single component S\'ersic
  fits in the WFPC2/F300W-band ($45^{0}$ hashing) and the ACS/F850LP
  band ($135^{0}$ hashing) for our 162 galaxies. A two sample KS test shows that there is $> 99\%$ probability of the two samples being drawn from a different parent population. (right) A scatter plot of S\'ersic index $n(F300W)$ vs $n(F850LP)$. Note the clumping at the bottom left (enclosed by the black rectangle), showing objects with $n(F300W)\leq0.5$ and $n(F850LP)\sim1.0$, corresponding to the peaks at $n\leq0.5$ in the F300W band and $n\sim1.0$ in the F850LP seen in the S\'ersic index distribution (left). The morphology of these objects is discussed in Section 5. The average errorbars are derived using the differences in the computed S\'ersic index $n$ using two different PSFs in the F300W band as shown in Fig.~\ref{compare2}.}
\label{n_comparision}
\end{figure*}

Fig.~\ref{n_comparision} (left) shows the distribution of the derived S\'ersic index $n$ for the 162 sources in the F300W band ($45^{0}$ hashing) as well as the F850LP band ($135^{0}$ hashing).
As is apparent, the two distributions are quite different, with the distribution in F300W band peaking at much smaller values of $n$, compared to the F850LP band. A two sample KS test confirms that there is $> 99\%$ probability of the two samples being drawn from a different parent population.  
Fig.~\ref{n_comparision} (right) shows a scatter plot of S\'ersic index $n(F300W)$ vs $n(F850LP)$. Note the clumping at the bottom left, showing objects with $n(F300W)\leq0.5$ and $n(F850LP)\sim1.0$, corresponding to the peaks in the S\'ersic index distribution at $n\leq0.5$ in the F300W band and $n\sim1.0$ in the F850LP band. The average errorbars have been derived using the differences in the computed S\'ersic index $n$ using two different PSFs in the F300W band as shown in Fig.~\ref{compare2}.

We notice that in the F300W band, the S\'ersic index $n$ is lower in a large number of cases compared the the F850LP band. This effect is mainly caused by galaxies with low values of $n(F300W)\leq0.5$ that have a corresponding $n(F850LP)\sim1.0$, while this effect does not seem to be important in galaxies with high values of $n$, as can be clearly seen in Fig.~\ref{n_comparision}(right).
This has serious consequences, as $n$ is widely used in the literature for identifying the morphological class of a galaxy. Especially, $n$ flatter than exponential is usually used for isolating galaxies with multiple cores or disturbed morphologies, possibly indicative of mergers (eg. Ravindranath et al. 2006). However, it is imperative to note here that it is impossible to isolate a {\em clean} sample of merger candidates using this criterion, as the objects that exhibit $n$ shallower than exponential profile include galaxies with patchy star forming HII regions and low surface brightness galaxies etc., in addition to true merging galaxies (see eg. the discussion in Ravindranath et al. 2006). Keeping the above mentioned caveat in mind, we find from Fig.~\ref{n_comparision} (left) that $n_{F300W}\leq 0.5$ for 57 objects, whereas $n_{F850LP}\leq 0.5$ for only 28 cases. This means that the use of F300W band, which probes the rest-frame UV light of the objects at these redshifts, increases the number of merger candidates by a factor $\sim2$ compared to the F850LP band (rest-frame optical). This factor is a less severe (but no less significant) $\sim1.4$, if we employ a $n\leq0.8$ as the cutoff for identifying merger candidates. 

We checked whether the objects with $n \leq 0.5$ in either of the two filters are systematically fainter than the whole sample, i.e. whether poorer SNR might be responsible for these objects having anomalously low $n$. We found that in the $F300W$ band, the median surface brightness for the subset of 57 objects with $n_{F300W}\leq 0.5$ is $24.08~mag/arcsec^2$ (compared to $24.03~mag/arcsec^2$ for the whole sample). This yields a SNR of $\sim$20 per pixel (compared to $\sim$21 for the whole sample). Similarly, in the $F850LP$ band, the median surface brightness for the subset of 28 objects with $n_{F850LP}\leq 0.5$ is $23.05~mag/arcsec^2$ (compared to $22.95~mag/arcsec^2$ for the whole sample). This yields an SNR of $\sim$67 per pixel (compared to $\sim$73 per pixel for the whole sample). Hence we see that the SNR for the subset of objects with extremely low values of $n$ in both the filters is very similar to that for the entire sample indicating that the parameters derived for these objects are likely to be just as robust as for the entire sample.

The fact that the differences in the derived value of $n$ in the two filters are dominated by galaxies with low values of $n(F300W)\leq0.5$ that have a value of $n(F850LP)\sim1.0$ leads us to believe that we might probing clumpy star forming HII regions in these objects in the $F300W$ band. 
The tacit assumption that we are making here is that clumpy star formation in galaxies will lead to a lower $n$ value in the rest-frame UV but not in the rest-frame optical (because star forming HII regions are blue). This would mean that objects with $n(F300W)\leq 0.5$ and $n(F850LP)\sim1.0$ are probably clumpy starforming galaxies which are intrinsically disky ($n\sim1.0$). While it is true that blue neighbours will tend to mimic HII regions in the rest-frame UV by producing a shallow S\'ersic profile, however the very fact that they contribute little or no flux in the optical (parametrized by a steeper $n(F850LP)\sim1.0$) means that they cannot be massive systems (as z band flux is a better tracer of stellar mass than UV flux) and will constitute {\em minor} mergers at best (see Rawat et al. (2008) for further discussion on how flux in redder wavebands is better suited for identifying massive major merger candidates). So objects with $n(F300W)\leq 0.5$ and $n(F850LP)\sim1.0$ are likely to be either clumpy star forming galaxies or minor mergers, both of which are essentially contaminants, if one is interested in identifying major mergers. In Section~\ref{individual_comments}, we include comments on the morphology of individual objects with $n(F300W)\leq 0.5$ and $0.7\leq n(F850LP) \leq1.3$ (the working definition of $n(F850LP)\sim1.0$) in detail to show that our assertion above is indeed correct.

\begin{figure*}[t!]
\includegraphics[height=8.5cm,clip]{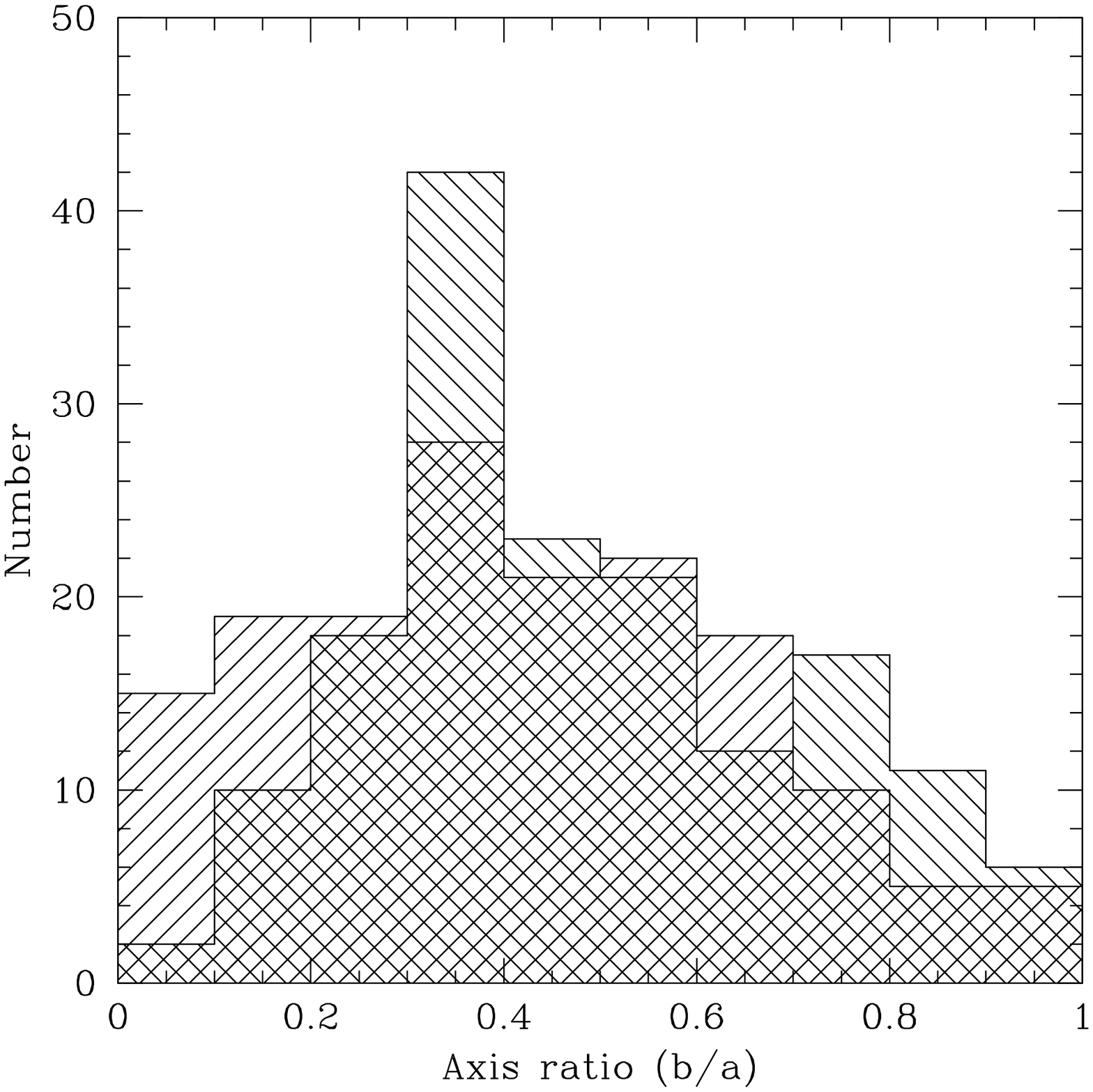}%
\includegraphics[height=8.5cm,clip]{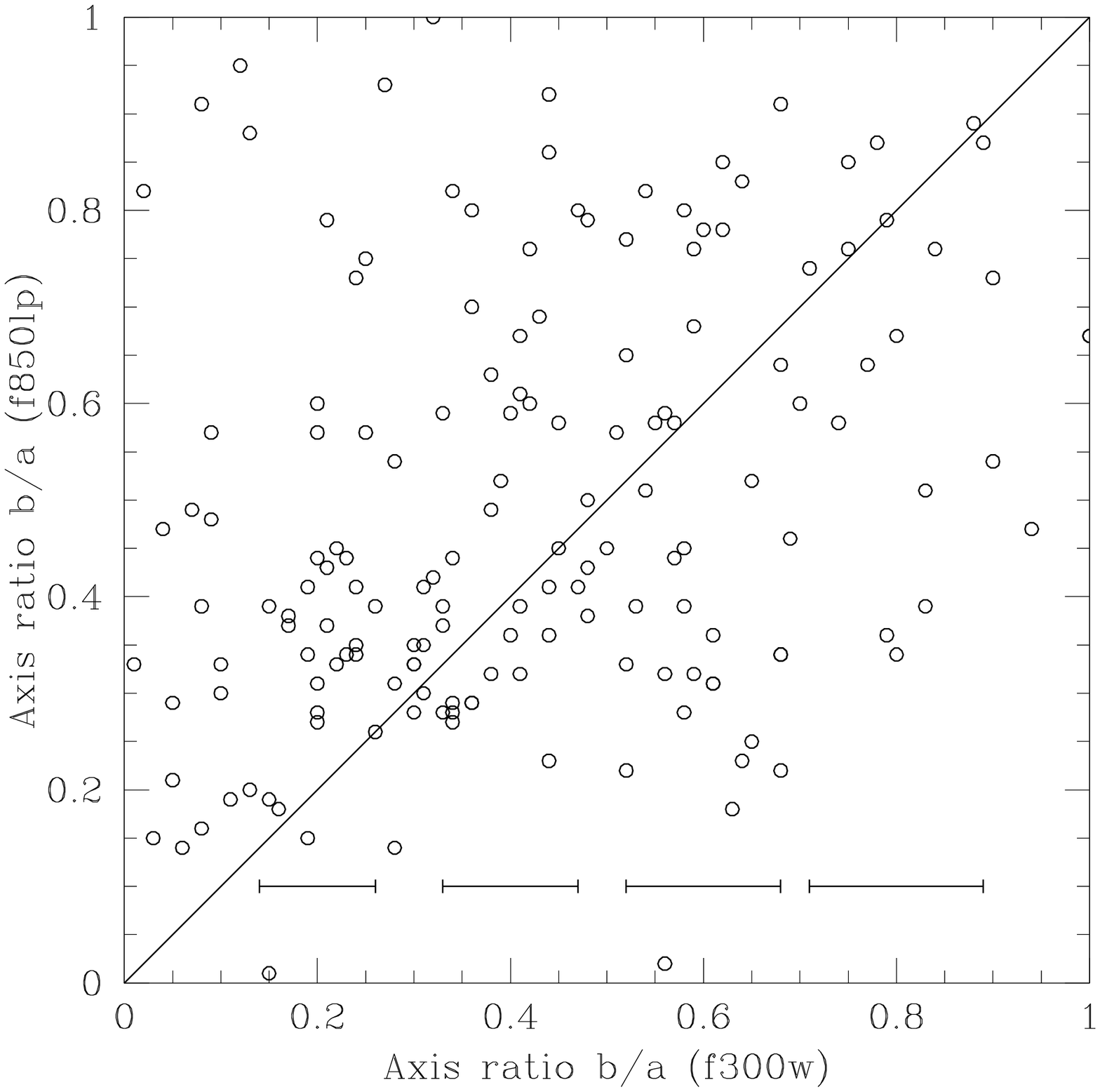}
\caption{(left) A comparison between the axis-ratio (b/a) derived using {\it Galfit}, single component S\'ersic
  fits in the WFPC2/F300W-band ($45^{0}$ hashing) and the ACS/F850LP
  band ($135^{0}$ hashing) for our 162 galaxies. The mean values of the axis ratio are $(b/a)_{F300W}=0.42\pm0.23$ and $(b/a)_{F850LP}=0.49\pm0.22$. A two sample KS test shows that there is $>99\%$ probability of the two samples being drawn from a different parent population. (right) A scatter plot of the axis ratio $b/a(F300W)$ vs $b/a(F850LP)$. Note the paucity of sources with $b/a(F850LP)\leq 0.2$. The average errorbars are derived using the differences in the computed axis-ratio $b/a$ using two different PSFs in the F300W band as shown in Fig.~\ref{compare2}.}
\label{q_comparision}
\end{figure*}

Similarly, Fig.~\ref{q_comparision} (left) shows the distribution of
the axis ratio (b/a) in the F300W band ($45^{0}$ hashing) as well as in the F850LP band ($135^{0}$ hashing). From the figure, we find that $(b/a)_{F300W}\leq 0.2$ for 34 objects, while $(b/a)_{F850LP}\leq 0.2$ for only 12 objects. This means that in the rest-frame UV, the number of high ellipticity ($e\geq0.8$) objects is higher by a factor $\sim2.8$ compared to the rest-frame optical. This might explain the reported claim in literature that high-z LBGs tend to show a significant skew towards higher ellipticities (e.g. Ravindranath et al. 2006). Also, the mean values of the axis ratio are $(b/a)_{F300W}=0.42\pm0.23$ and $(b/a)_{F850LP}=0.49\pm0.22$. Again, a two sample KS test shows that there is $>99\%$ probability of the two samples being drawn from a different parent population.
Fig~\ref{q_comparision} (right) shows a scatter plot of the axis ratio $(b/a)_{F300W}$ vs $(b/a)_{F850LP}$.
Here we see that in the F300W band, the axis-ratio $(b/a)$ is lower (i.e ellipticity $e=(1-b/a)$ is higher) in a large number of cases compared to the F850LP band. The average errorbars are derived using the differences in the computed axis-ratio $b/a$ using two different PSFs in the F300W band as shown in Fig.~\ref{compare2}.

\begin{figure*}[t]
\includegraphics[height=8.5cm,clip]{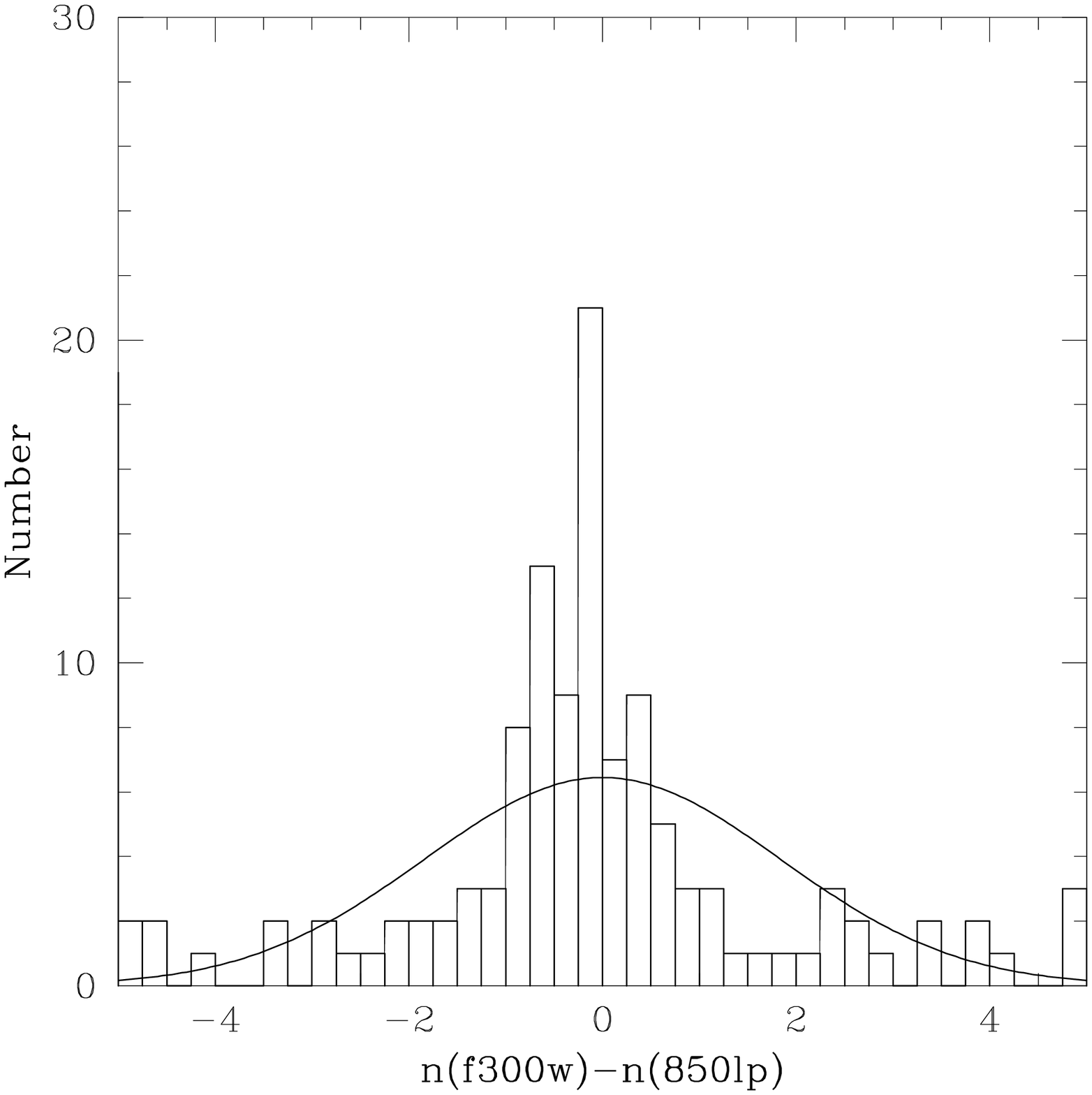}%
\includegraphics[height=8.5cm,clip]{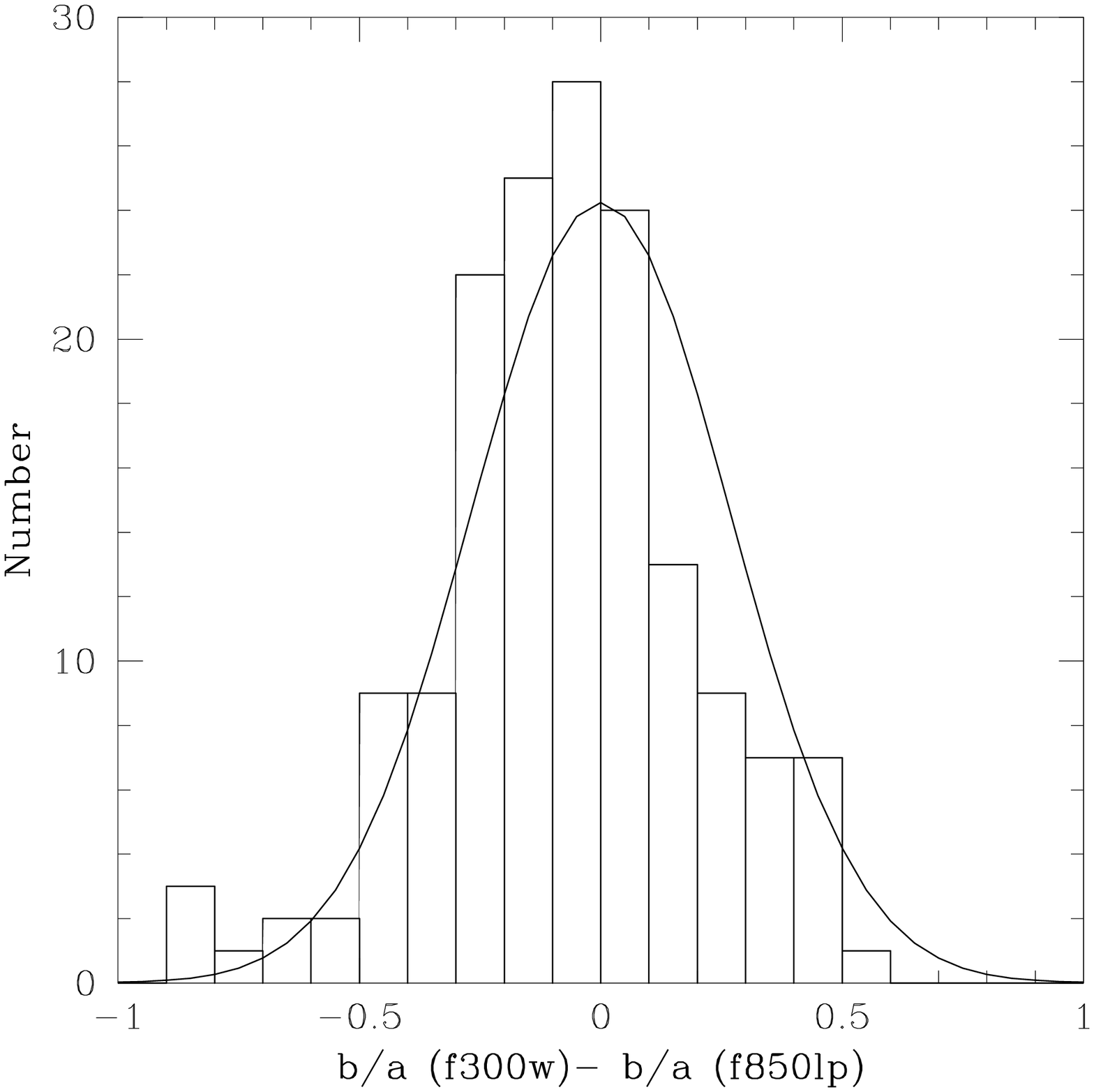}
\caption{(left) A histogram of the difference between $n(F300W)$ and $n(F850LP)$ for our sample galaxies. The smooth curve is the normal distribution, centered on the origin and having the same variance as the distribution of the sample galaxies. Notice the sharp departure from pure gaussianity around $-1.0\leq n(F300W)-n(F850LP)\leq0.0$. These objects correspond to the clump of objects around $n(F300W)\leq0.5$ and $n(F850LP)\sim1.0$ shown in Fig.~\ref{n_comparision}(b). (right) A histogram of the difference between $b/a(F300W)$ and $b/a(F850LP)$ for the same sample. Again, the smooth curve is the normal distribution, centered on the origin and having the same variance as the distribution of the sample galaxies. Both the distributions are found to have negative skew (left skewed), with the distribution having more area under the left tail than expected from a normal distribution. }
\label{n_q_diff}
\end{figure*}

Another way to visualize the difference in the derived quantities $n$ and $b/a$ in UV vs optical is presented in Fig.~\ref{n_q_diff}. Fig.~\ref{n_q_diff} (left) shows a histogram of the {\it difference} between the derived values of S\'ersic index $n(F300W)$ and $n(F850LP)$ for our sample objects. The smooth curve is the normal distribution, centered on the origin and having the same variance as the distribution of the sample galaxies. 
The peak close to zero corresponds to objects having same value of $n$ in both the filters. However the dispersion about the peak is {\it not} symmetrical. 
Notice the sharp departure from pure gaussianity around $-1.0\leq n(F300W)-n(F850LP)\leq0.0$. These objects correspond to the clump of objects around $n(F300W)\leq0.5$ and $n(F850LP)\sim1.0$ shown in Fig.~\ref{n_comparision}(right).
We quantified this by calculating a measure of skewness$\footnote[4]{We would like to inform the reader that while we had to bin the data for the purpose of plotting in Fig.~\ref{n_q_diff}, the calculation of the skew was done from first principles and no binning was used.}$ (Fisher Gamma 1) for this distribution, defined as $\gamma_1 = m_3/m_2^{3/2}$, where $m_2$ and $m_3$ are the second and third moments of the distribution respectively. We find the value of the skew to be $\gamma_1=-0.10$. This negative value of skew $\gamma_1$ indicates that the distribution has more area under the left tail than expected from a normal distribution. This implies that $n(F300W)-n(F850LP)\leq 0.0$ for a larger number of cases than expected by pure chance (normal distribution), i.e. the S\'ersic index $n$ is lower in the UV compared to the optical in a large number of cases. This difference is mainly dominated by galaxies with low values of $n(F300W)\leq0.5$ that have a value of $n(F850LP)\sim1.0$, and is not affected by galaxies with high values of $n$.

Similarly, Fig.~\ref{n_q_diff} (right) shows a histogram of the difference between $b/a(F300W)$ and $b/a(F850LP)$. Again, the smooth curve is the normal distribution, centered on the origin and having the same variance as the distribution of the sample galaxies. The value of skew in this case is found to be $\gamma_1=-0.30$ (left skewed). Again, a negative value of $\gamma_1$ implies that the distribution has more area under the left tail than expected from a normal distribution and that the axis ratio $b/a$ is lower in the UV compared to the optical in a larger number of cases than expected by random chance (normal distribution).

Since we are using objects which have a wide range of photometric redshifts, we are
probing different rest-frame wavelengths for different objects 
with the same filter. For example, the F300W band will probe
rest-frame 2000\angst\ (NUV) for a redshift 0.5 object, whereas the
same filter will probe rest frame 1200\angst\ (FUV) for a redshift 1.5
object. We checked whether this could be affecting our interpretation
of Fig.~\ref{n_comparision} \& \ref{q_comparision} by restricting our sample to only objects
with $0.5 \leq z_{photo} \leq 1.0$. This yields a sample of 58
objects. Using a calculation similar to the one explained above, we 
find that $n_{F300W}\leq 0.5$ for 22 objects, whereas $n_{F850LP}\leq
0.5$ for only 9 objects. This yields a factor $\sim$2.4 overestimation 
of the number of merger candidates in the rest-frame UV (1500\angst -
2000\angst) compared to the rest-frame optical (4250\angst -
5700\angst). This compares favorably with the factor $\sim$2 we had 
determined using the entire sample of 162 objects. This factor is
found to be $\sim1.9$, if we employ a $n\leq0.8$ as the cutoff for
identifying merger candidates, again comparable to the factor we
obtained using the entire sample. It is encouraging to see that our 
results are robust and change little irrespective of the fact as to
whether or not we apply the redshift cutoff. Part of the reason is
that the redshift distribution of our sample objects anyway peaks 
between 0.5 and 1.0 (Fig.~\ref{photoz}) with a median of 1.02, so that 
in most cases we are probing rest-frame 1500\angst - 2000\angst.
Note the fact that our result is {\it{stronger}} (albeit with a smaller sample size) if we use only objects with $0.5 \leq z_{photo} \leq 1.0$. This means that the use of objects without regard to their redshift dilutes the result (as expected) and must therefore be treated as a {\it{lower limit}}.

\section{Comments on individual sources}
\label{individual_comments}

We shortlist objects with $n(F300W)\leq0.5$ and $0.7\leq n(F850LP) \leq1.3$ to check our assertion that these objects are actually patchy starforming disk dominated galaxies, rather than major merger candidates. There are 17 objects satisfying the above mentioned criteria. Fig.~\ref{fig9} shows the $u-z$ color map stamps (right) of these objects, along with $F300W$ (left) and $F850LP$ (center) imaging graylevel stamps. The color bar in each color map stamp shows the $u-z$ color range from 0 to 4. A description of each object is presented below. 
\begin{description}

\item[{\bf  J033233.90-274237.9}] This is a smooth face on disk galaxy as evidenced visually from the F850LP band image with a $n(F850LP)=1.03$ and a red nucleus seen clearly in the optical image as well as in the $u-z$ color map. There is a blue HII region seen in the F300W band image westwards from the nucleus. It is this patchy star formation which is responsible for a significantly lower $n(F300W)=0.30$. The F850LP band image confirms that there is no sign of any merger activity.

\item[{\bf J033244.13-274234.1  }] This seems like an edge on disk galaxy with $n(F850LP)=0.97$. The light distribution in the F300W band is more diffuse, leading to a lower value of $n(F300W)=0.38$. 

\item[{\bf J033235.98-274217.7  }] This is a marginal case, with $n(F300W)=0.47$ only slightly smaller than $n(F850LP)$. The light distribution is slightly more diffuse in the F300W band compared to the F850LP band. This shows up as blue structure at the North-East edge of the galaxy as seen in the color map. This might be contributing to the marginally lower value of $n(F300W)$ compared to the optical.

\item[{\bf  J033237.68-274219.3  }] This is a rather diffuse disk galaxy, with a distinct nucleus seen in the F850LP band and $n(F850LP)=0.90$. The F300W band light distribution is rather patchy with no clear {\it center} evident. This is responsible for the low value of $n(F300W)=0.18$.

\item[{\bf  J033236.34-274202.6  }] This is a patchy star forming galaxy with several blue HII regions seen clearly in the F300W band image as well as in the color map (particularly one due north-west of the center). The smooth underlying stellar population is revealed in the F850LP band image with $n(F850LP)=1.29$. The patchy star formation is responsible for the very low value of $n(F300W)$.

\item[{\bf  J033242.23-274130.4  }] This is a linear chain type galaxy with $n(F850LP)=0.71$. The light distribution is more diffuse in F300W band leading to a lower value of $n(F300W)$. The galaxy as a whole is rather blue with $u-z=0.38$.

\item[{\bf  J033240.57-274131.2  }] This is a featureless disk galaxy with $n(F850LP)=0.91$. A nucleus is clearly discernible in the F850LP band, though one is not clearly identifiable in the F300W band. This diffuse distribution of light in the F300W band might be responsible for the low value of $n(F300W)=0.25$. There is no sign of any merger activity.

\item[{\bf  J033242.24-274131.1 }] This is a linear chain type galaxy. The F300W band light distribution is more diffuse compared to the F850LP band, leading to a lower value of $n(F300W)$.

\item[{\bf  J033239.04-274132.4 }] This objects shows clear evidence of patchy star formation in the F300W band as well as in the color map. Multiple HII regions can be seen leading to a low value of $n(F300W)=0.16$. The F850LP band light distribution is much smoother with a close to exponential light profile ($n(F850LP)=1.03$).

\item[{\bf  J033242.77-274144.6 }] This is an edge on disk galaxy with $n(F850LP)=0.81$. This is a marginal case, since the $n(F300W)$ is only slightly smaller than $n(F850LP)$, owing probably to the slightly more diffuse light distribution in the F300W band.

\item[{\bf  J033231.49-274158.0 }] This is a rather compact galaxy with a S\'ersic profile slightly steeper than an exponential $n(F850LP)=1.28$. The F300W band light distribution is more diffuse, showing up as a blue halo conspicuous to the north-west of the galaxy center. In particular, there is no evidence of any merger activity.

\item[{\bf  J033238.34-274120.4 }] This is an elongated galaxy with $n(F850LP)=1.05$. The patchy light distribution in the F300W band is reminiscent of an irregular, with an ill defined center and parameterized by a very low value of $n(F300W)$.

\item[{\bf  J033239.25-274059.3 }] This looks like a diffuse, low surface brightness disk galaxy with $n(F850LP)=0.96$. A bright HII region (offset from the center) is clearly seen in the F300W band image as well as in the color map, which might be responsible for the low value of $n(F300W)$. No sign of any merger activity in the F850LP band.

\item[{\bf  J033241.88-274059.9 }] This is a rather red ($u-z=3.34$), compact and featureless galaxy. The F300W band light distribution is rather diffuse, probably leading to a low value of $n(F300W)$. In particular, there are no signs of any merger activity.

\item[{\bf  J033241.64-274103.4 }] This is an edge on disk galaxy with $n(F850LP)=1.19$. The F300W band light distribution is more fragmented: notice the relatively redder center with bluer edges due north-west and south-east of the center in the color map. Again, there are no signs of any merger activity.

\item[{\bf  J033238.75-274027.8 }] This is a compact galaxy with a steeper than exponential light profile $n(F850LP)=1.10$. The F300W band light distribution is fragmented, leading to a very low value of $n(F300W)$.

\item[{\bf  J033241.42-274044.8 }] This is a very diffuse galaxy, reminiscent of an Irregular. This is a marginal case with the $n(F300W)$ only slightly lower than the $n(F850LP)$.

\end{description}

As can be seen from the individual description above and from Fig.~\ref{fig9}, most of the objects with $n(F300W)\leq 0.5$ and $n(F850LP)\sim1.0$ are probably clumpy starforming galaxies with an underlying disk ($n\sim1.0$) of older stellar population. Many objects are isolated and compact and the low value of $n(F300W)$ can be attributed to diffuse/patchy light distribution in the F300W band rather than any merger activity. As mentioned earlier, while it is true that blue neighbours will tend to mimic HII regions in the rest-frame UV by producing a shallow S\'ersic profile, however the very fact that they contribute little or no flux in the optical (parametrized by a steeper $n(F850LP)\sim1.0$) means that they cannot be massive systems (as z band flux is a better tracer of stellar mass than UV flux) and will constitute {\em minor} mergers at best. This analysis should go a long way in convincing the reader that using S\'ersic profile fitting in the rest-frame UV is fraught with severe biases and should be used with caution.

\section{Summary and conclusion}

We have compared the rest-frame UV vs rest-frame optical morphologies
of intermediate redshift ($z_{\rm median}=1.02$) UV bright galaxies, selected using the deep HST/WFPC2
U-band HUDF parallels image in combination with the HST/ACS F850LP GOODS dataset. 
The UV bright galaxies in this case are at lower redshifts and have lower 
luminosities than high-{\it z} LBGs (de Mello et al. 2006).
We performed single component S\'ersic fits in both WFPC2/F300W and
ACS/F850LP bands for the 162 objects and find that: 

1) In the rest-frame UV, the S\'ersic index $n$ is lower in a large number of cases compared to the rest-frame optical. This is possibly due to the fact that 
rest-frame UV light is sensitive to star-forming HII regions which are fragmented and patchy in nature leading to a light distribution which is shallower than the underlying galaxy light distribution seen in the rest-frame optical. 
This difference is mainly caused by galaxies with low values of $n(F300W)\leq0.5$ that have a value of $n(F850LP)\sim1.0$, and is not affected by galaxies with high values of $n$, as can be clearly seen in Fig.~\ref{n_comparision}(right).
This has serious consequences as the S\'ersic index $n$ is widely used for identifying the morphological class of a galaxy, with $n$ flatter than exponential being used for identifying merger candidates. In particular, we find that the use of rest-frame UV overestimates the number of merger candidates by $\sim$40\%-100\% compared to the rest-frame optical, depending upon the cutoff in $n$ employed for identifying merger candidates.
This shows that the so-called morphological K-correction is a serious consideration in constraining galaxy morphology at intermediate redshifts, as shown by Conselice et al.~\cite{conselice2008}.
 Using $n\leq0.8$ as the criterion for identifying merger candidates, we recommend a correction factor of $\sim1.4$ to the result of Ravindranath et al.~\cite{ravindranath2006} for calculating the fraction of LBGs at $z\sim3$ which are likely to be merger candidates.   

%

2) We also find that in the rest-frame UV, the number of high ellipticity ($e\geq0.8$) objects is higher by a factor $\sim2.8$ compared to the rest-frame optical. 
This might explain the reported results in literature that high-{\it z} LBGs tend to show a skew towards higher ellipticities (Ravindranath et al. 2006), as most such work has been done using rest-frame UV datasets. We also note that the axis ratio $b/a$ is lower, i.e. ellipticity is higher in rest-frame UV compared to the rest-frame optical. The mean values of the axis ratio are $(b/a)_{F300W}=0.42\pm0.23$ and $(b/a)_{F850LP}=0.49\pm0.22$.

This indicates that the reported dominance of elongated morphologies among high-{\it z} LBGs
might just be a bias related to the use of rest-frame UV datasets.
However, we cannot conclusively exclude the
possibility that LBGs might still have intrinsically different or elongated morphologies. 

Our results are in agreement with the work of Papovich et al.~\cite{papovich2005}, who found that $z\sim1$ galaxies show strong transformation between their rest-frame UV and optical morphologies, using NICMOS observations of HDF-N.
It has been reported (Papovich et
al. 2005, Conselice et al. 2005) that at redshifts $2.0\leq z \leq
3.0$, the morphology of a galaxy remains essentially unchanged from
rest-frame UV to rest-frame optical. However, one should not neglect
the effect produced by the $(1+z)^{-4}$ surface brightness dimming, 
which at higher redshifts suppresses the (relatively low surface
brightness) underlying galaxian light (by $\sim6$ magnitudes at
$z\sim3.0$), so that the only parts of a galaxy that are left visible 
even in the rest-frame optical are the high surface brightness HII
regions (while the underlying galaxy remains unconstrained). 
Future medium depth survey like GOODS (depth $\sim$10 orbits) using the Near-IR channel of the upcoming camera, WFC3 on HST might {\it{fail}} to probe the underlying rest-frame
optical galaxian light at $z\sim3.0$ simply due to $(1+z)^{-4}$
surface brightness dimming effects. Ultra-deep images (depth $\sim$150 orbits) with the Near-IR channel of WFC3 will be
required to constrain the underlying galaxian light at $z\sim3.0$ and
will allow similar analysis of the morphology of galaxies to high-{\it z}.

\acknowledgments A.R. would like to thank CSIR for PhD. funding. Special thanks to Swara Ravindranath for fruitful discussions and comments. We would like to thank the referee for his/her insightful comments and constructive criticisms that have greatly improved the quality of this work.

\clearpage

\begin{figure*}[t!] \centering

\resizebox{0.315\textwidth}{!}{\includegraphics*[0.6cm,0.6cm][10.5cm,10.5cm]{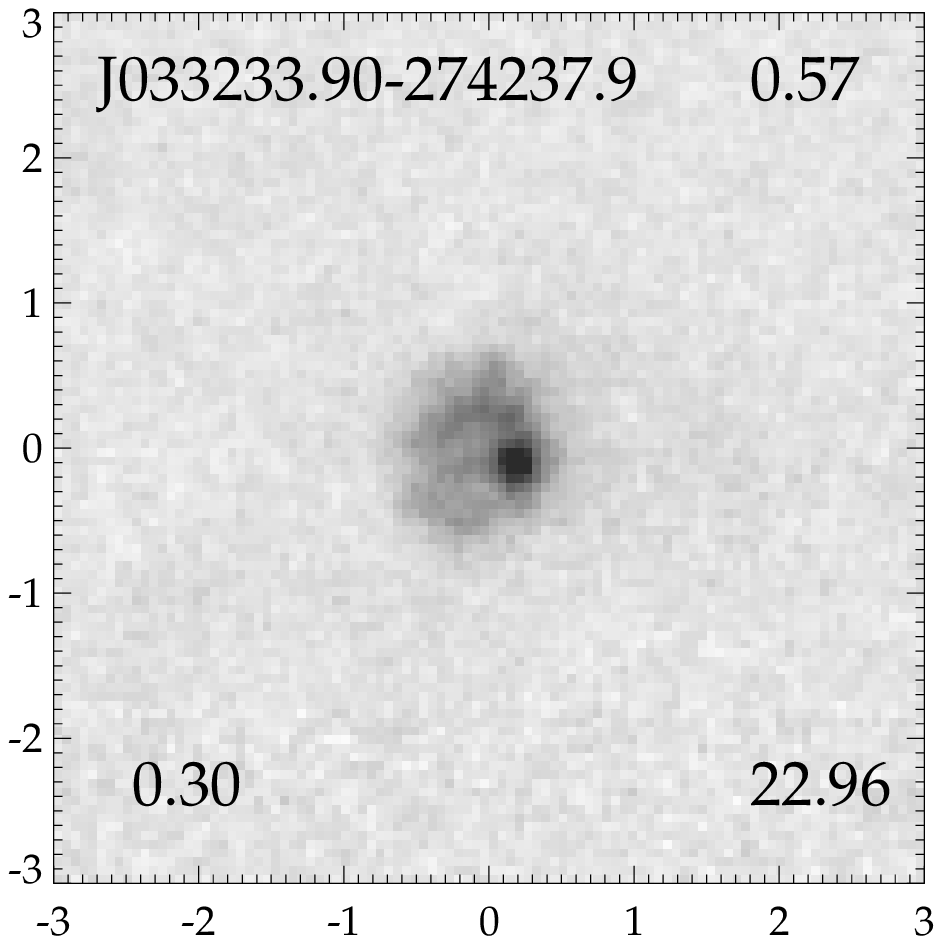}}%
\resizebox{0.31\textwidth}{!}{\includegraphics*[0cm,0cm][11cm,11cm]{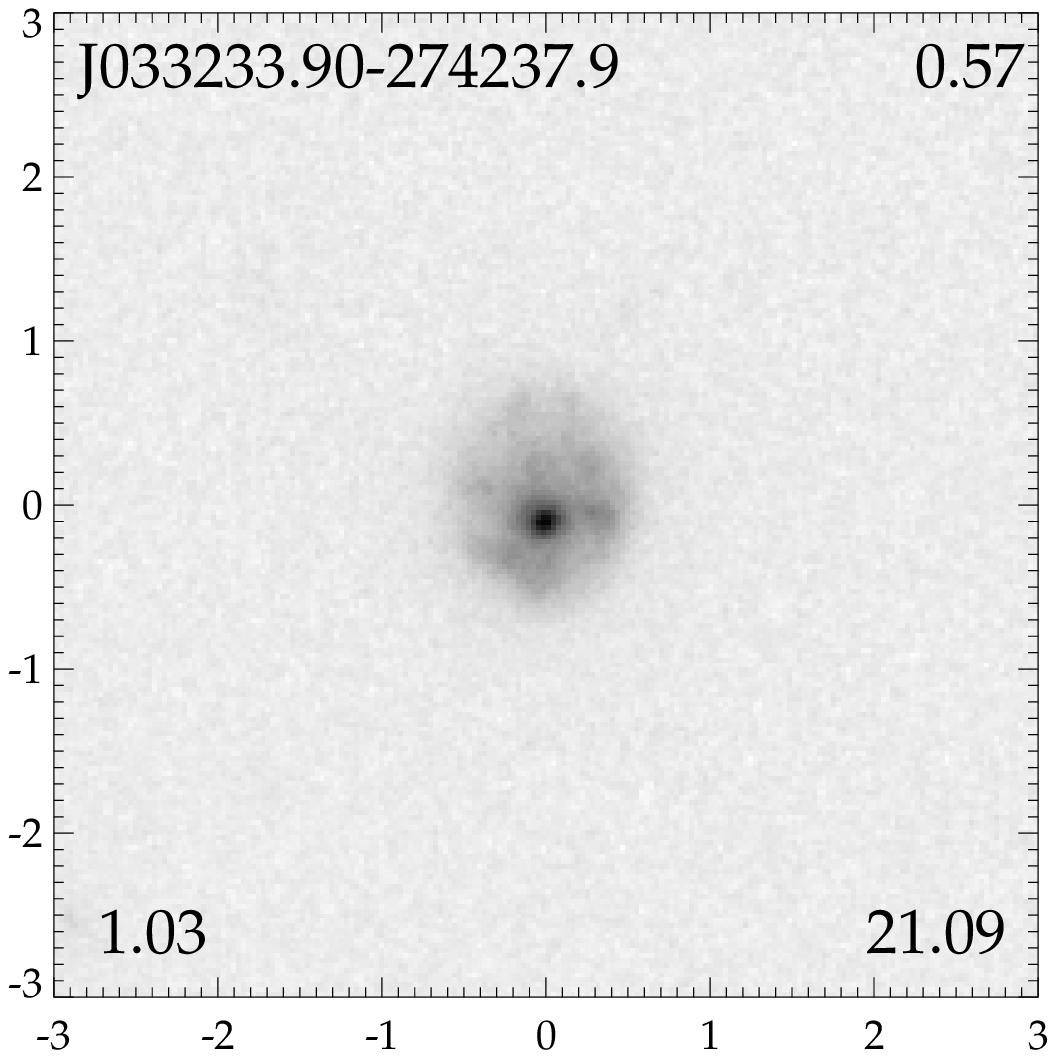}}%
\resizebox{0.4\textwidth}{!}{\includegraphics*[0cm,0.2cm][13.5cm,11cm]{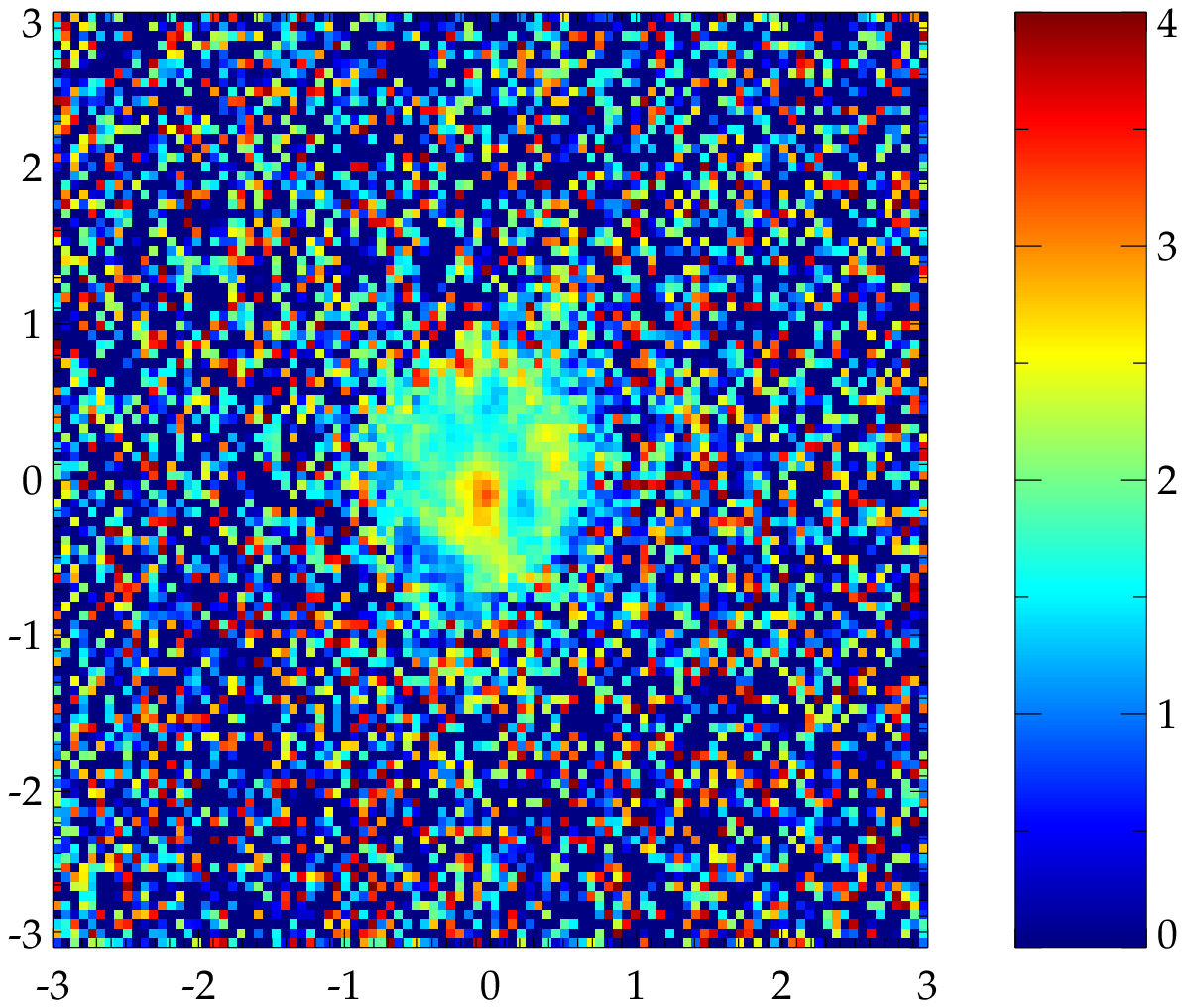}}

\resizebox{0.315\textwidth}{!}{\includegraphics*[0.6cm,0.6cm][10.5cm,10.5cm]{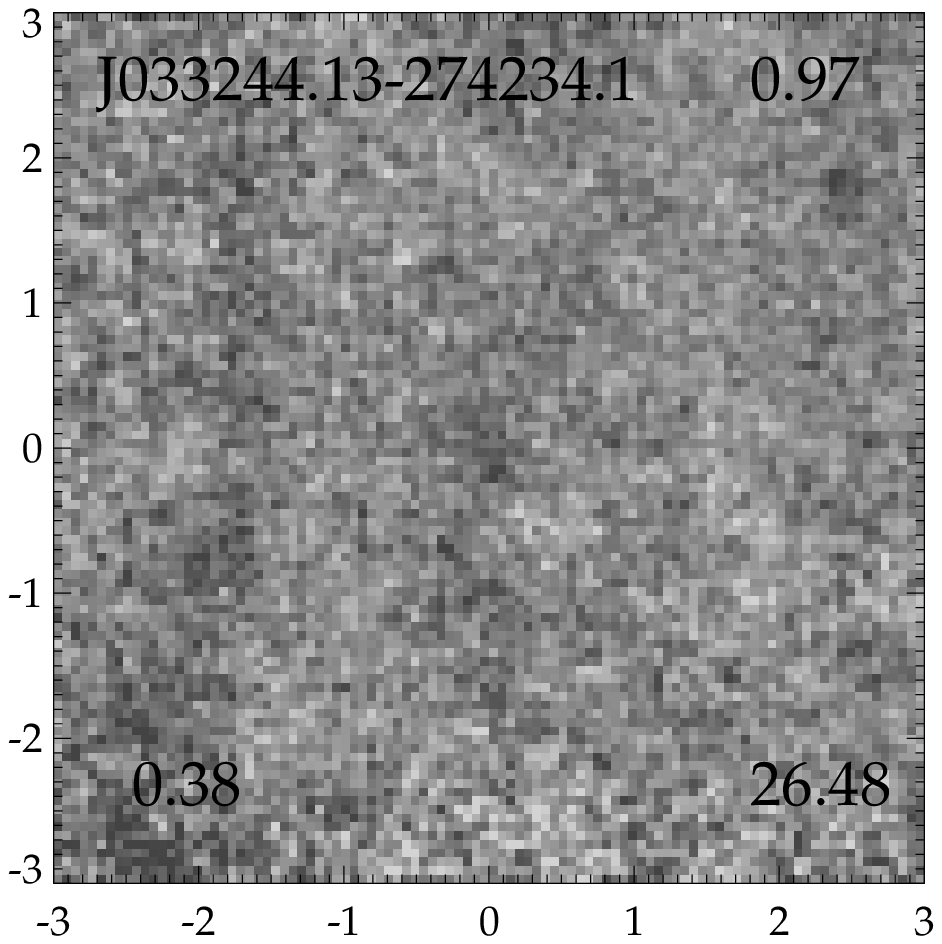}}%
\resizebox{0.31\textwidth}{!}{\includegraphics*[0cm,0cm][11cm,11cm]{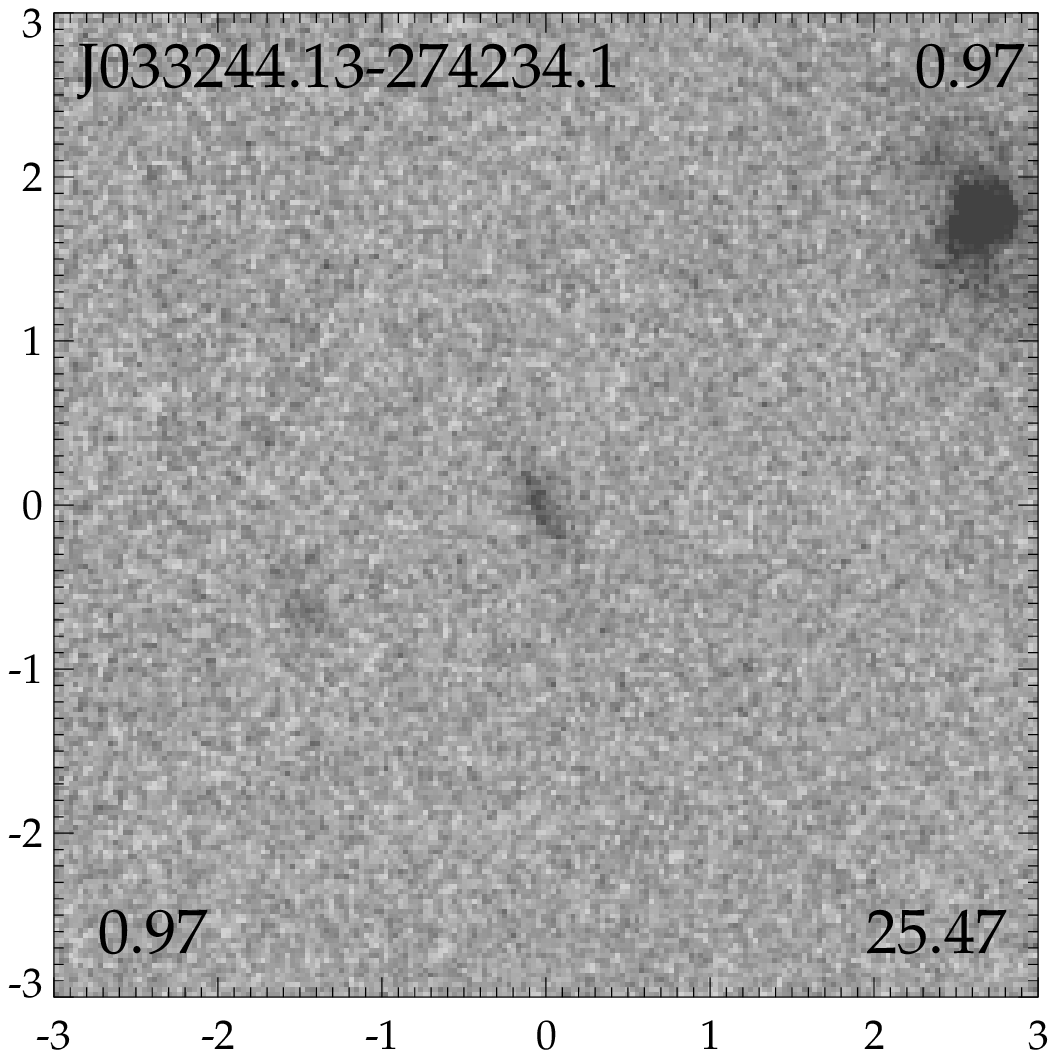}}%
\resizebox{0.4\textwidth}{!}{\includegraphics*[0cm,0.2cm][13.5cm,11cm]{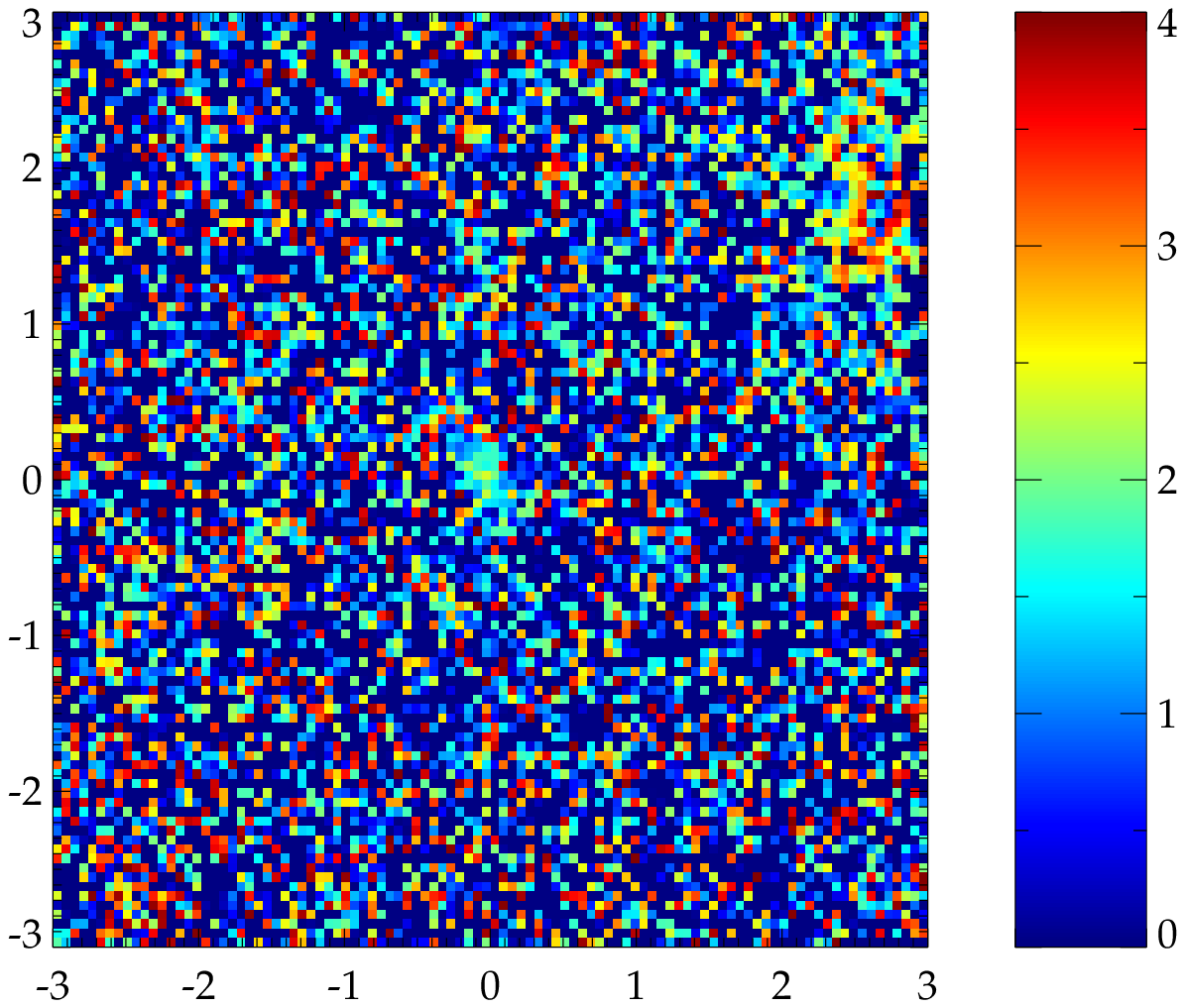}}

\resizebox{0.315\textwidth}{!}{\includegraphics*[0.6cm,0.6cm][10.5cm,10.5cm]{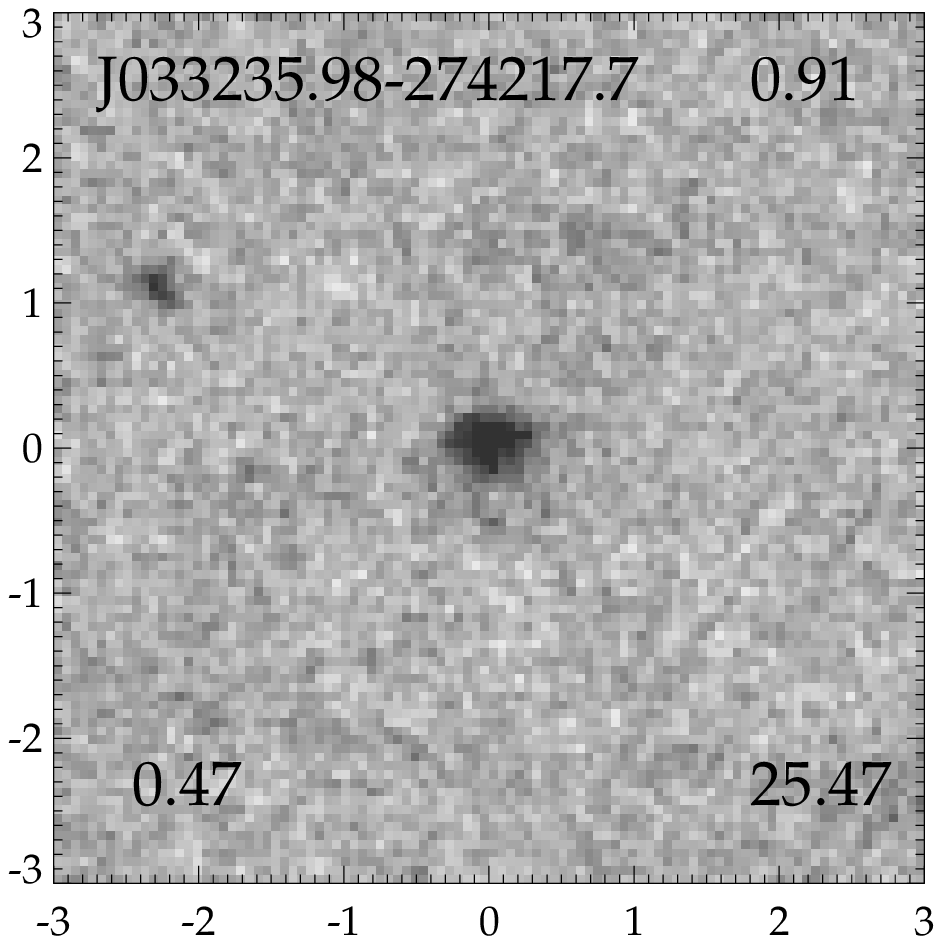}}%
\resizebox{0.31\textwidth}{!}{\includegraphics*[0cm,0cm][11cm,11cm]{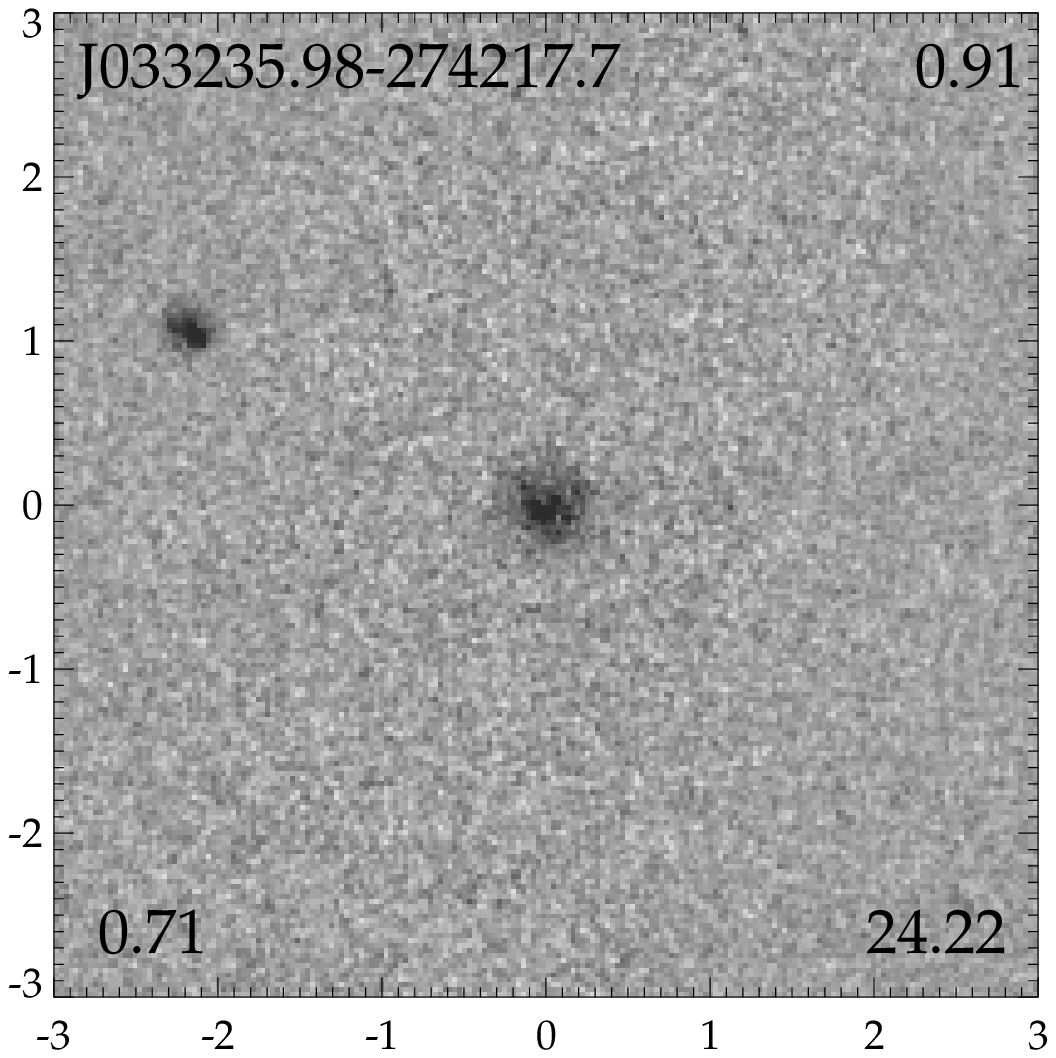}}%
\resizebox{0.4\textwidth}{!}{\includegraphics*[0cm,0.2cm][13.5cm,11cm]{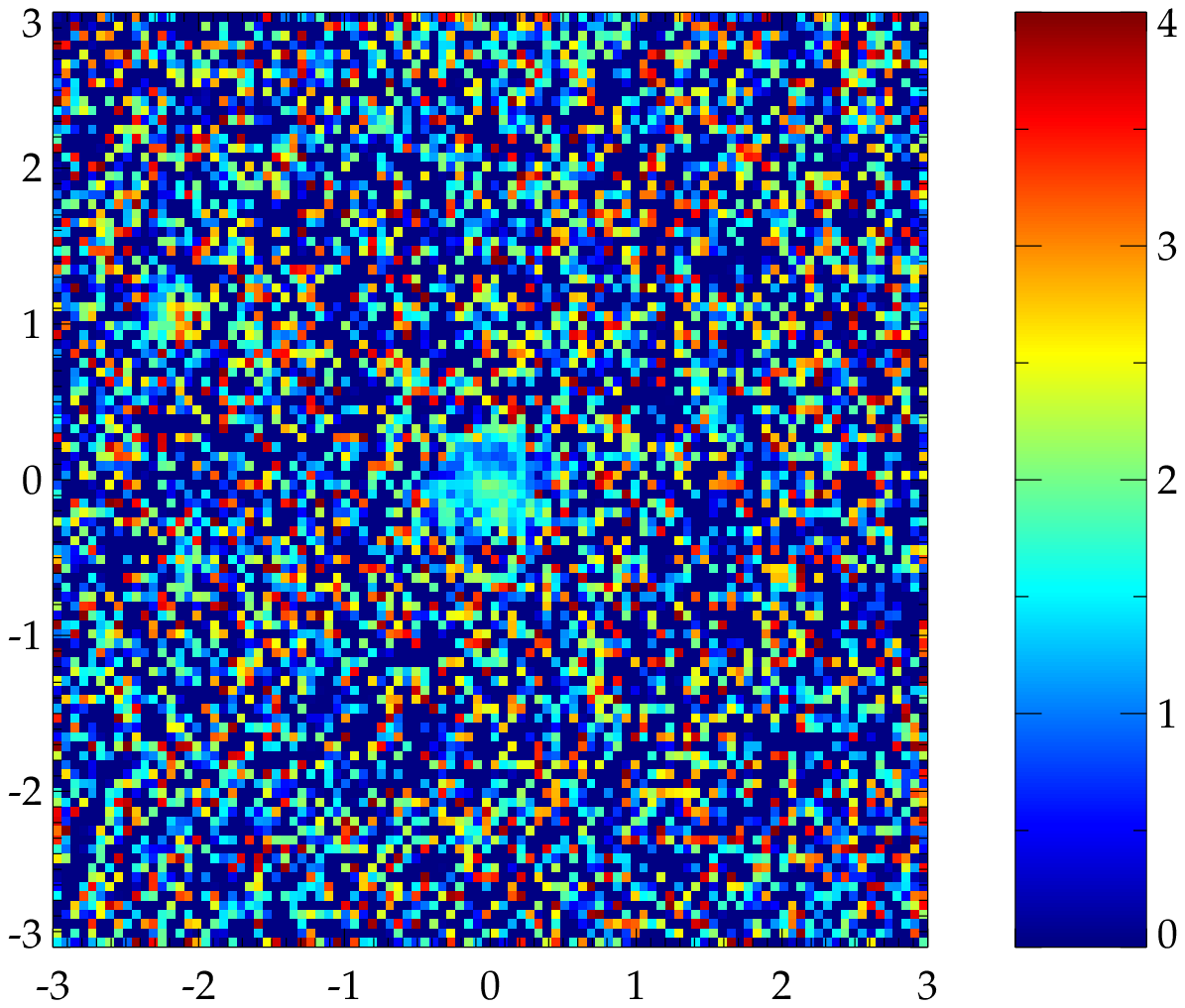}}

\resizebox{0.315\textwidth}{!}{\includegraphics*[0.6cm,0.6cm][10.5cm,10.5cm]{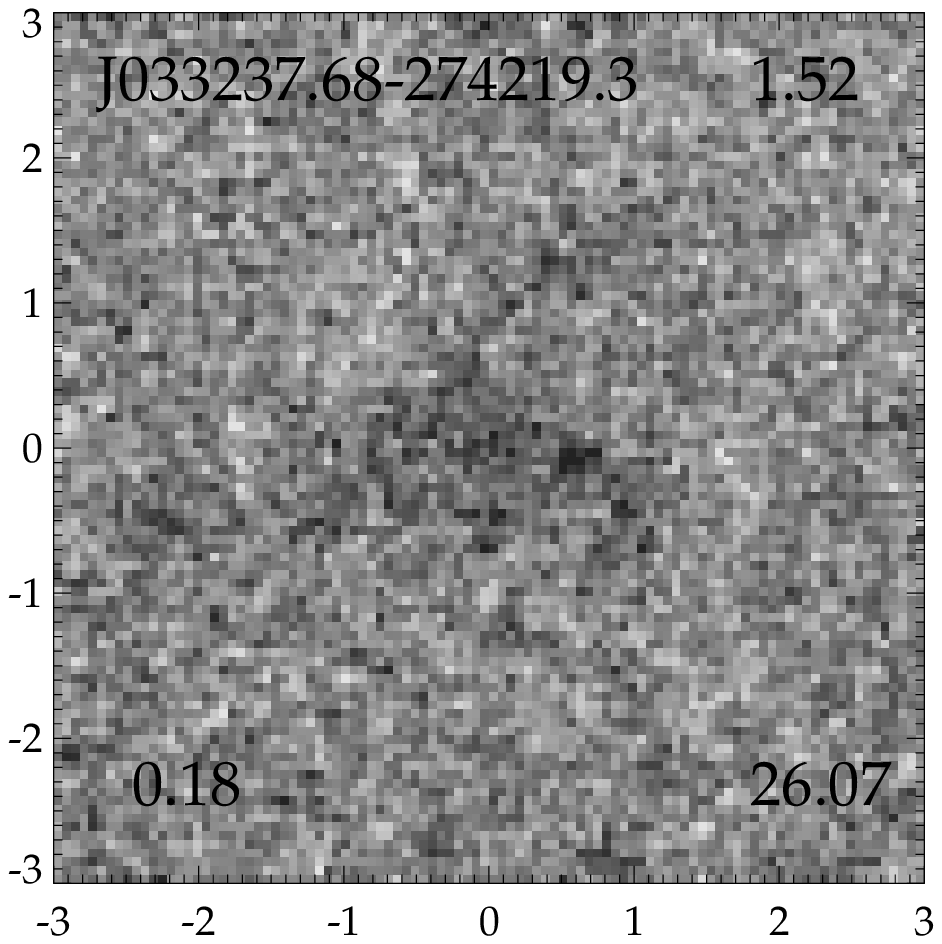}}%
\resizebox{0.31\textwidth}{!}{\includegraphics*[0cm,0cm][11cm,11cm]{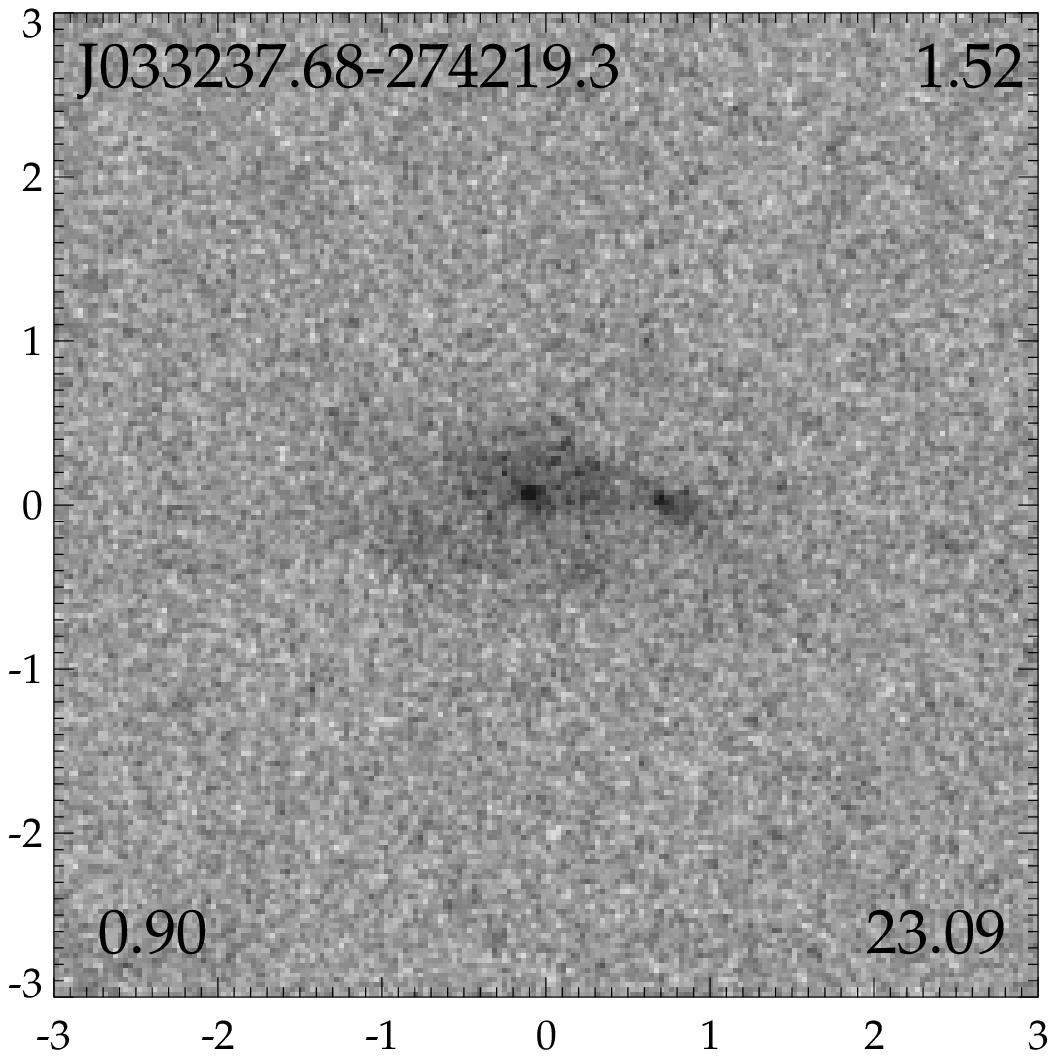}}%
\resizebox{0.4\textwidth}{!}{\includegraphics*[0cm,0.2cm][13.5cm,11cm]{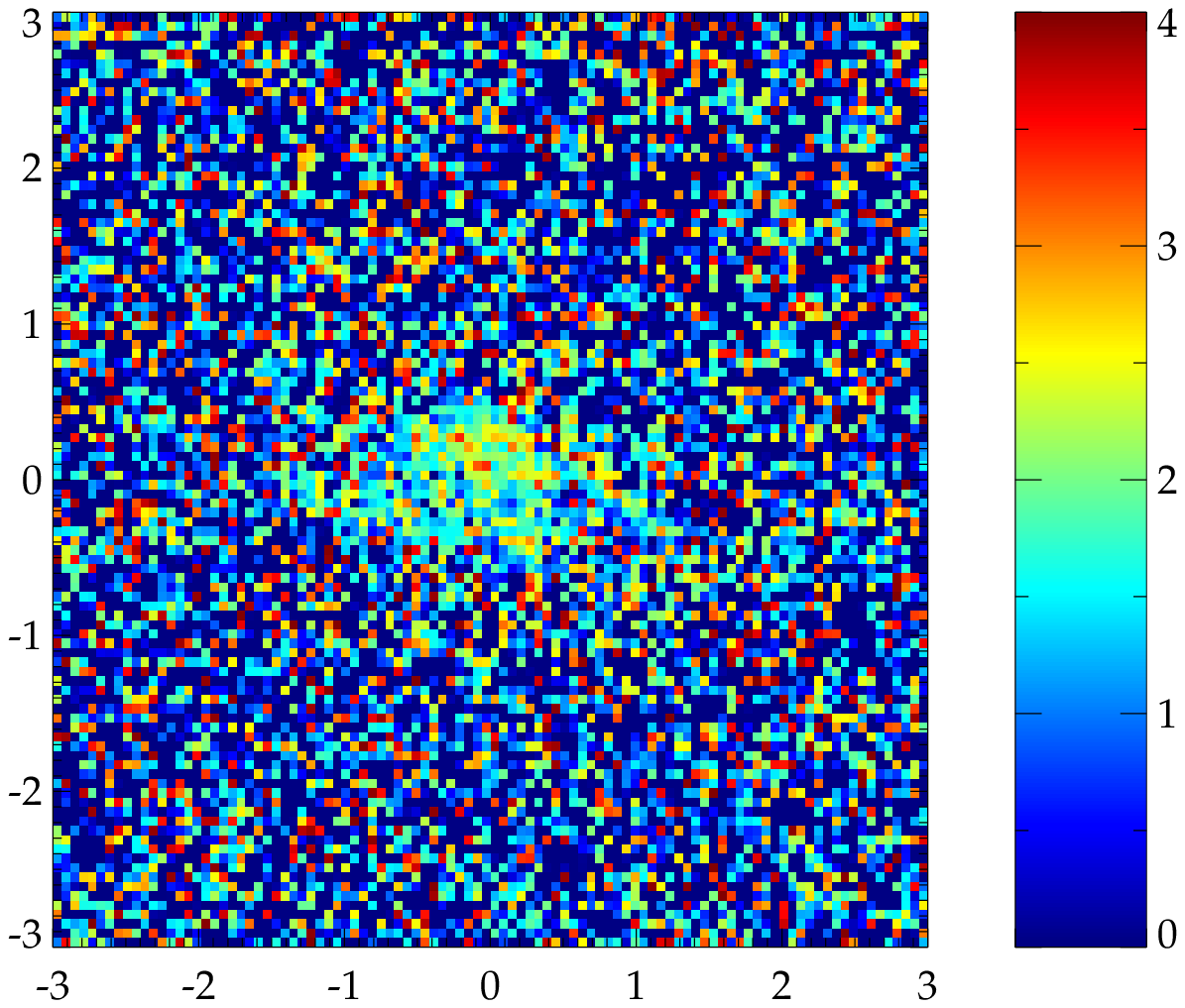}}

\caption{F300W, F850LP and the $u-z$ color map images of objects with $n(F300W)\leq0.5$ and $0.7\leq n(F850LP) \leq1.3$. Explanation is given at the end of this figure.}
\label{fig9}
\end{figure*}

\addtocounter{figure}{-1}

\begin{figure*}[t!] \centering

\resizebox{0.315\textwidth}{!}{\includegraphics*[0.6cm,0.6cm][10.5cm,10.5cm]{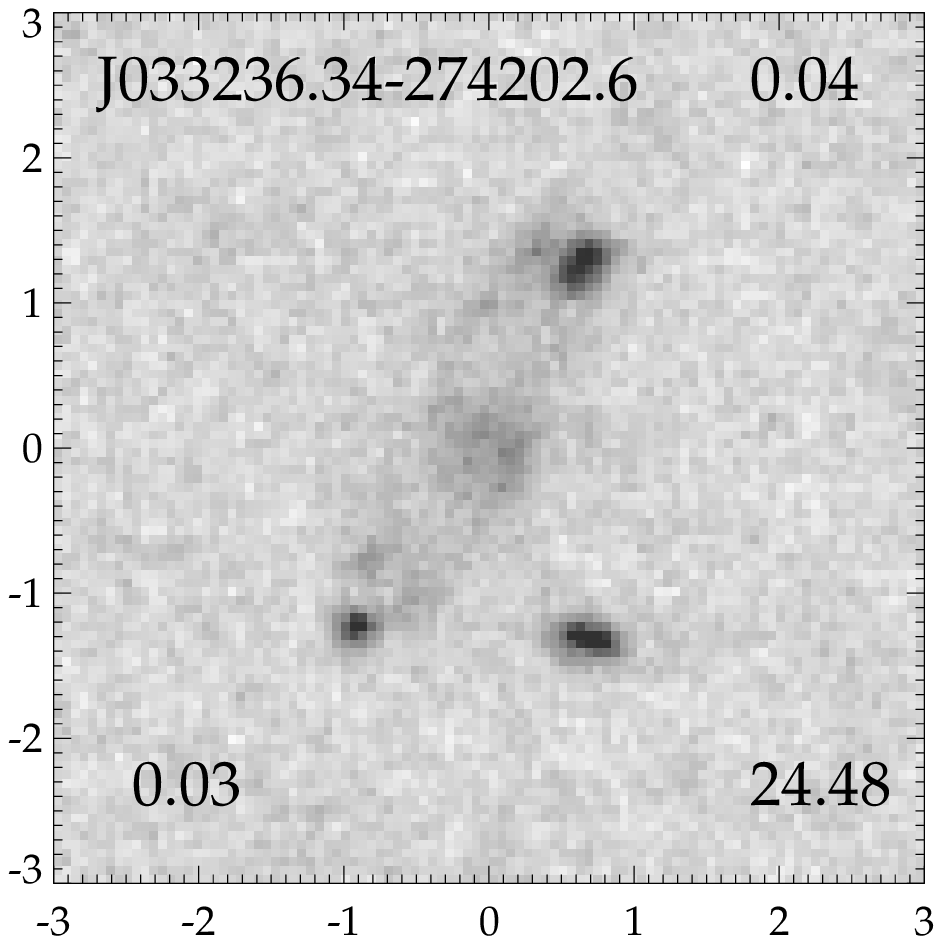}}%
\resizebox{0.31\textwidth}{!}{\includegraphics*[0cm,0cm][11cm,11cm]{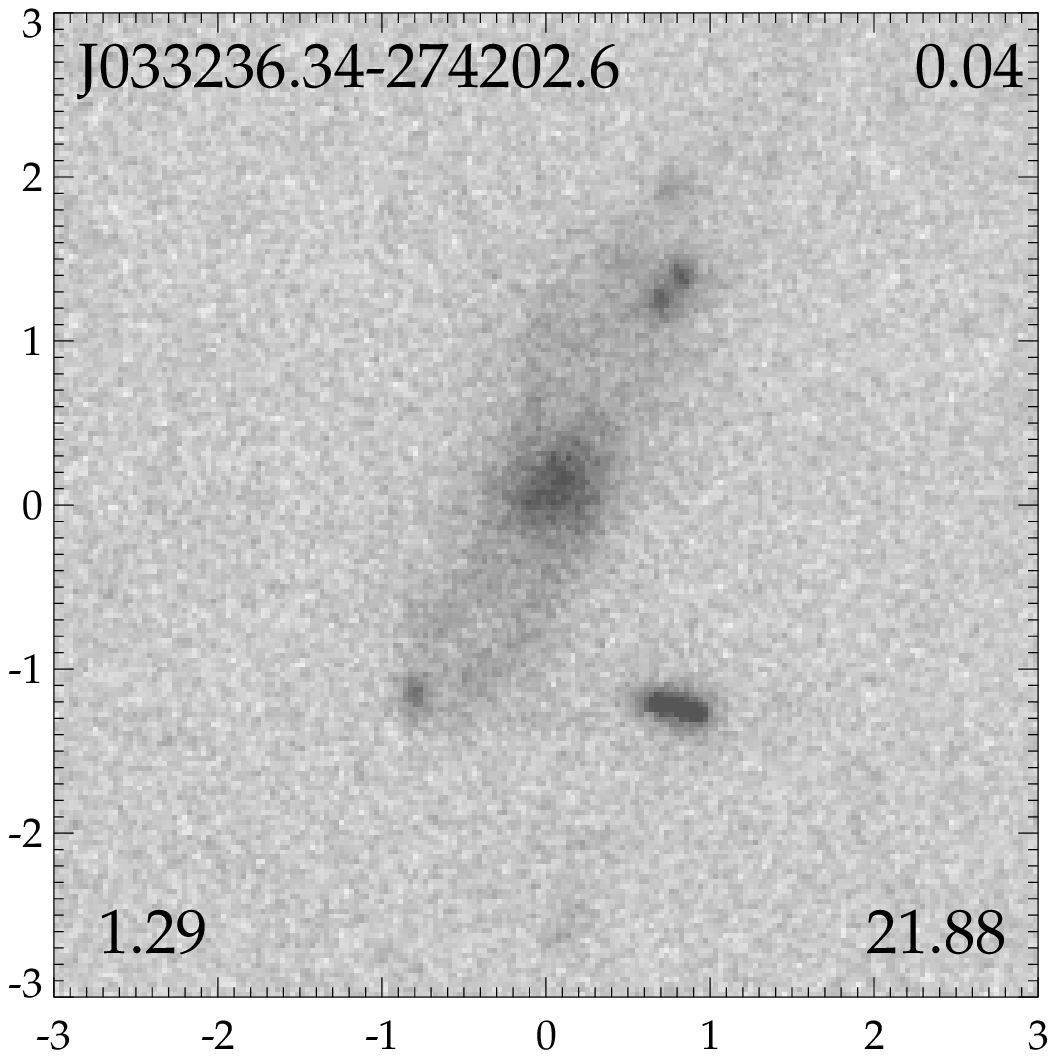}}%
\resizebox{0.4\textwidth}{!}{\includegraphics*[0cm,0.2cm][13.5cm,11cm]{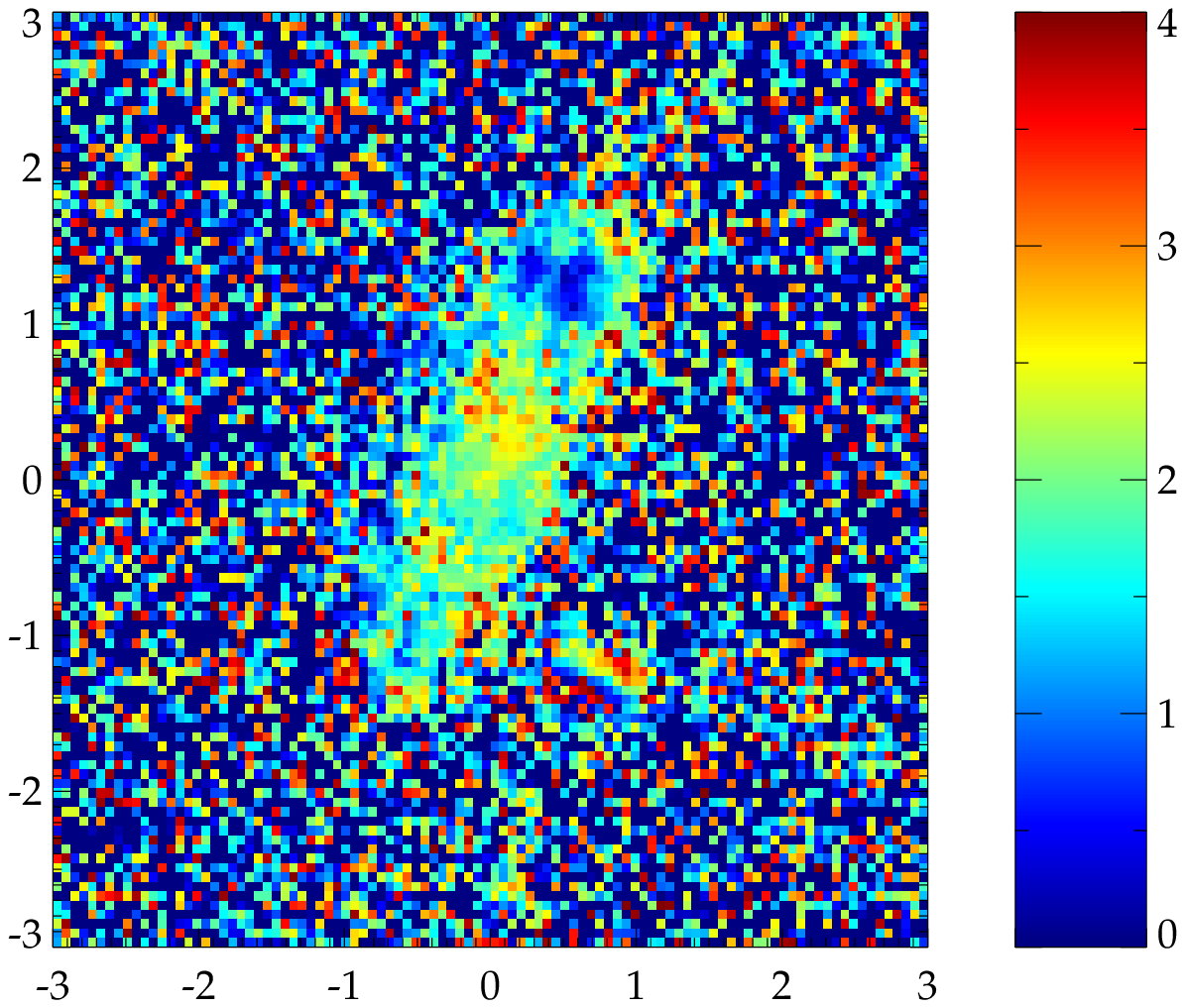}}

\resizebox{0.315\textwidth}{!}{\includegraphics*[0.6cm,0.6cm][10.5cm,10.5cm]{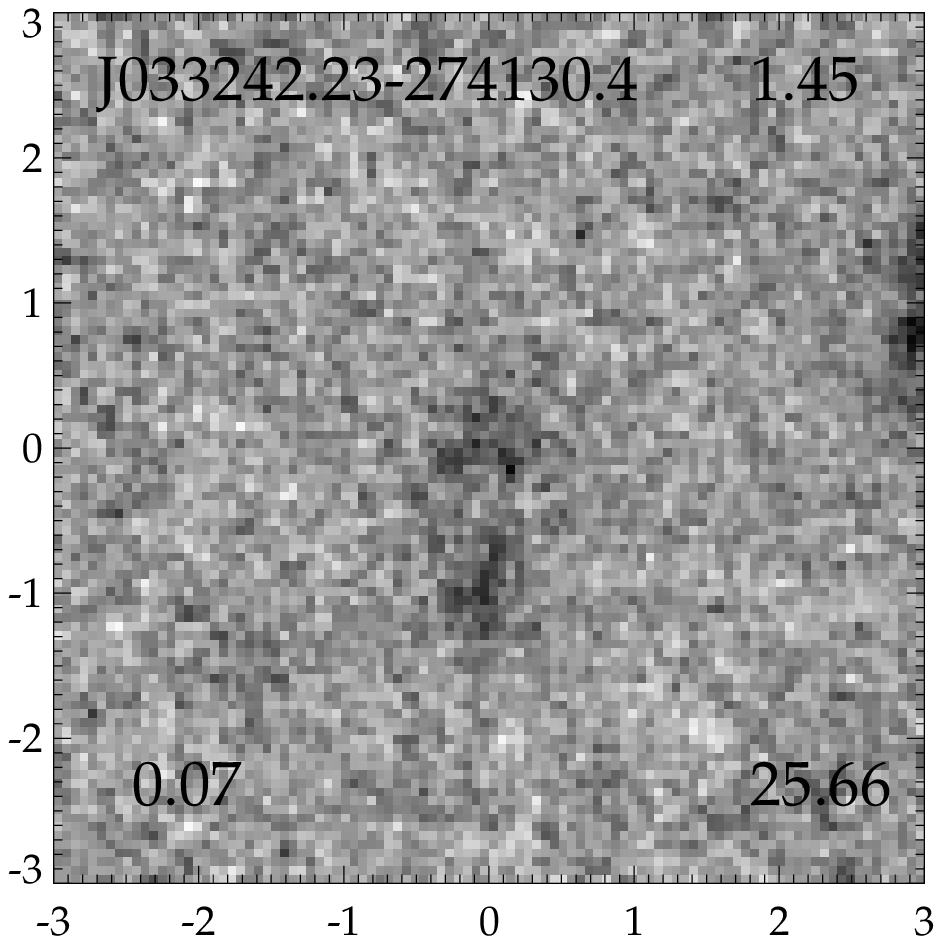}}%
\resizebox{0.31\textwidth}{!}{\includegraphics*[0cm,0cm][11cm,11cm]{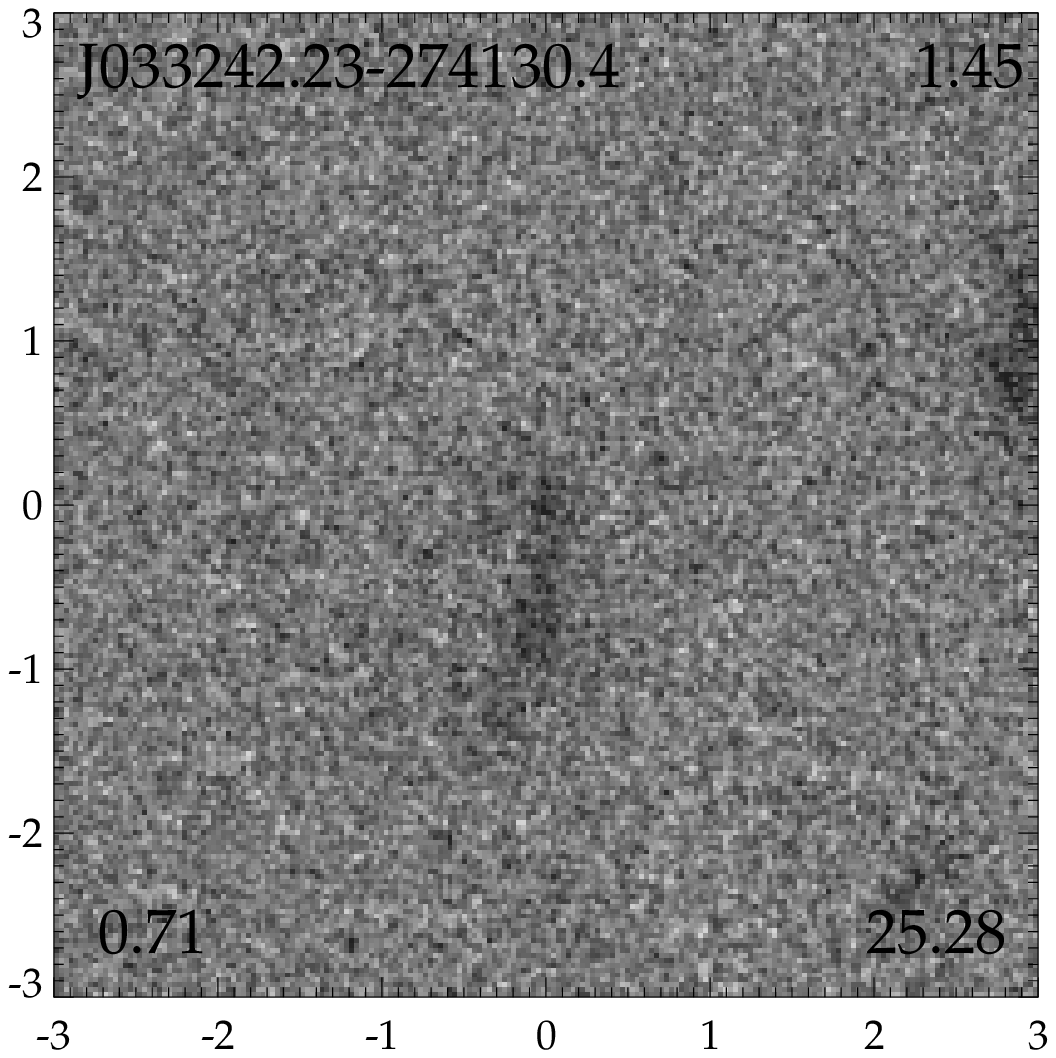}}%
\resizebox{0.4\textwidth}{!}{\includegraphics*[0cm,0.2cm][13.5cm,11cm]{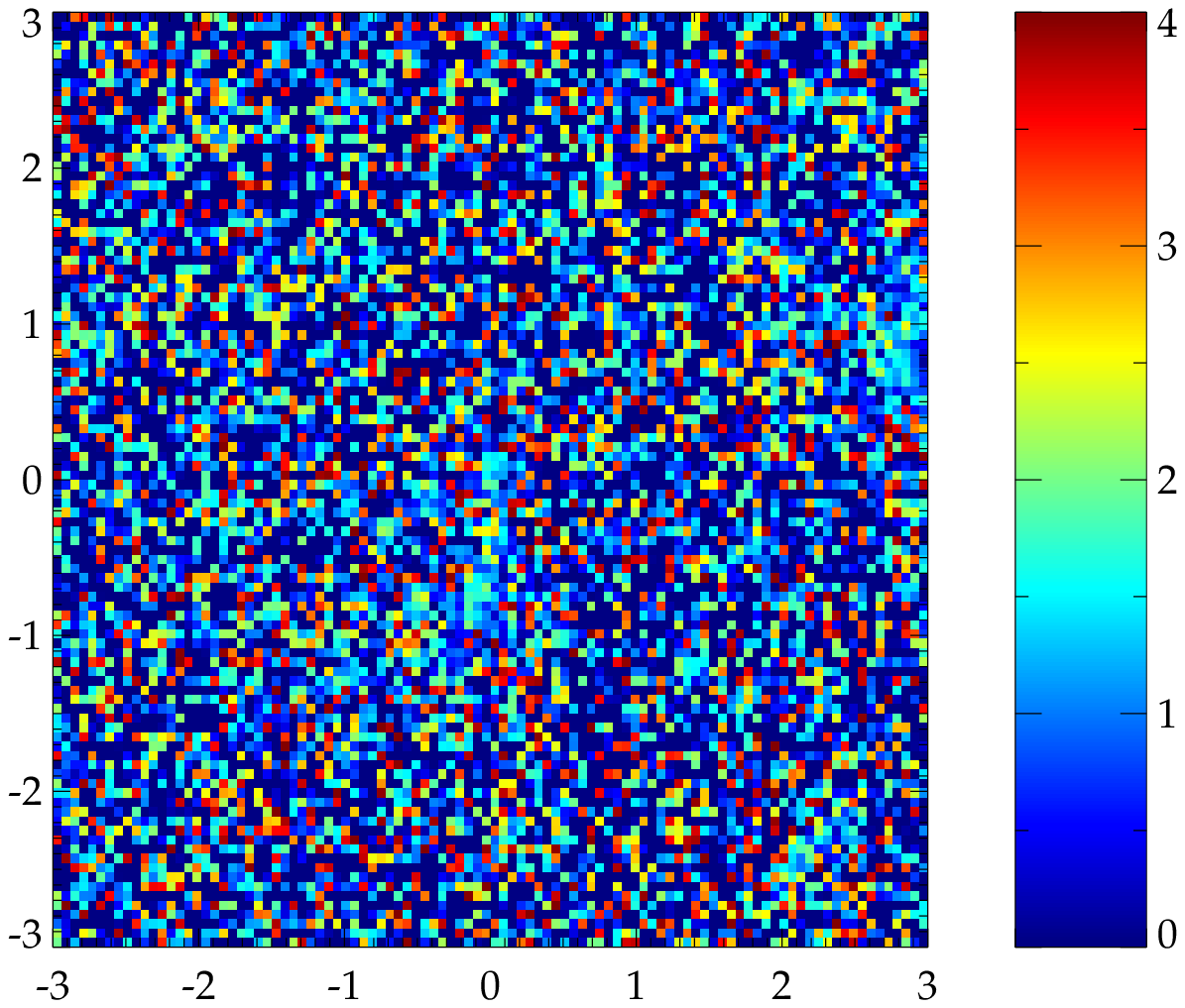}}

\resizebox{0.315\textwidth}{!}{\includegraphics*[0.6cm,0.6cm][10.5cm,10.5cm]{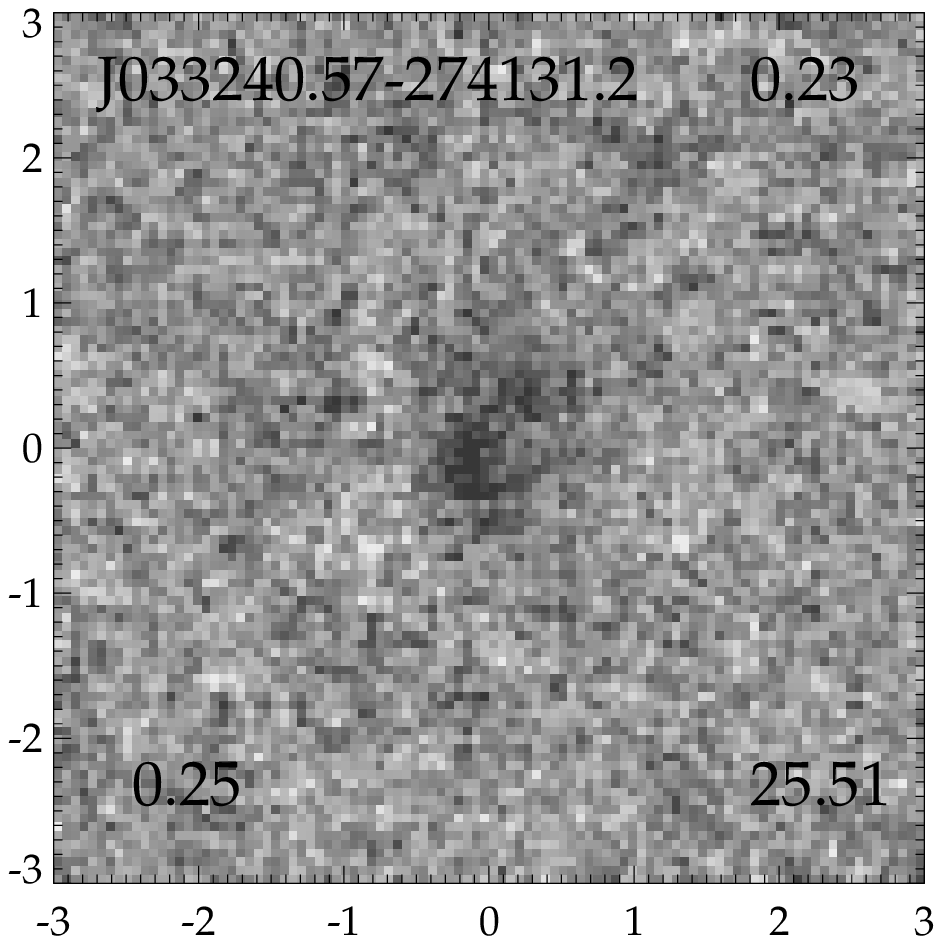}}%
\resizebox{0.31\textwidth}{!}{\includegraphics*[0cm,0cm][11cm,11cm]{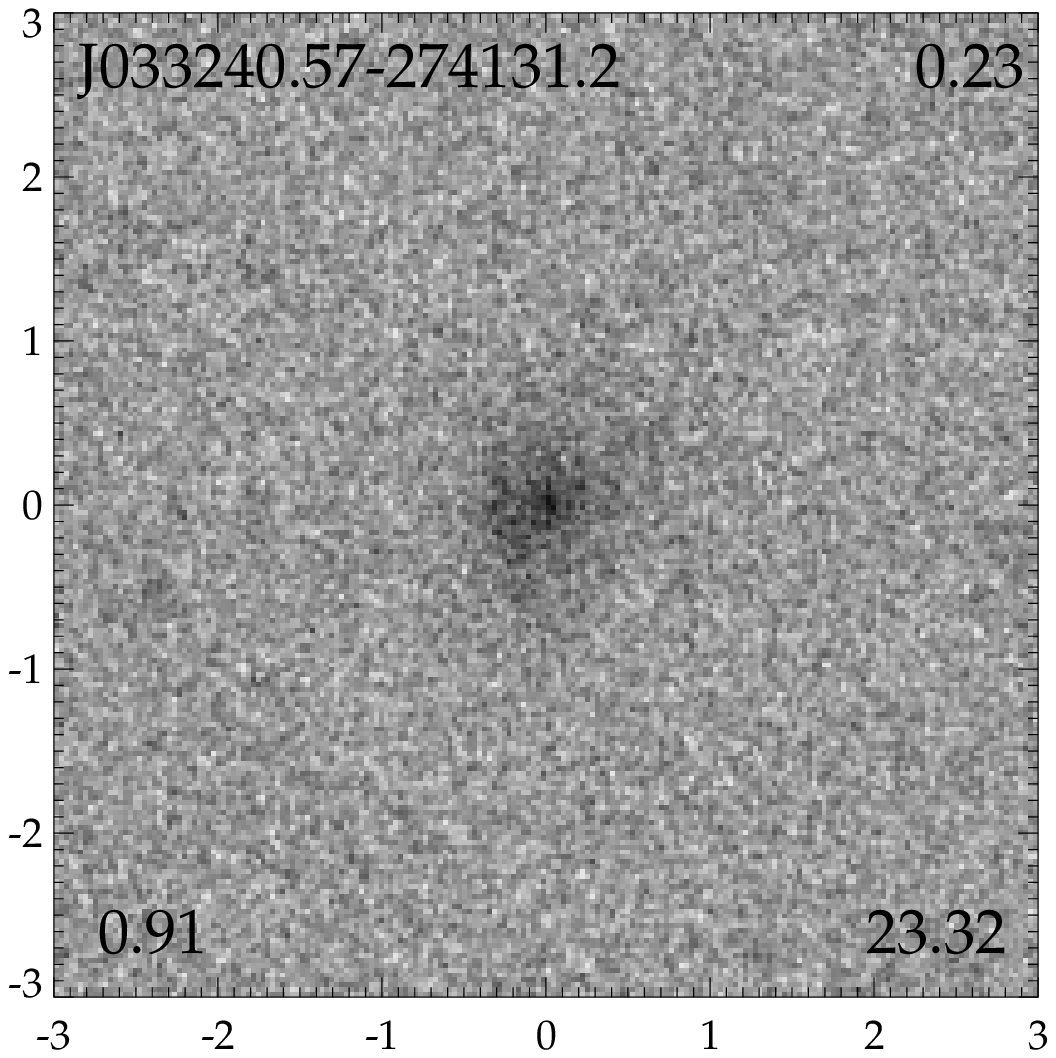}}%
\resizebox{0.4\textwidth}{!}{\includegraphics*[0cm,0.2cm][13.5cm,11cm]{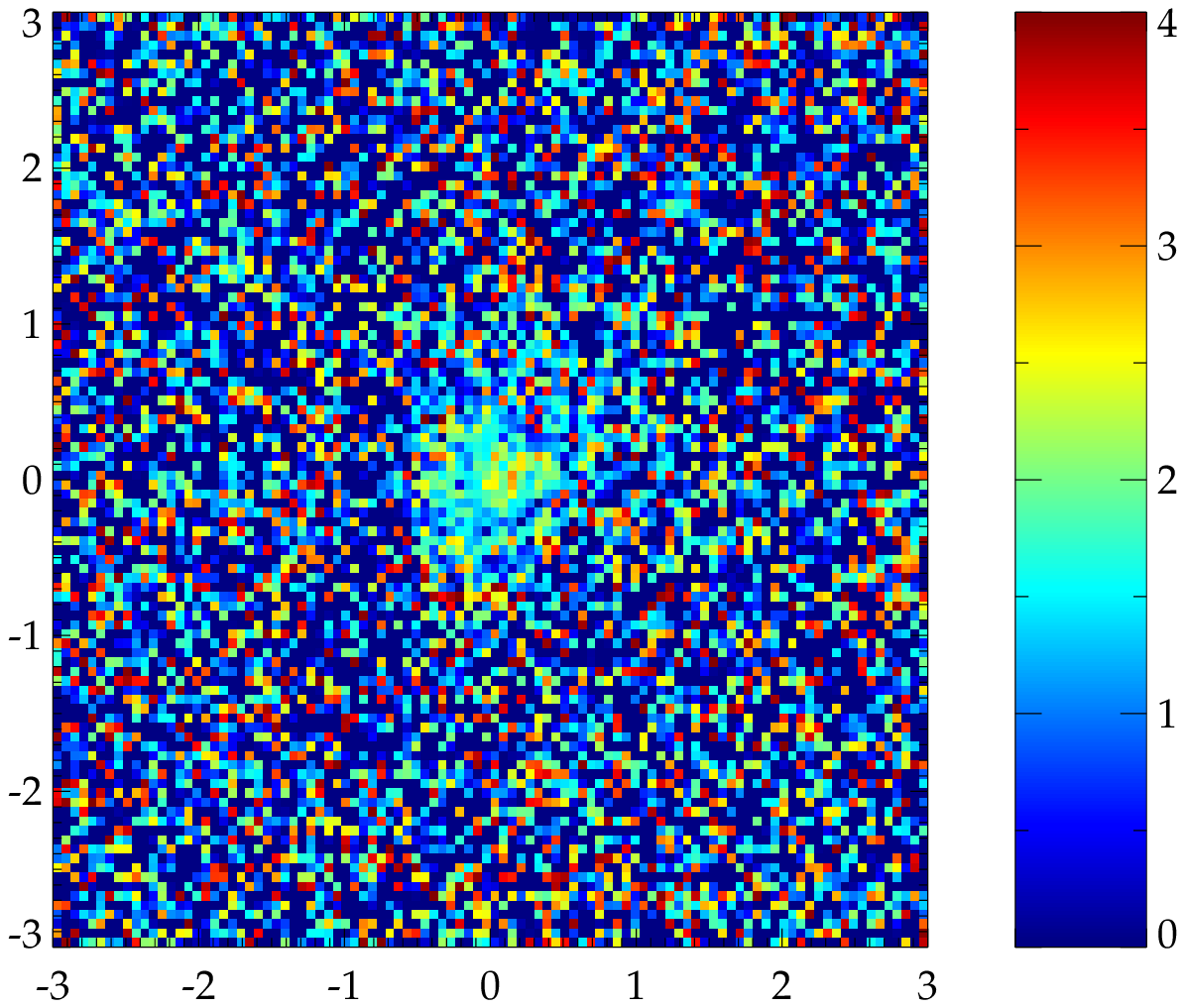}}

\resizebox{0.315\textwidth}{!}{\includegraphics*[0.6cm,0.6cm][10.5cm,10.5cm]{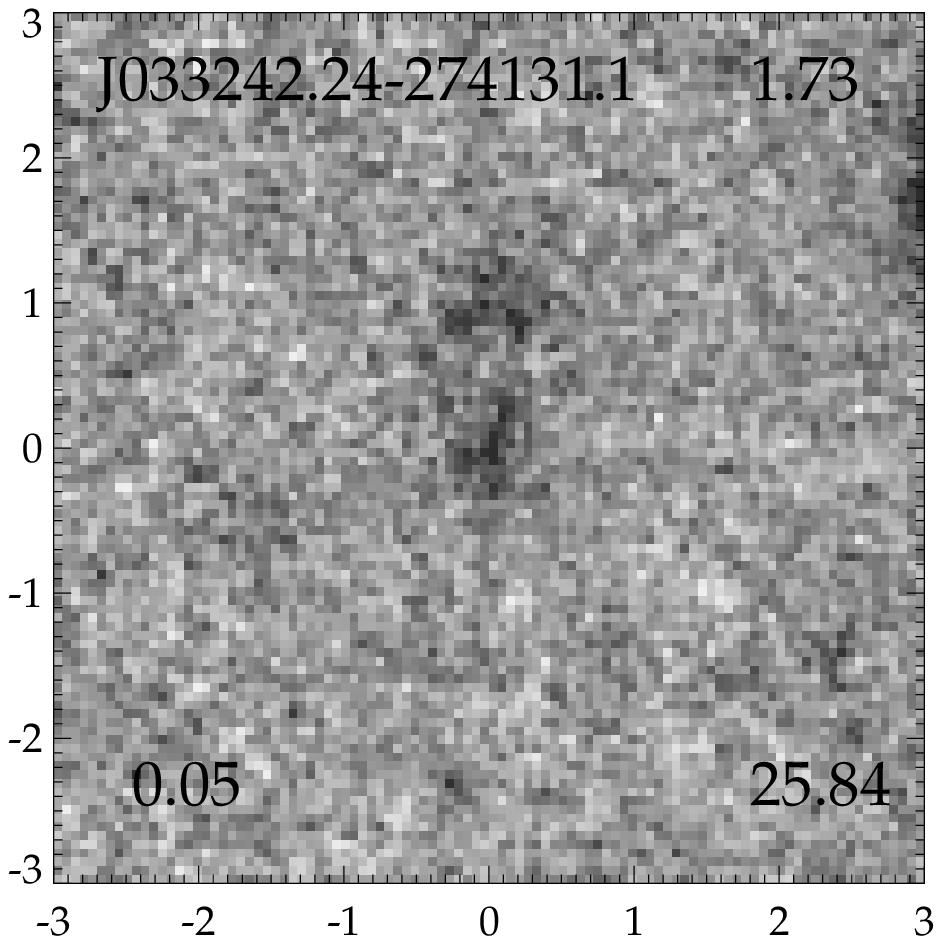}}%
\resizebox{0.31\textwidth}{!}{\includegraphics*[0cm,0cm][11cm,11cm]{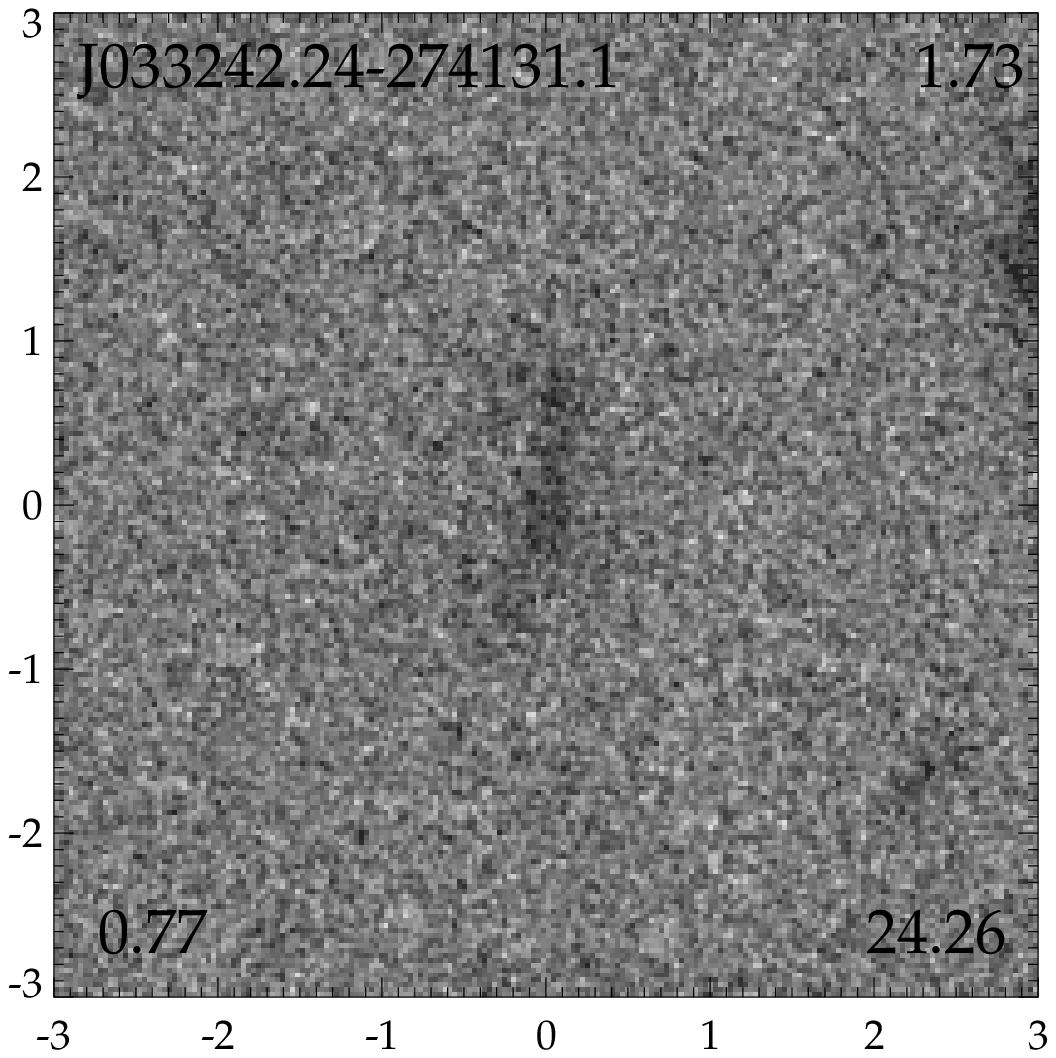}}%
\resizebox{0.4\textwidth}{!}{\includegraphics*[0cm,0.2cm][13.5cm,11cm]{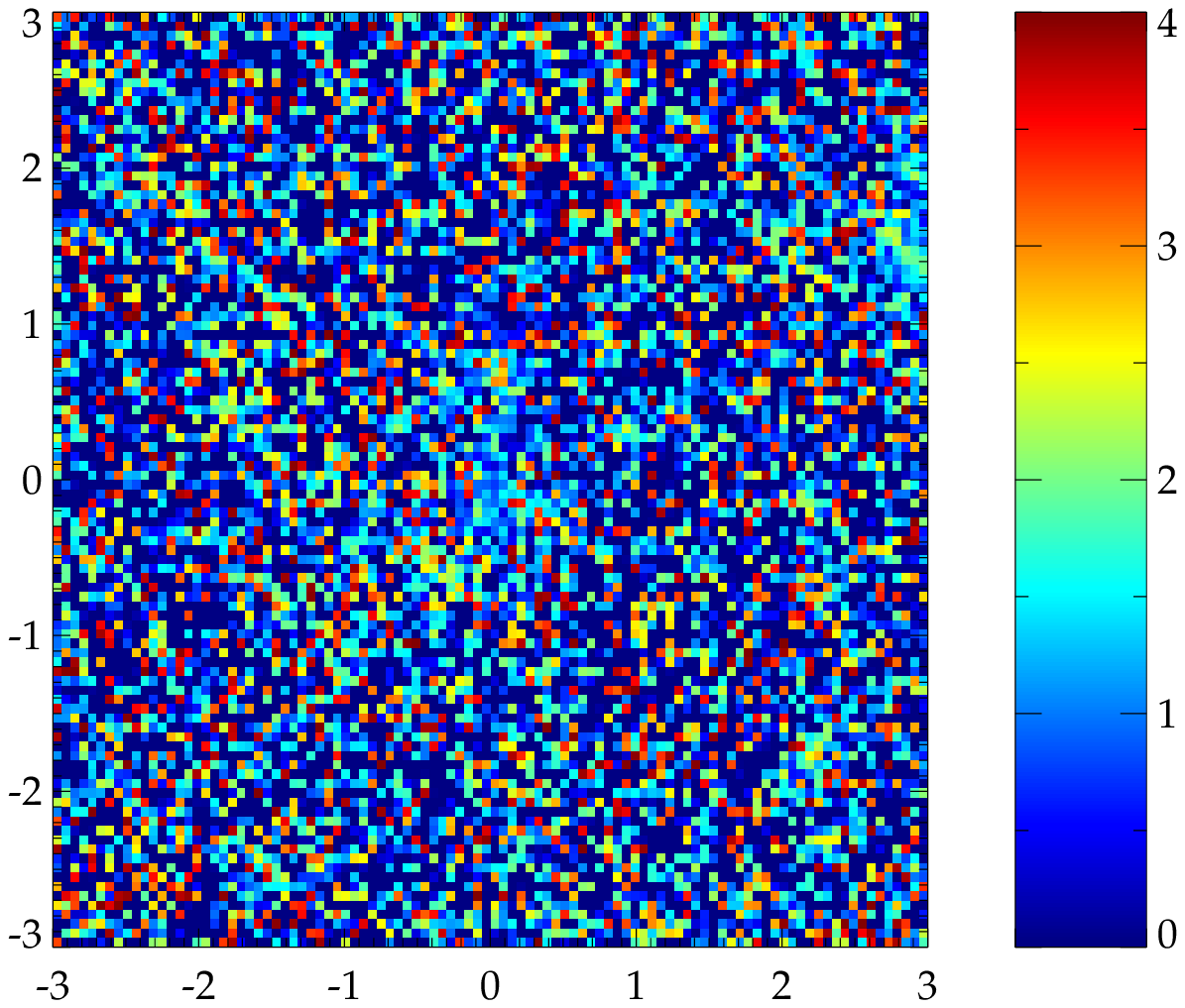}}

\caption{Continued.}   

\end{figure*}

\addtocounter{figure}{-1}

\begin{figure*}[t!] \centering

\resizebox{0.315\textwidth}{!}{\includegraphics*[0.6cm,0.6cm][10.5cm,10.5cm]{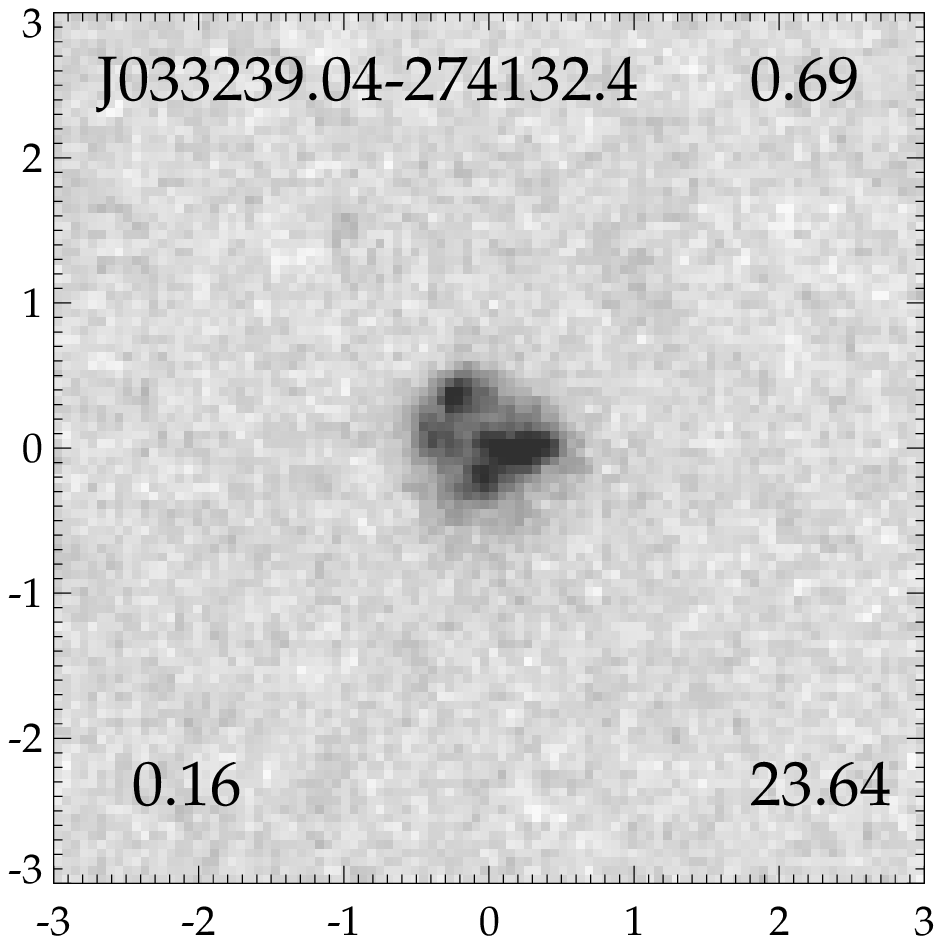}}%
\resizebox{0.31\textwidth}{!}{\includegraphics*[0cm,0cm][11cm,11cm]{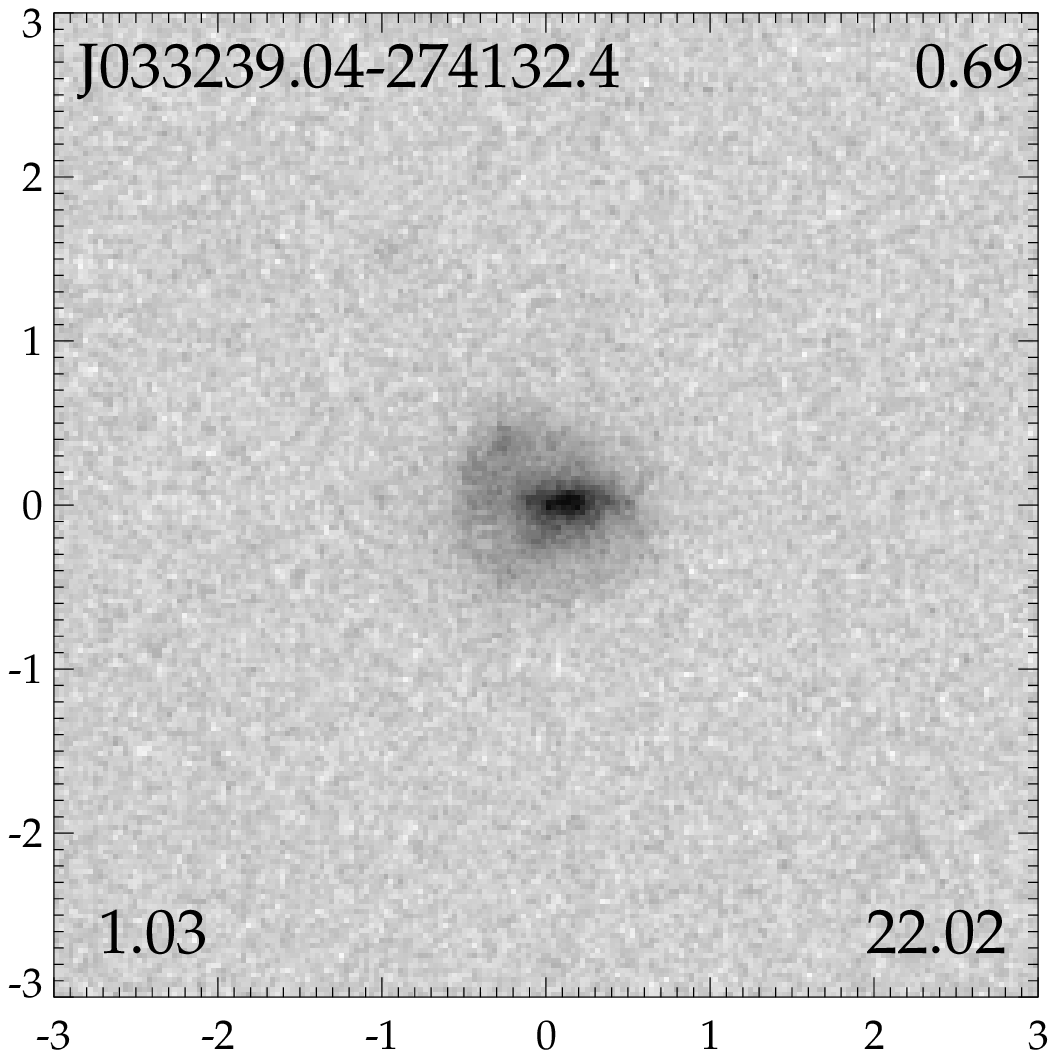}}%
\resizebox{0.4\textwidth}{!}{\includegraphics*[0cm,0.2cm][13.5cm,11cm]{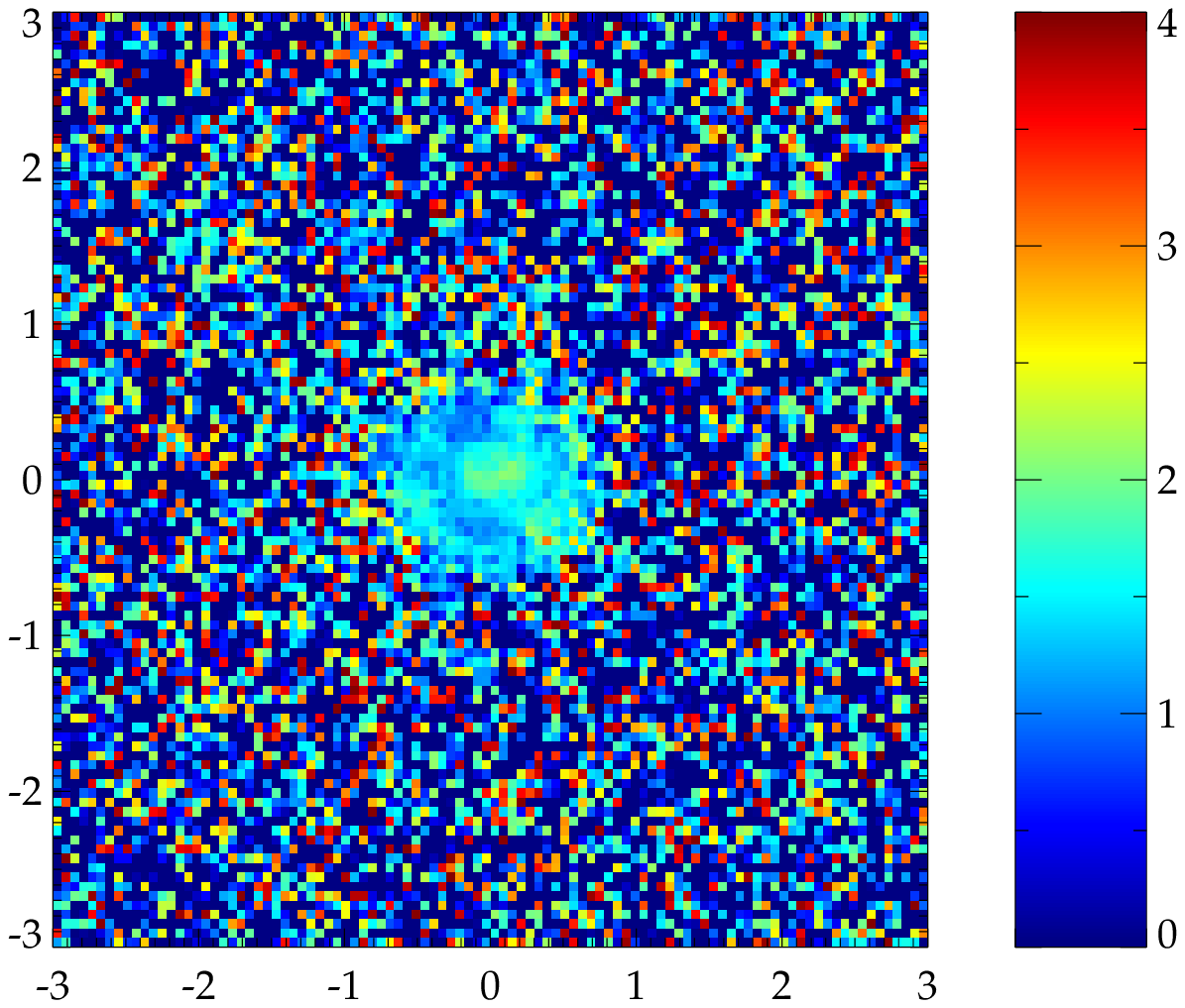}}

\resizebox{0.315\textwidth}{!}{\includegraphics*[0.6cm,0.6cm][10.5cm,10.5cm]{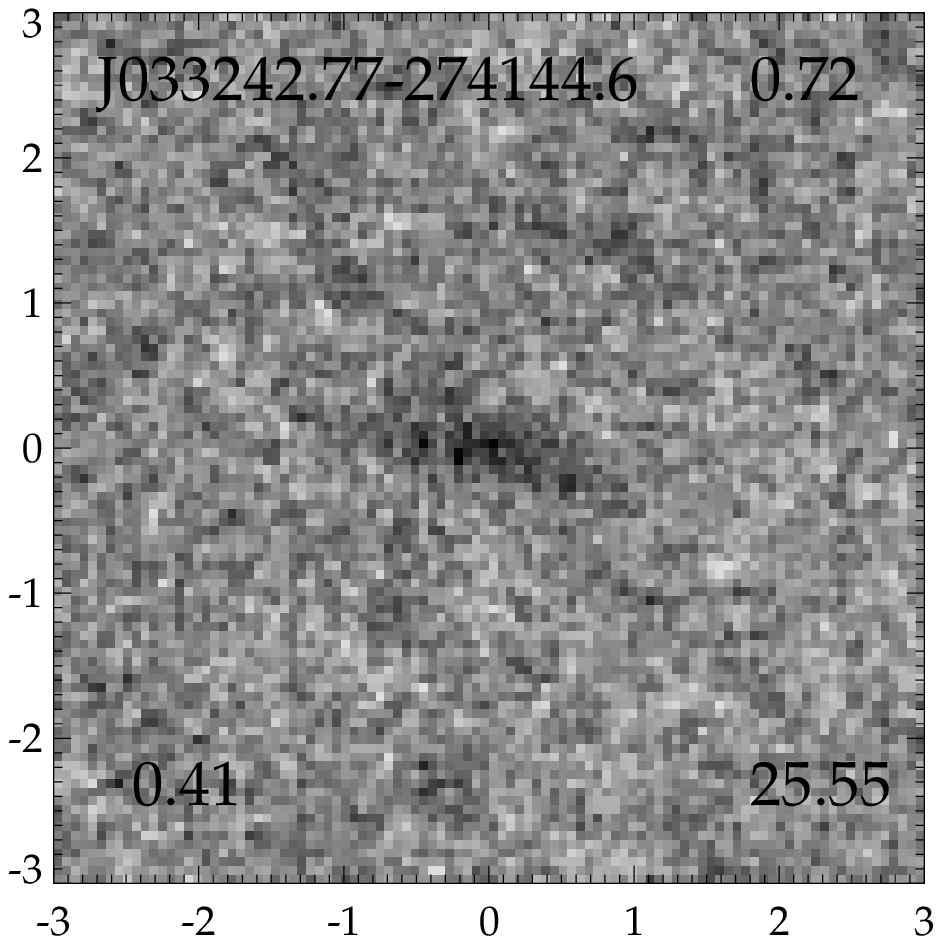}}%
\resizebox{0.31\textwidth}{!}{\includegraphics*[0cm,0cm][11cm,11cm]{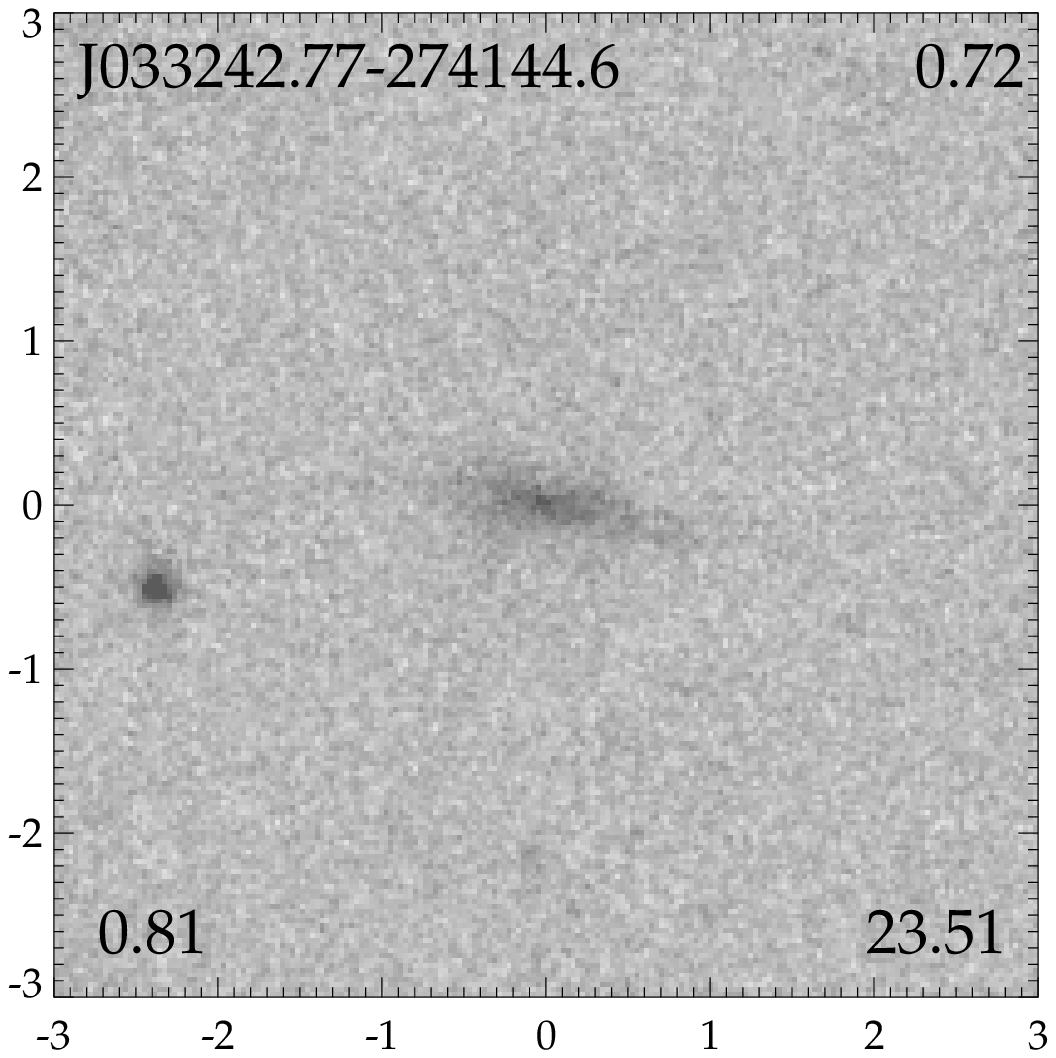}}%
\resizebox{0.4\textwidth}{!}{\includegraphics*[0cm,0.2cm][13.5cm,11cm]{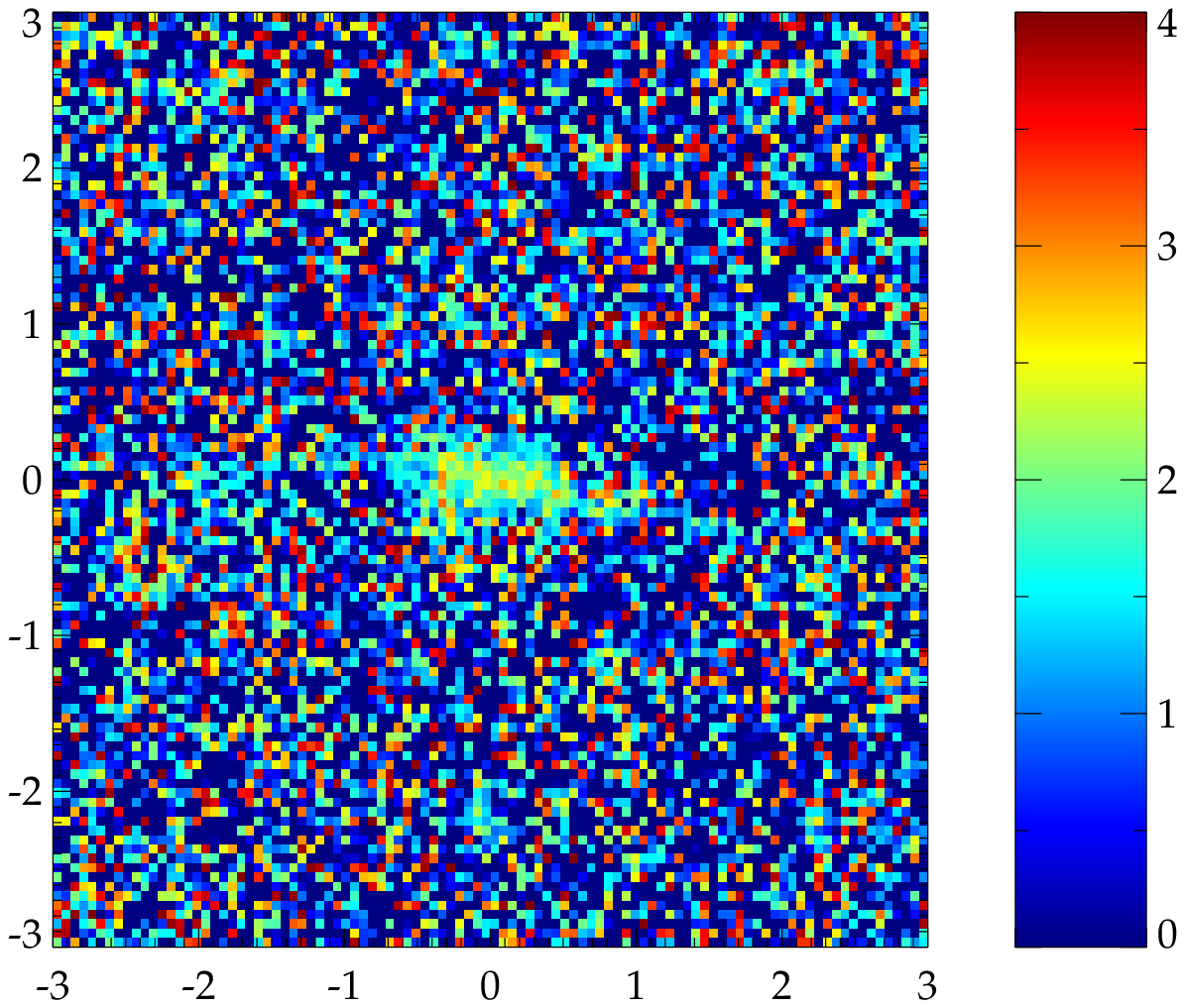}}

\resizebox{0.315\textwidth}{!}{\includegraphics*[0.6cm,0.6cm][10.5cm,10.5cm]{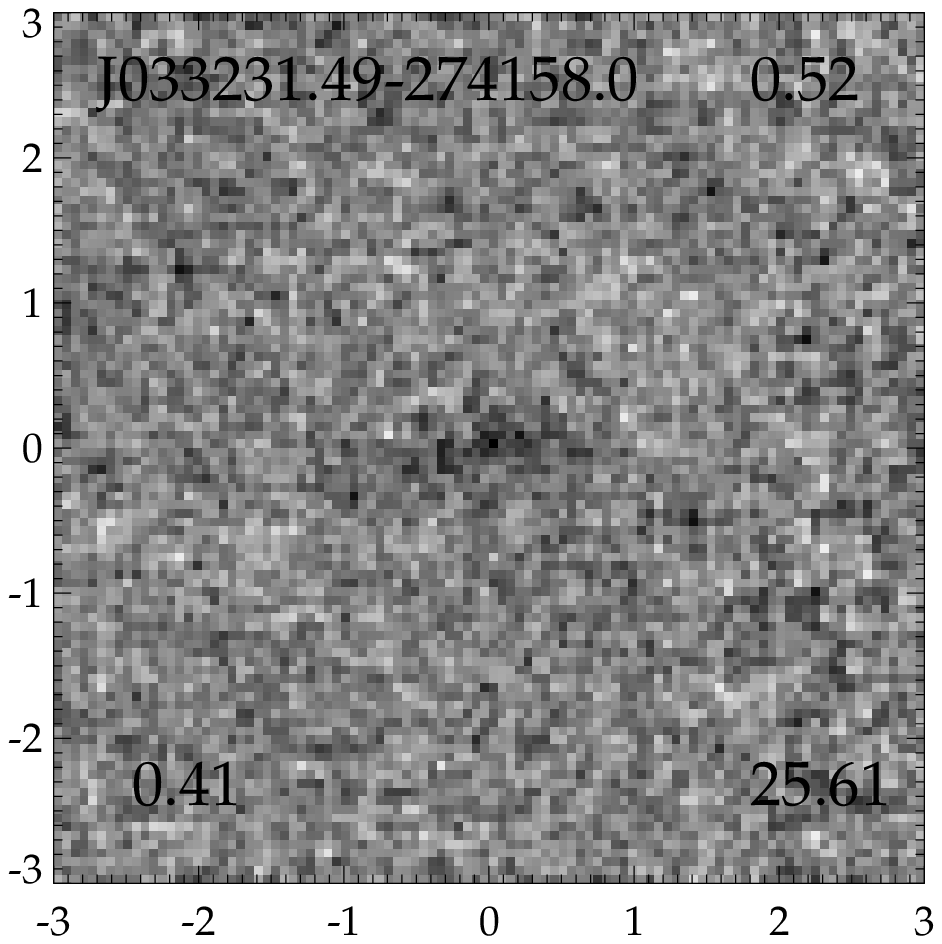}}%
\resizebox{0.31\textwidth}{!}{\includegraphics*[0cm,0cm][11cm,11cm]{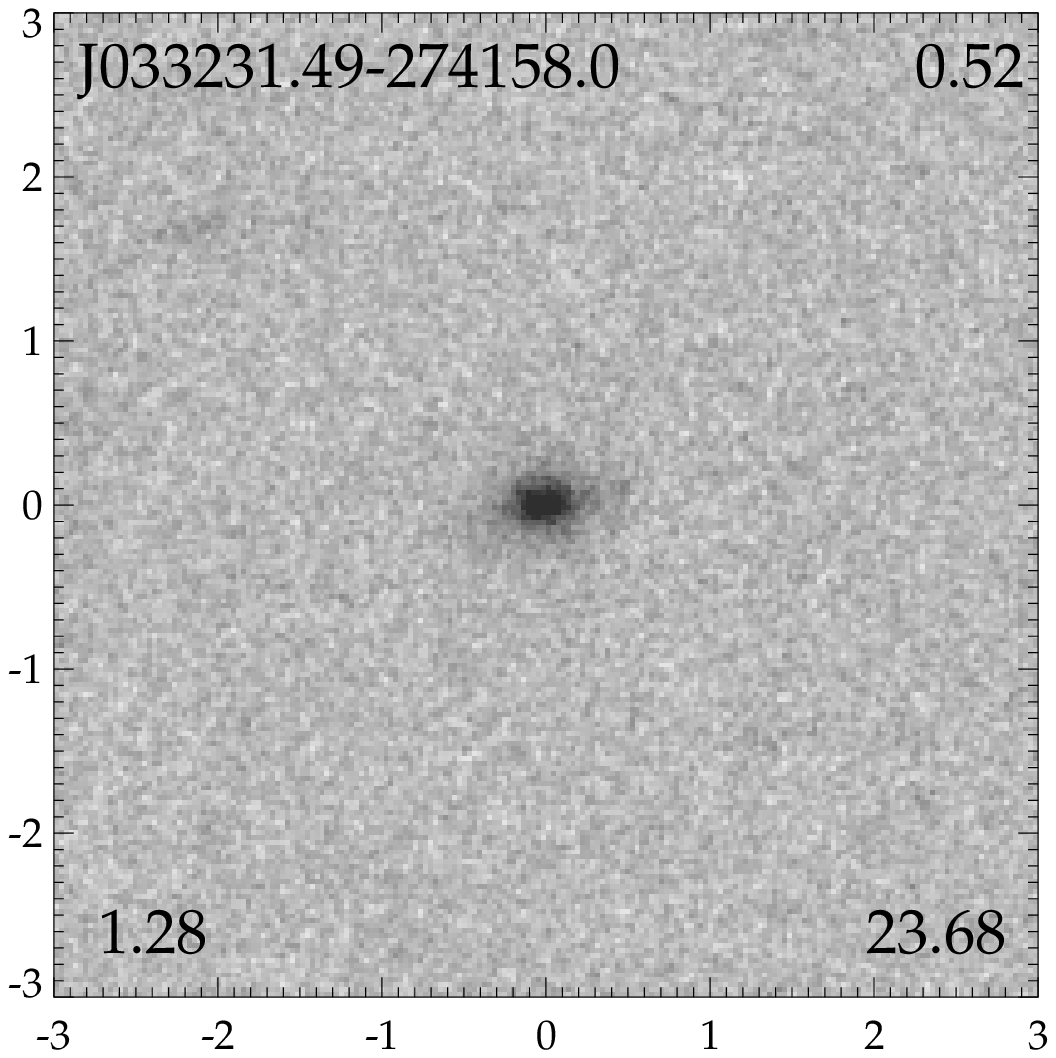}}%
\resizebox{0.4\textwidth}{!}{\includegraphics*[0cm,0.2cm][13.5cm,11cm]{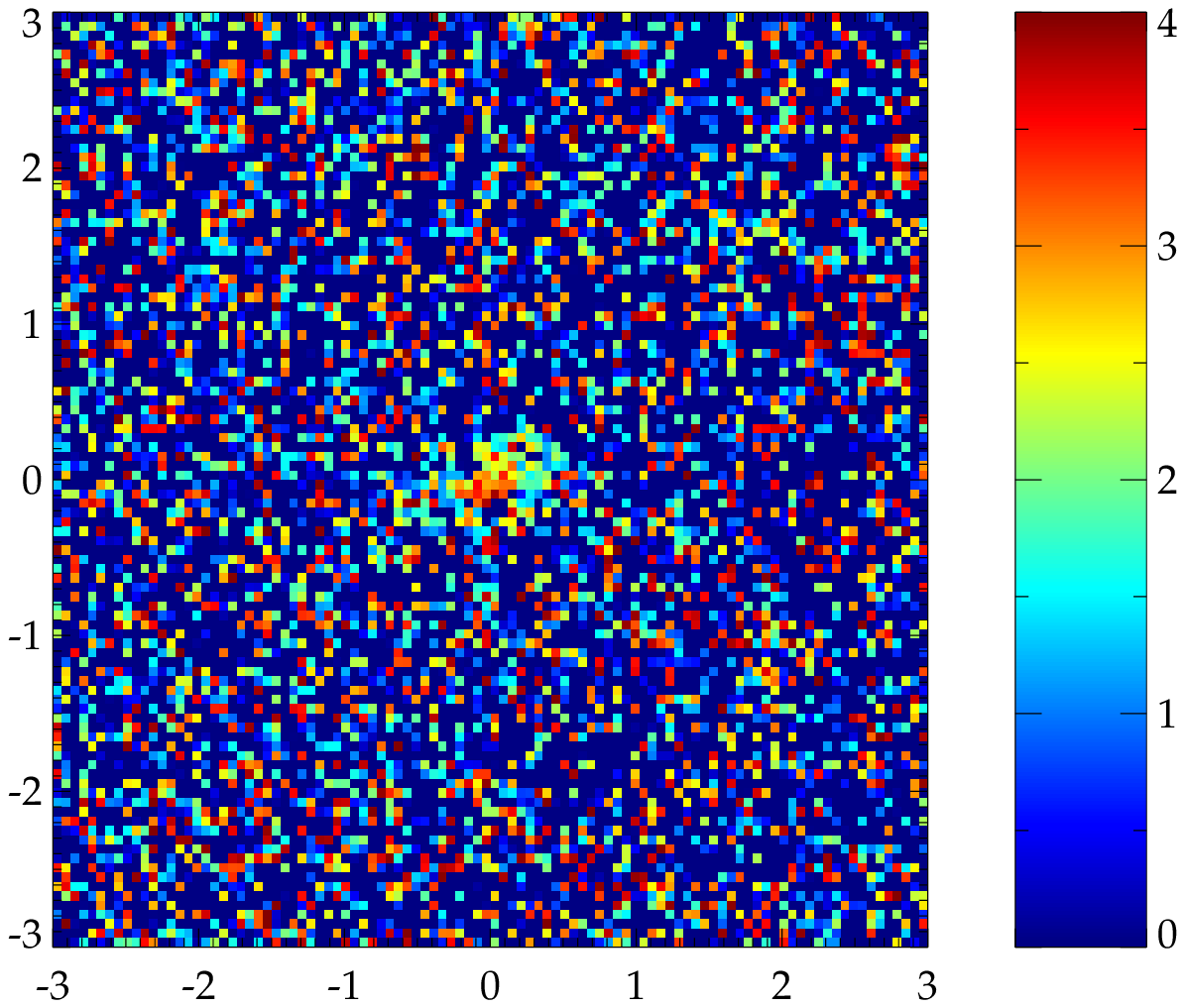}}

\resizebox{0.315\textwidth}{!}{\includegraphics*[0.6cm,0.6cm][10.5cm,10.5cm]{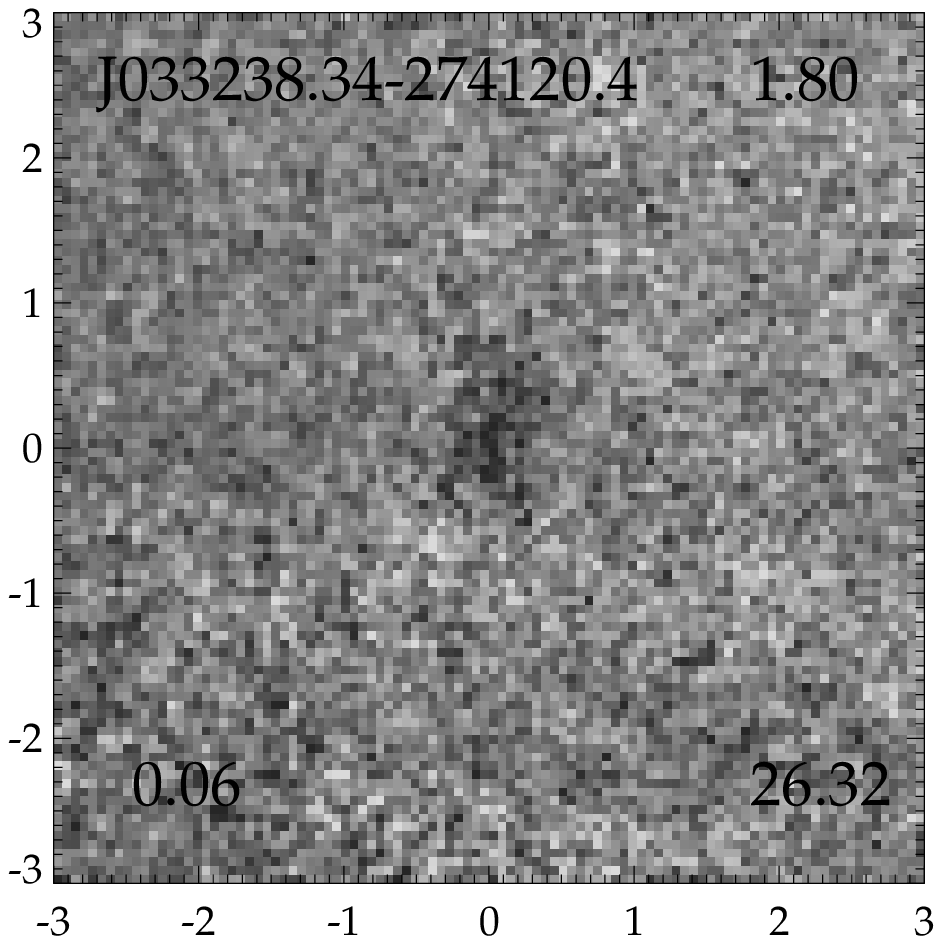}}%
\resizebox{0.31\textwidth}{!}{\includegraphics*[0cm,0cm][11cm,11cm]{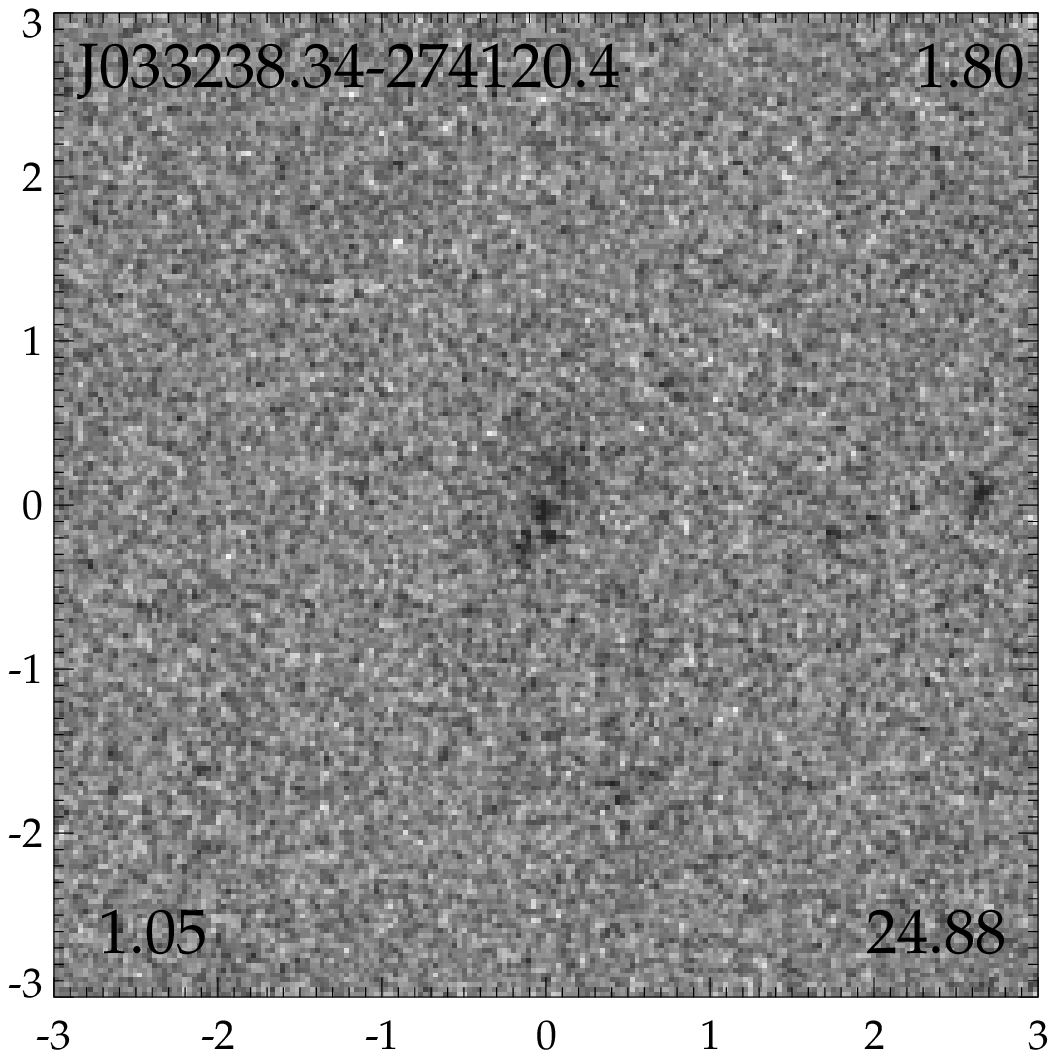}}%
\resizebox{0.4\textwidth}{!}{\includegraphics*[0cm,0.2cm][13.5cm,11cm]{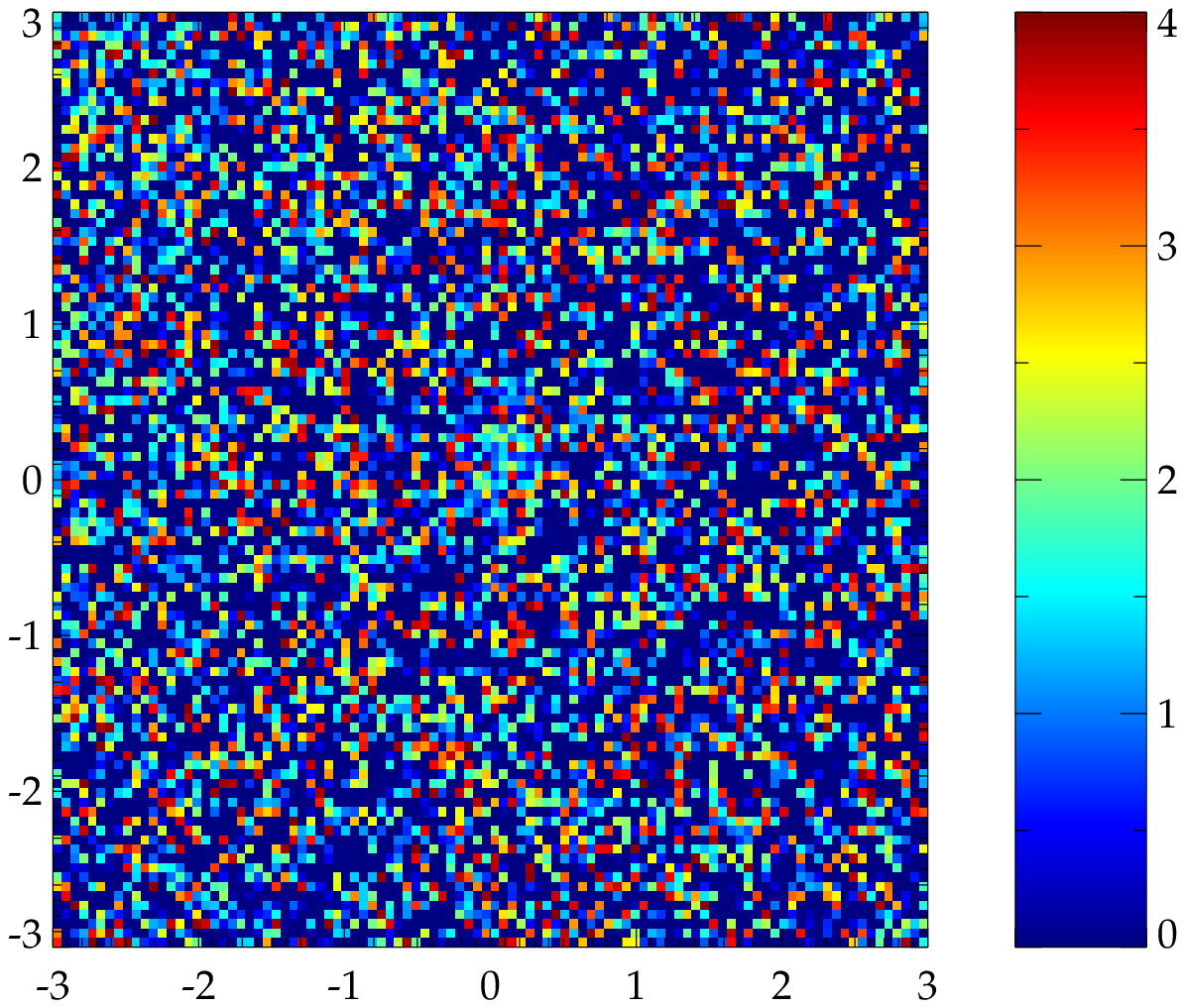}}

\caption{Continued.}   

\end{figure*}

\addtocounter{figure}{-1}

\begin{figure*}[t!] \centering

\resizebox{0.315\textwidth}{!}{\includegraphics*[0.6cm,0.6cm][10.5cm,10.5cm]{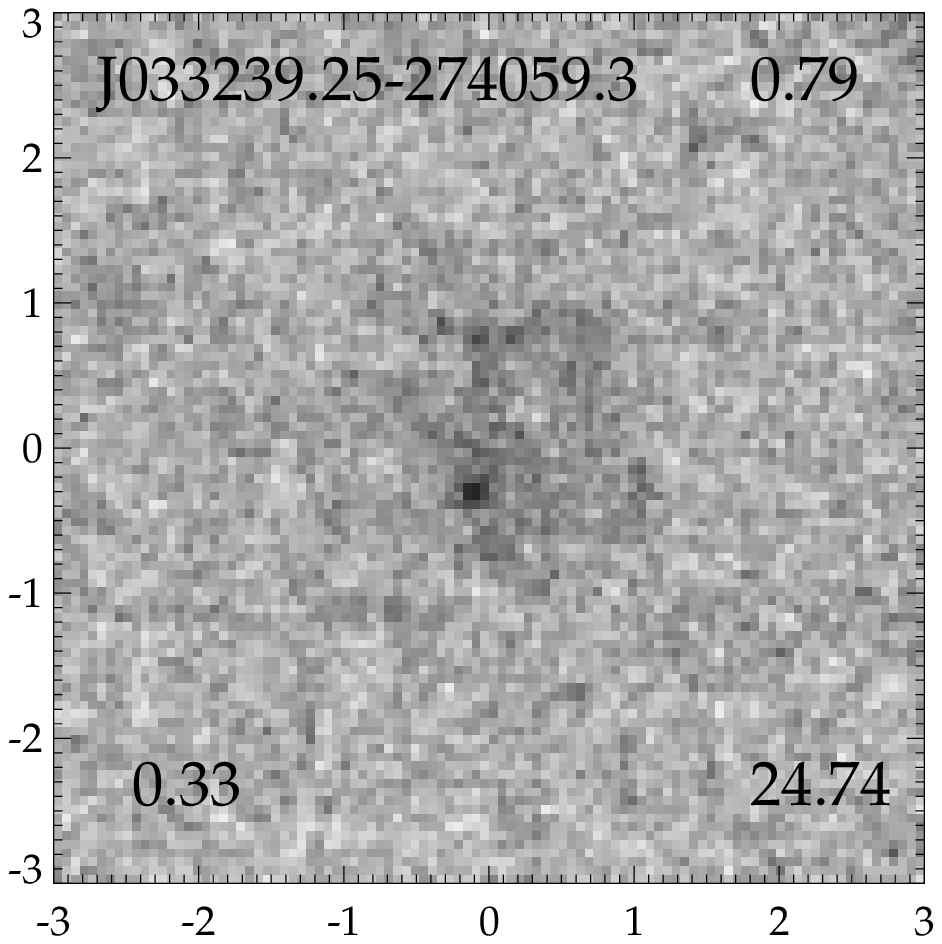}}%
\resizebox{0.31\textwidth}{!}{\includegraphics*[0cm,0cm][11cm,11cm]{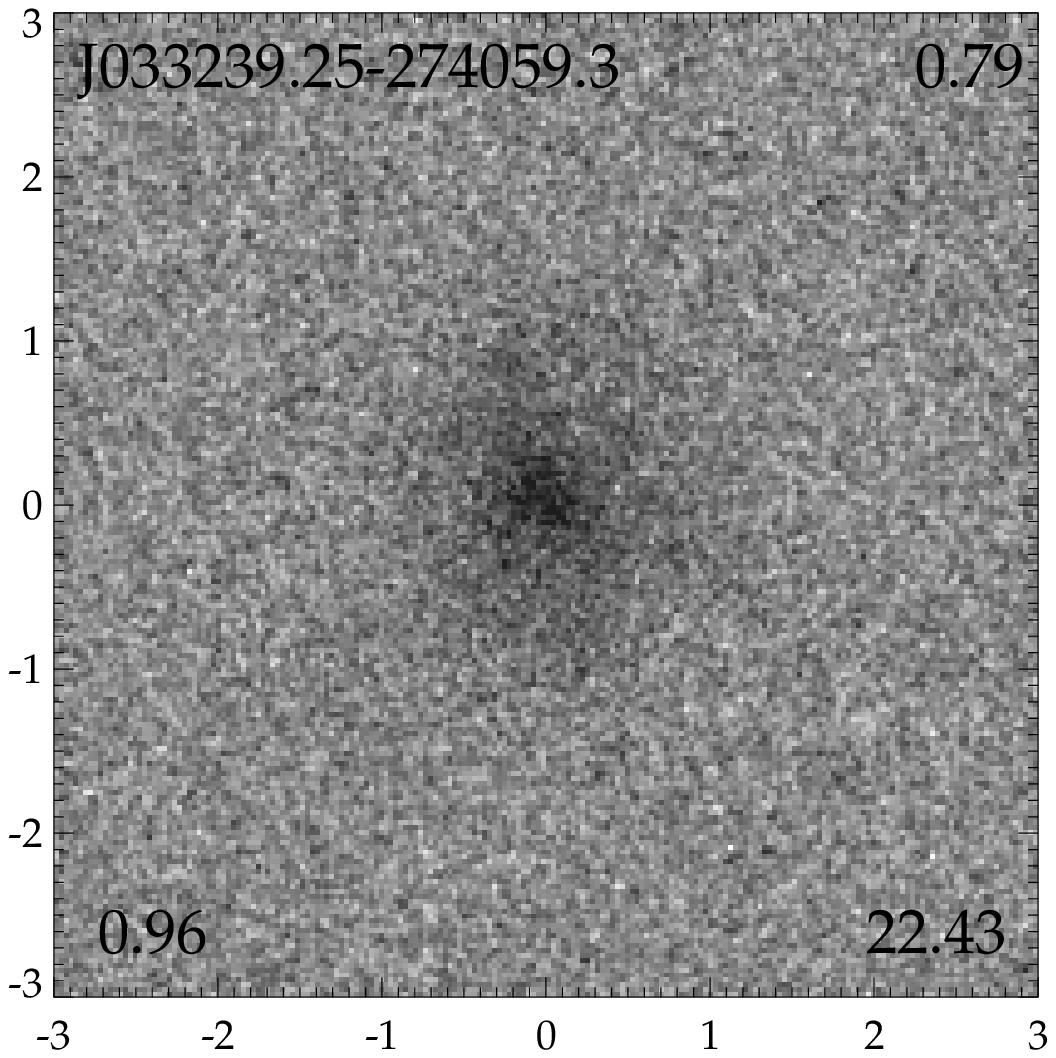}}%
\resizebox{0.4\textwidth}{!}{\includegraphics*[0cm,0.2cm][13.5cm,11cm]{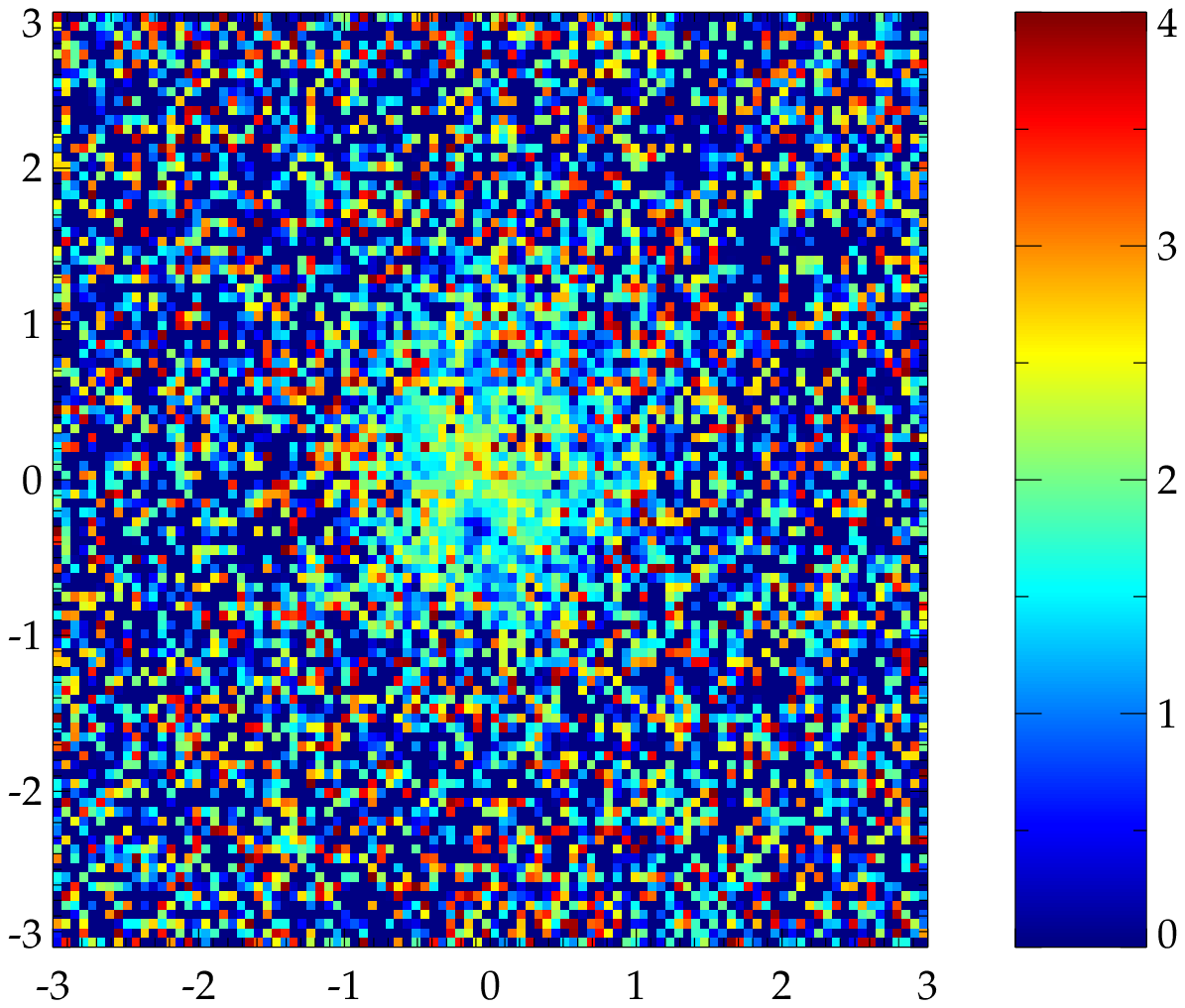}}

\resizebox{0.315\textwidth}{!}{\includegraphics*[0.6cm,0.6cm][10.5cm,10.5cm]{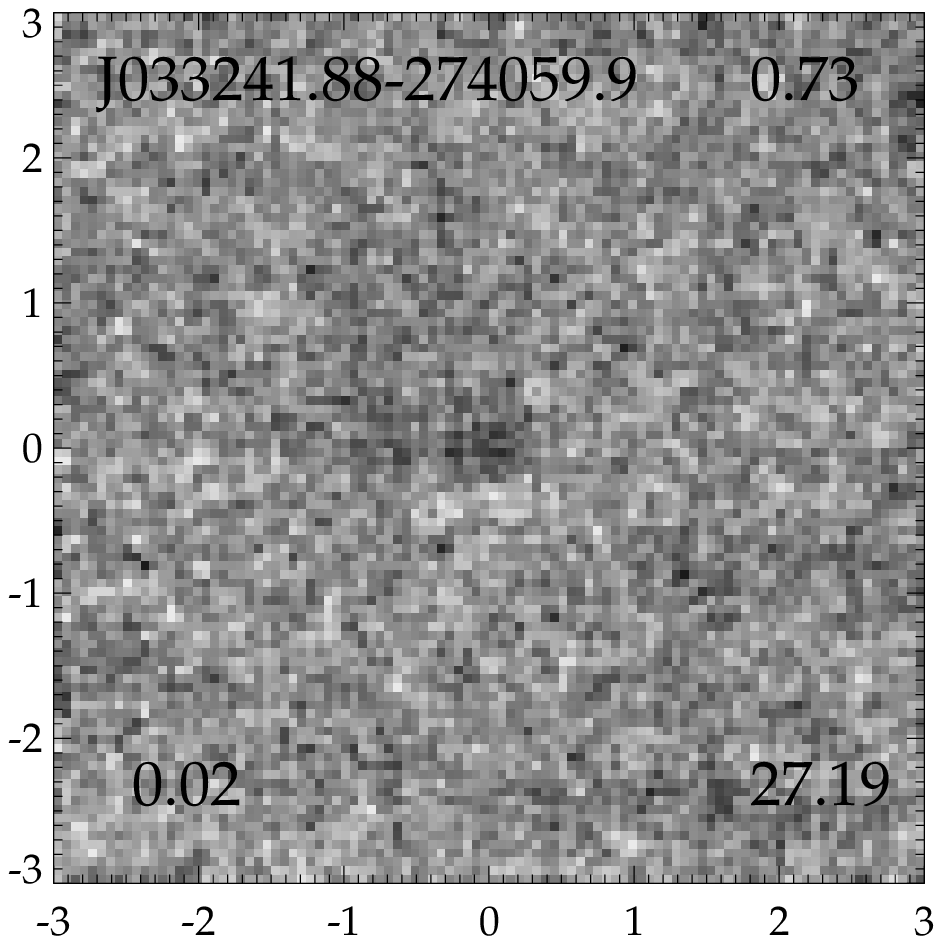}}%
\resizebox{0.31\textwidth}{!}{\includegraphics*[0cm,0cm][11cm,11cm]{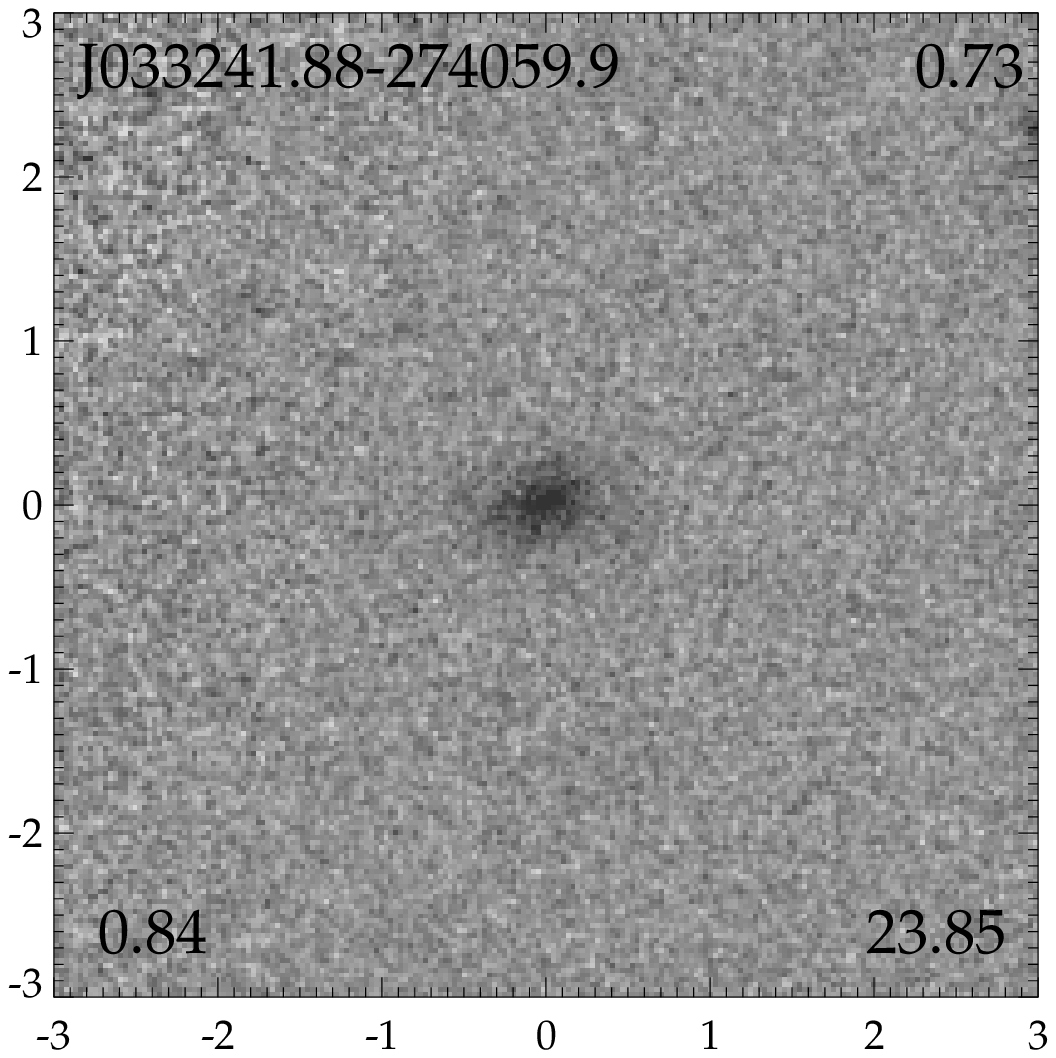}}%
\resizebox{0.4\textwidth}{!}{\includegraphics*[0cm,0.2cm][13.5cm,11cm]{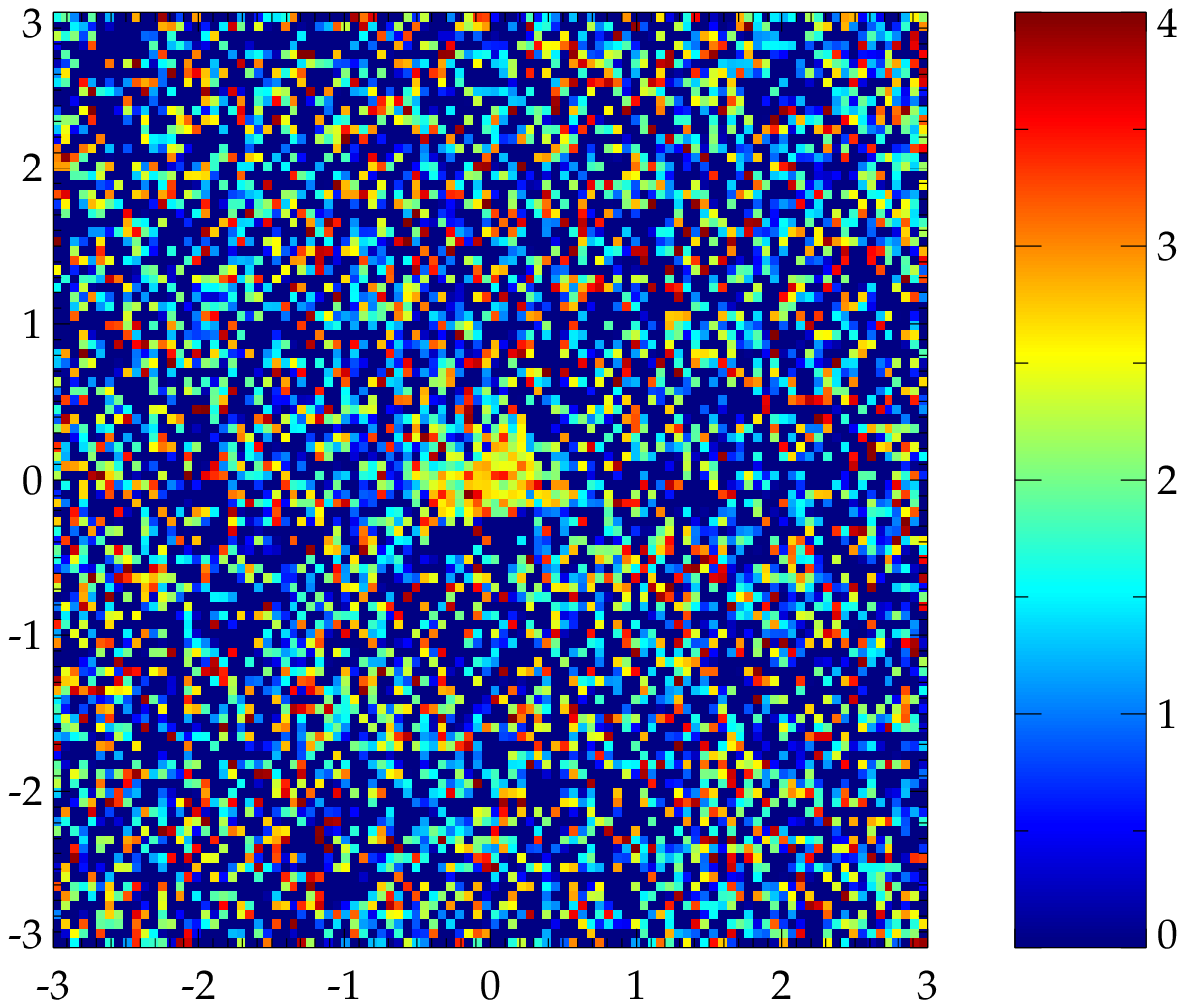}}

\resizebox{0.315\textwidth}{!}{\includegraphics*[0.6cm,0.6cm][10.5cm,10.5cm]{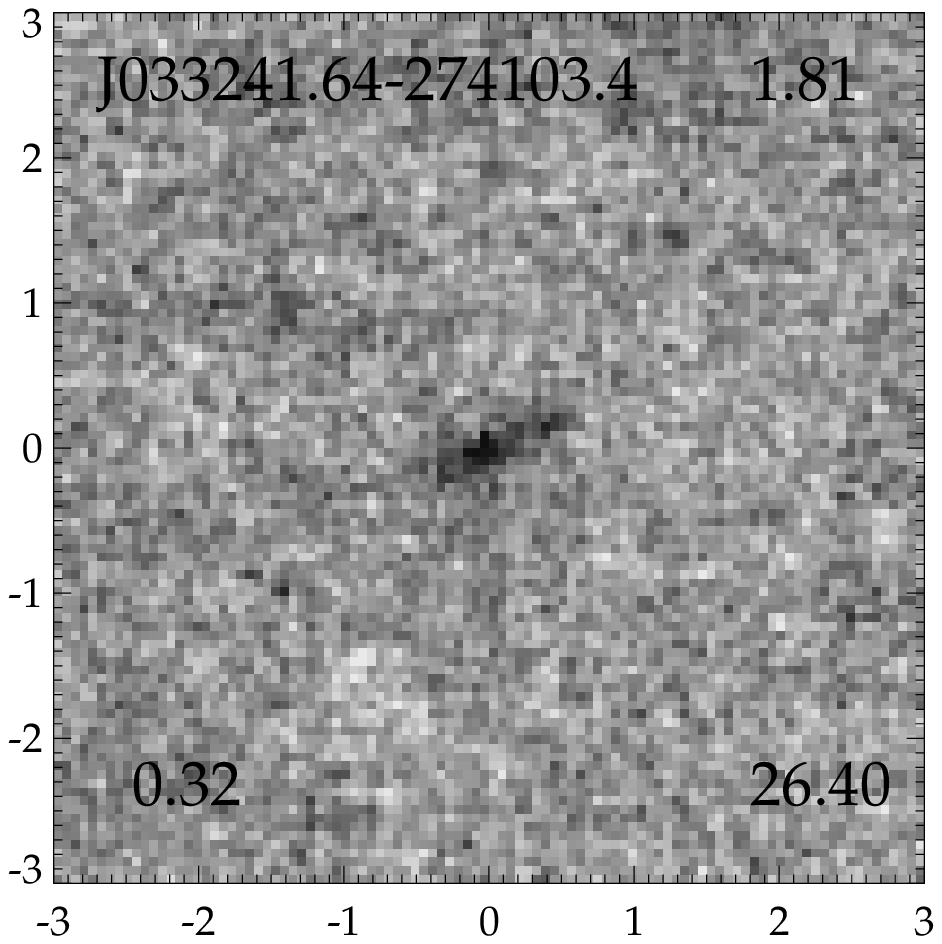}}%
\resizebox{0.31\textwidth}{!}{\includegraphics*[0cm,0cm][11cm,11cm]{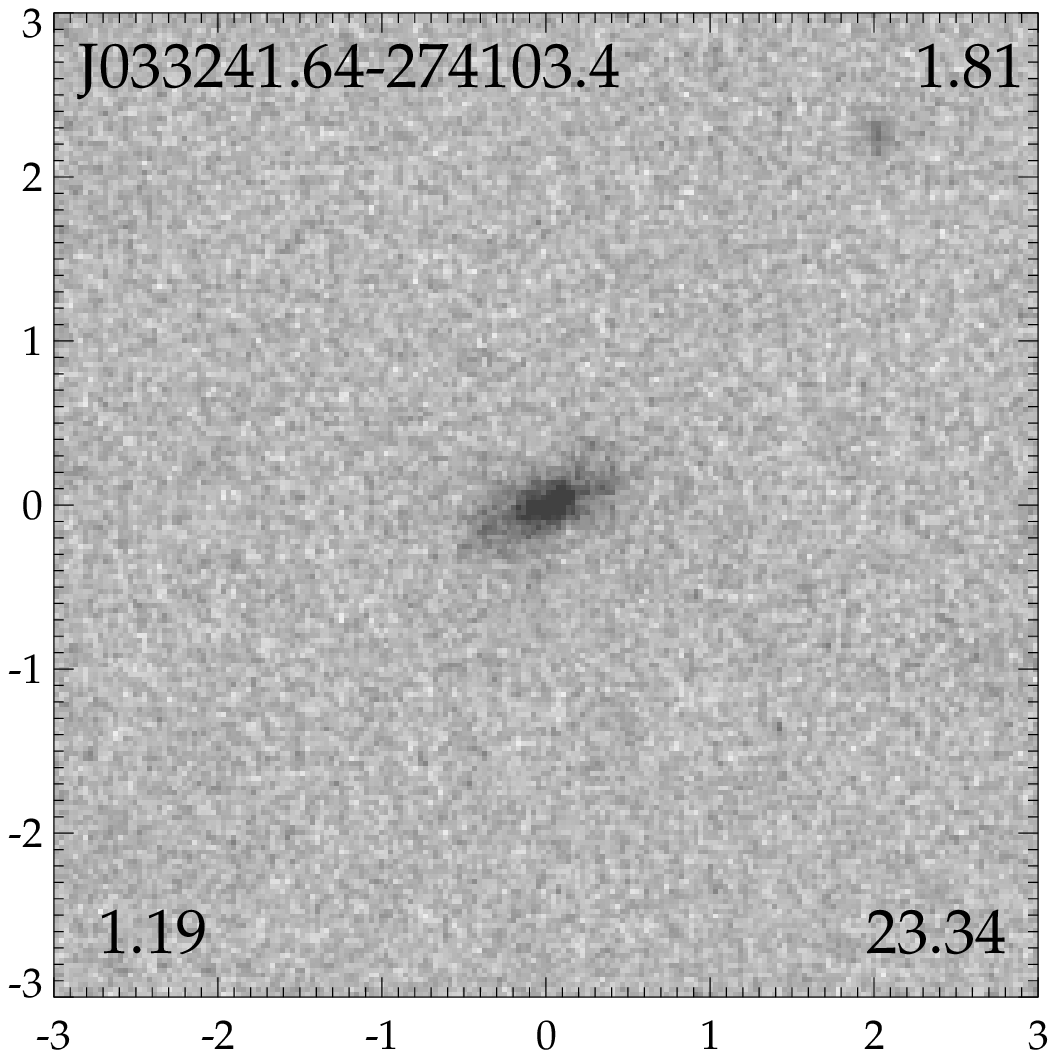}}%
\resizebox{0.4\textwidth}{!}{\includegraphics*[0cm,0.2cm][13.5cm,11cm]{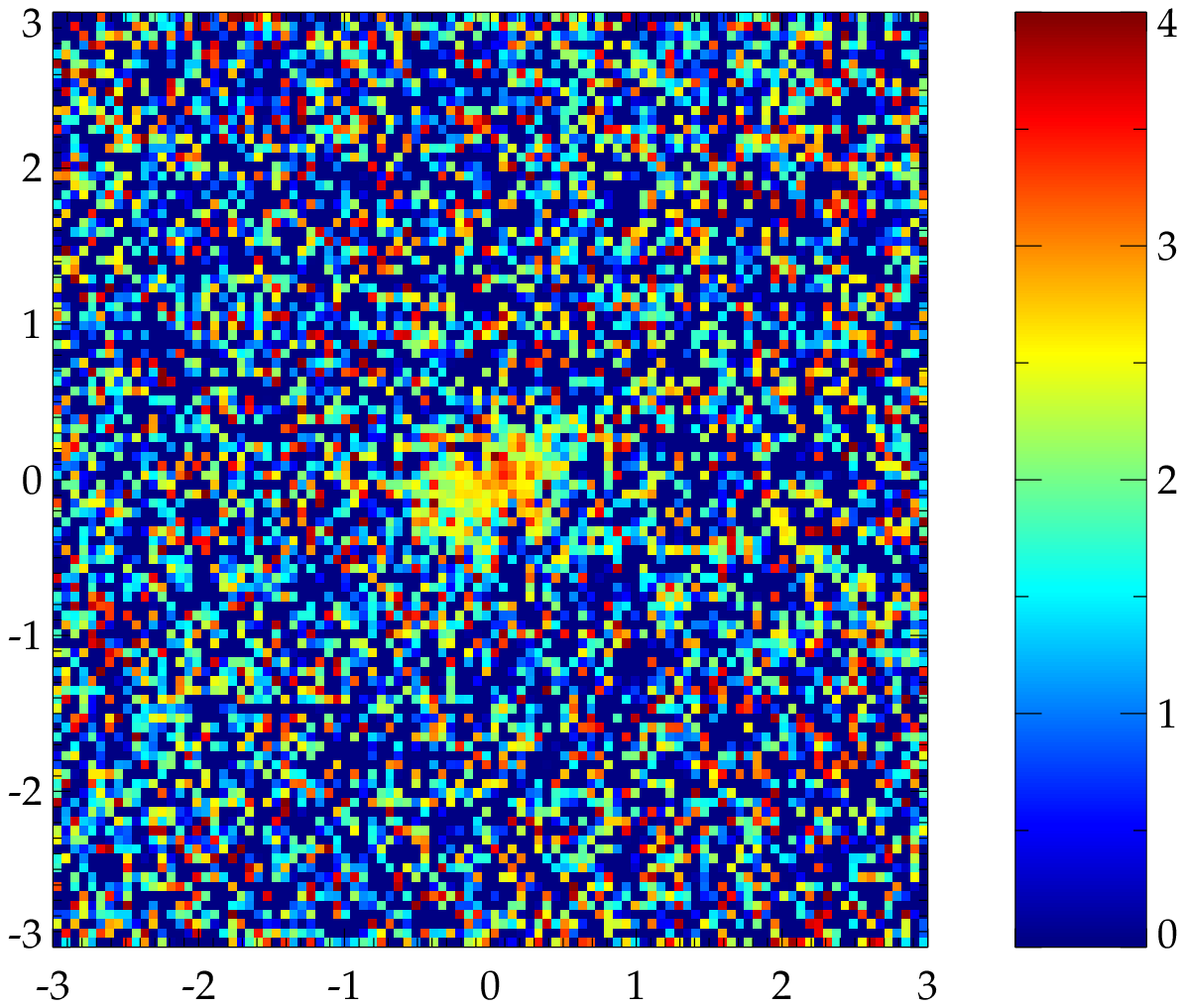}}

\resizebox{0.315\textwidth}{!}{\includegraphics*[0.6cm,0.6cm][10.5cm,10.5cm]{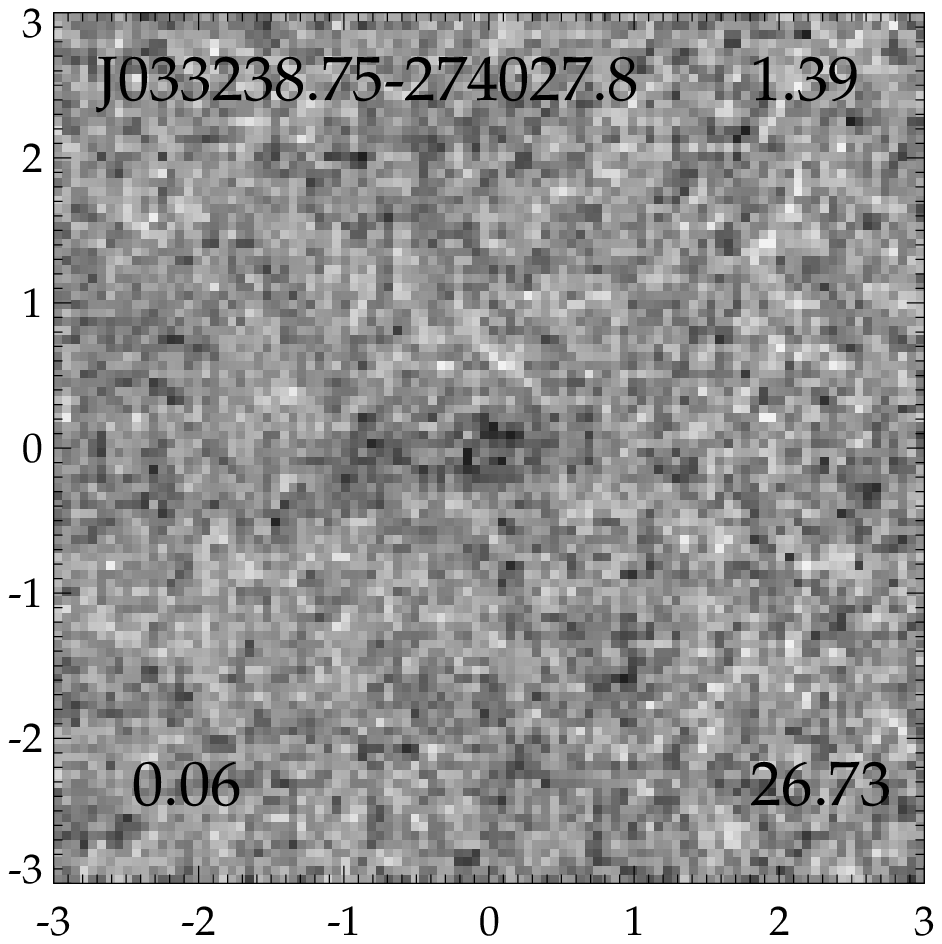}}%
\resizebox{0.31\textwidth}{!}{\includegraphics*[0cm,0cm][11cm,11cm]{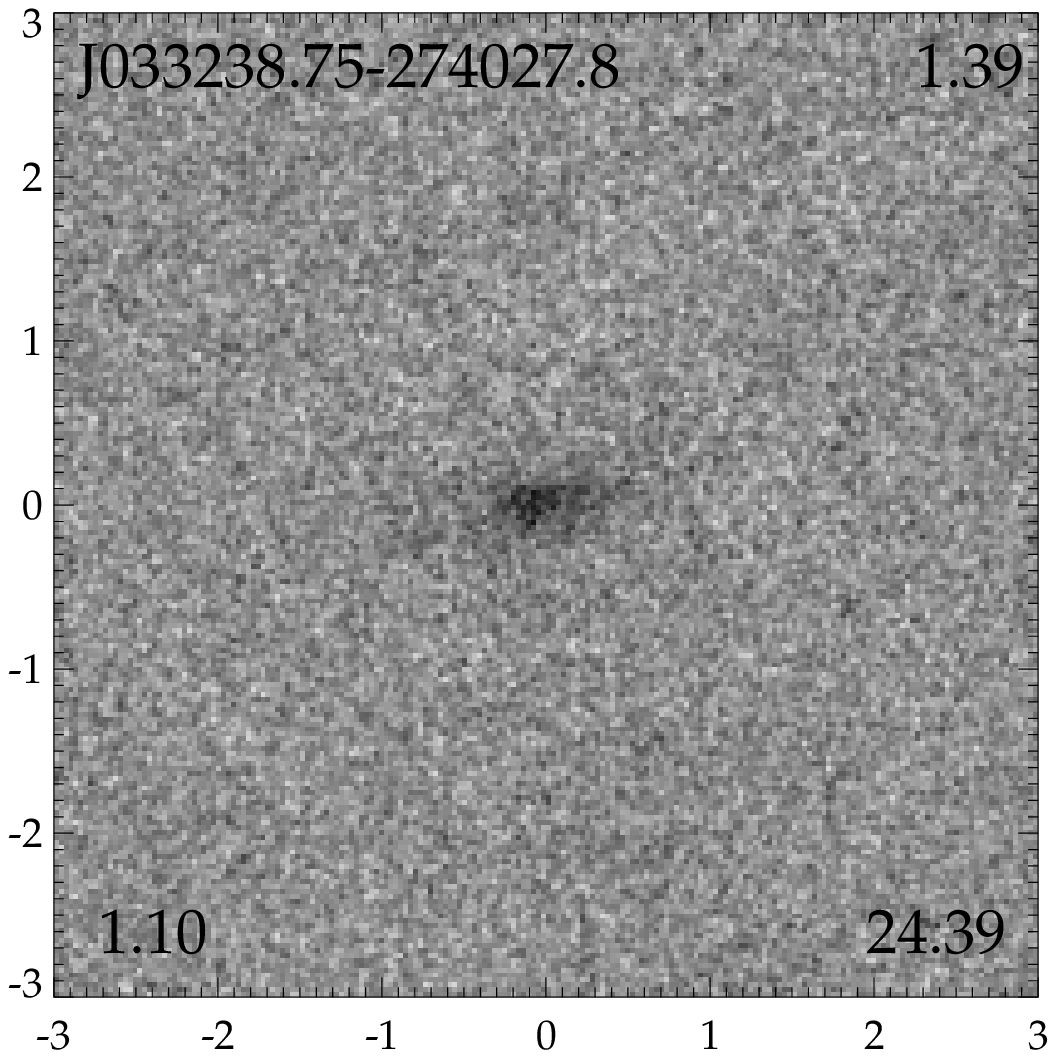}}%
\resizebox{0.4\textwidth}{!}{\includegraphics*[0cm,0.2cm][13.5cm,11cm]{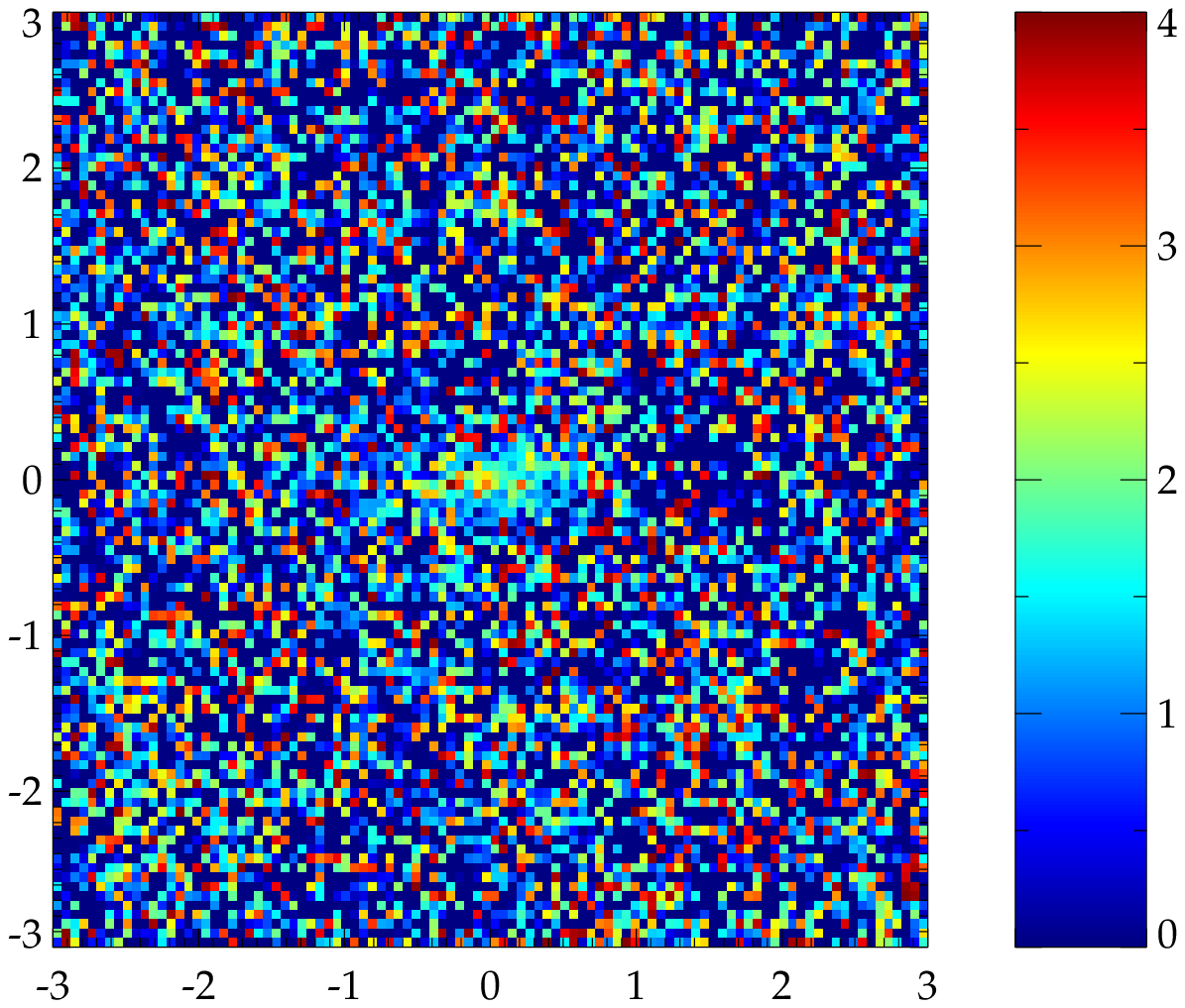}}

\caption{Continued.}   

\end{figure*}

\addtocounter{figure}{-1}

\begin{figure*}[t!] \centering

\resizebox{0.315\textwidth}{!}{\includegraphics*[0.6cm,0.6cm][10.5cm,10.5cm]{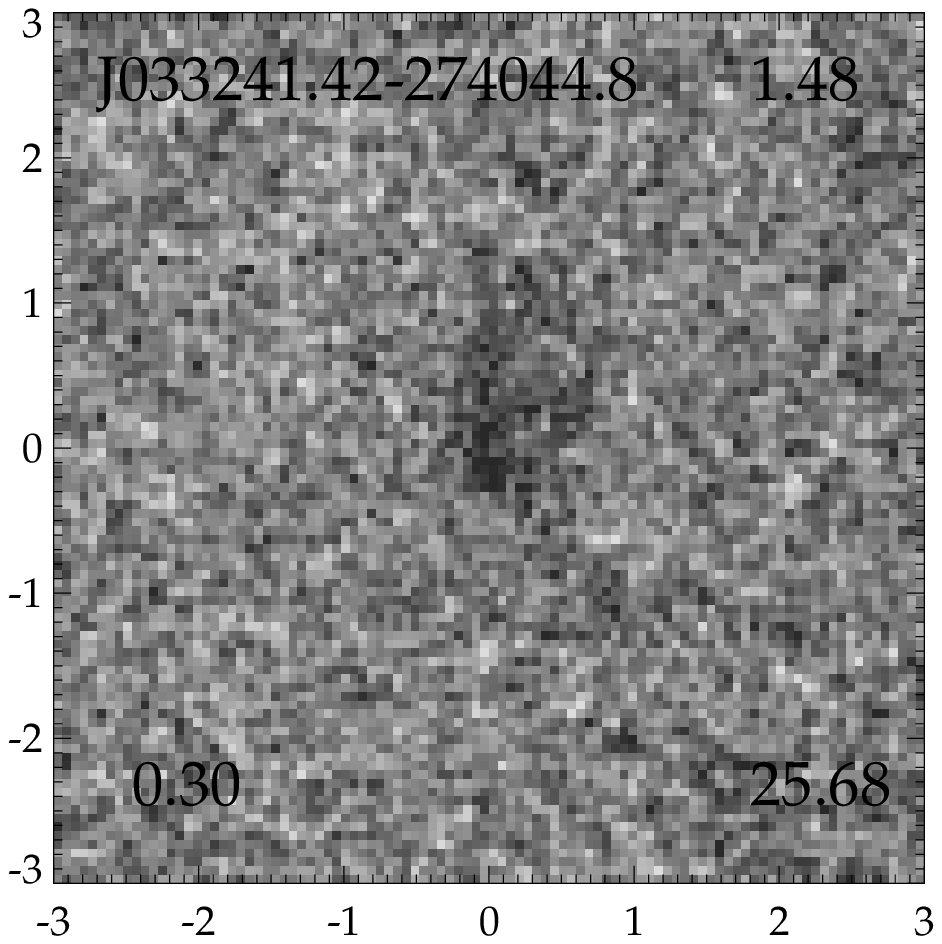}}%
\resizebox{0.31\textwidth}{!}{\includegraphics*[0cm,0cm][11cm,11cm]{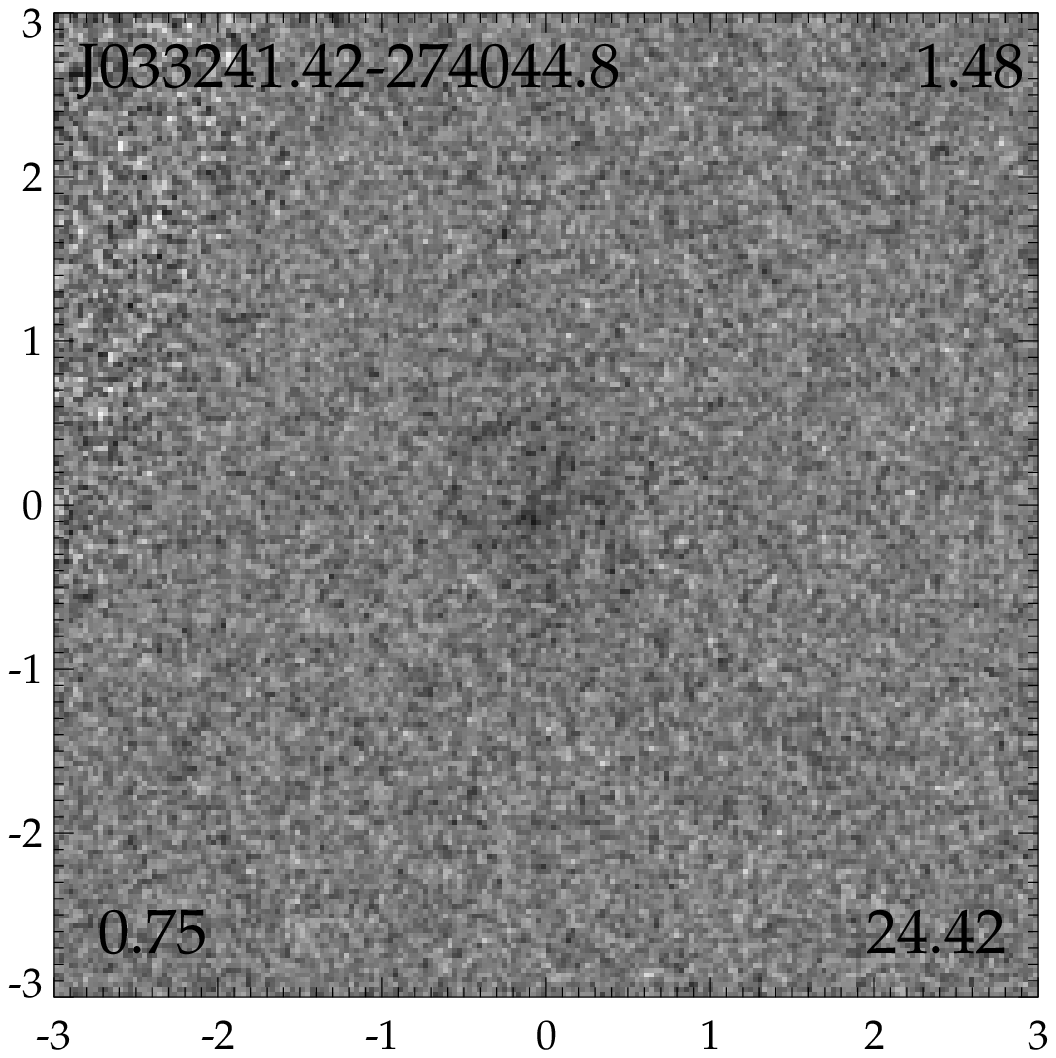}}%
\resizebox{0.4\textwidth}{!}{\includegraphics*[0cm,0.2cm][13.5cm,11cm]{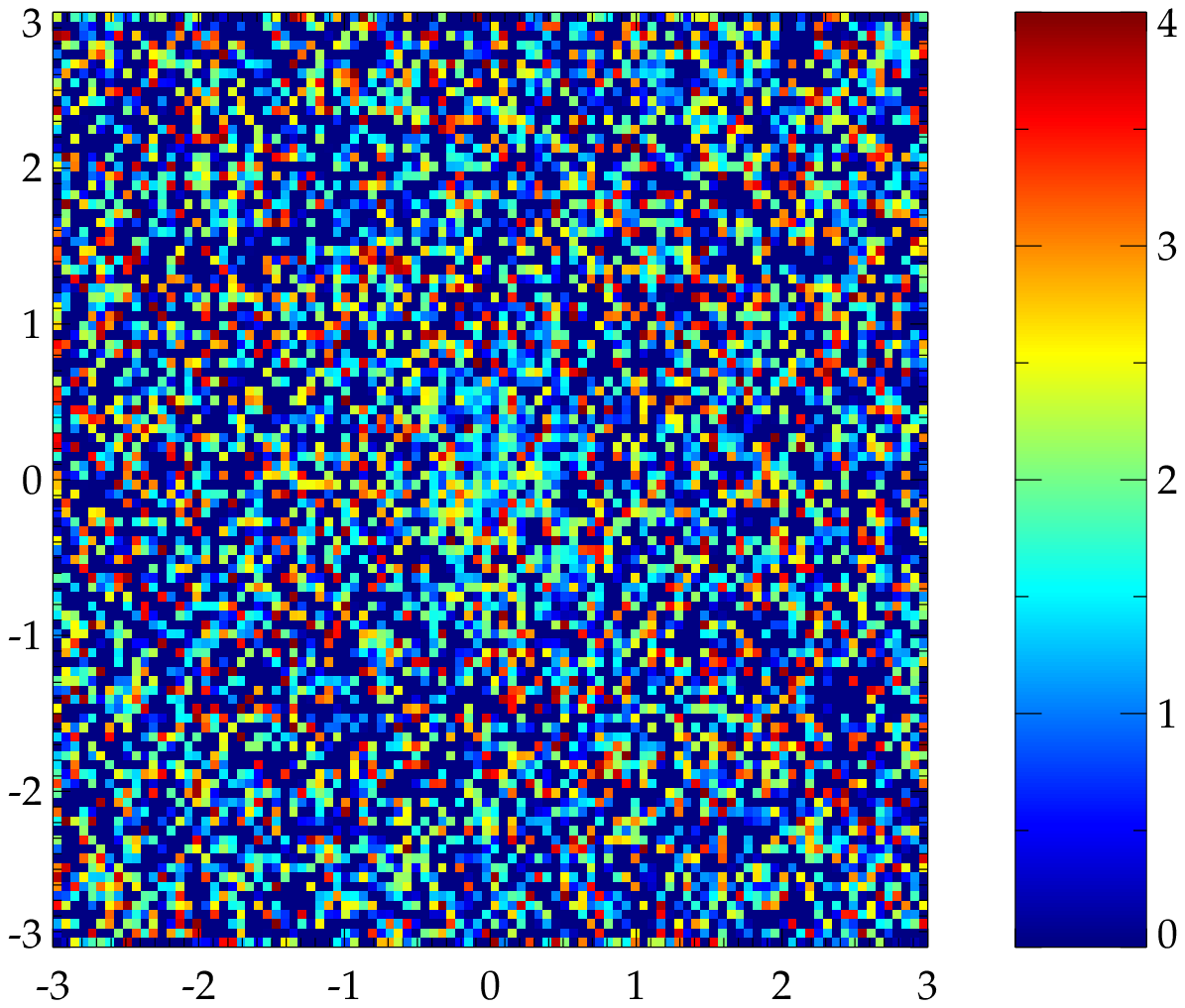}}

\caption{F300W, F850LP and the $u-z$ color map images of objects with $n(F300W)\leq0.5$ and $0.7\leq n(F850LP) \leq1.3$. In each row, the left cutout is a F300W filter HST/WFPC2 image from the Hubble-UDF
   parallels, the center cutout is the F850LP HST/ACS image of
   the same object from the GOODS survey, while the right cutout is the $u-z$ color map. Labels in the F300W and F850LP band cutouts are the same as in Fig.~\ref{fits}. Each image is 6 arcsec $\times$ 6 arcsec. North is up, East is to the left.}   

\end{figure*}

\clearpage

\end{document}